\documentclass[journal,12pt,onecolumn,draftclsnofoot,]{IEEEtran}
\usepackage{amsmath,amsfonts}
\usepackage{algorithmic}
\usepackage{array}
\usepackage[caption=false,font=normalsize,labelfont=sf,textfont=sf]{subfig}
\usepackage{textcomp}
\usepackage{stfloats}
\usepackage{url}
\usepackage{verbatim}
\usepackage{graphicx}
\usepackage{xcolor}
\usepackage{tipa}
\def\BibTeX{{\rm B\kern-.05em{\sc i\kern-.025em b}\kern-.08em
    T\kern-.1667em\lower.7ex\hbox{E}\kern-.125emX}}
\usepackage{balance}
\begin{document}
\title{A Tutorial on Clinical Speech AI Development: From Data Collection to Model Validation}
\author{Si-Ioi Ng, Lingfeng Xu, Ingo Siegert, Nicholas Cummins, Nina R. Benway, \\ Julie Liss, Visar Berisha
\thanks{Si-Ioi Ng, Lingfeng Xu, Julie Liss and Visar Berisha are with Arizona State University, Tempe, Arizona, USA. Ingo Siegert is with Otto von Guericke University, Magdeburg, Germany. Nicholas Cummins is with King’s College London, United Kingdom. Nina R. Benway is with University of Maryland, College Park, MD, USA}}

\markboth{Journal of \LaTeX\ Class Files,~Vol.~18, No.~9, September~2020}%
{How to Use the IEEEtran \LaTeX \ Templates}

\maketitle
\vspace{-3.5mm}
\begin{abstract}
There has been a surge of interest in leveraging speech as a marker of health for a wide spectrum of conditions. The underlying premise is that any neurological, mental, or physical deficits that impact speech production can be objectively assessed via automated analysis of speech. Recent advances in speech-based Artificial Intelligence (AI) models for diagnosing and tracking mental health, cognitive, and motor disorders often use supervised learning, similar to mainstream speech technologies like recognition and verification. However, clinical speech AI has distinct challenges, including the need for specific elicitation tasks, small available datasets, diverse speech representations, and uncertain diagnostic labels. As a result, application of the standard supervised learning paradigm may lead to models that perform well in controlled settings but fail to generalize in real-world clinical deployments.
With translation into real-world clinical scenarios in mind, this tutorial paper provides an overview of the key components required for robust development of clinical speech AI. 
Specifically, this paper will cover the design of speech elicitation tasks and protocols most appropriate for different clinical conditions, collection of data and verification of hardware, development and validation of speech representations designed to measure clinical constructs of interest, development of reliable and robust clinical prediction models, and ethical and participant considerations for clinical speech AI. The goal is to provide comprehensive guidance on building models whose inputs and outputs link to the more interpretable and clinically meaningful aspects of speech, that can be interrogated and clinically validated on clinical datasets, and that adhere to ethical, privacy, and security considerations by design. 
\end{abstract}


\section{Introduction}
Speaking is a deceptively complicated and sensitive activity. We must think of the words to convey our message, organize the words in compliance with the rules of a language, and activate the muscles that allow us to produce clear and understandable speech. This process requires coordination across multiple regions of the brain and precise activation of more than 100 muscles. If there is a disturbance to any of these regions of the brain or the physical apparatus itself, it becomes apparent in the resulting speech. 
The potential for extracting clinically-rich information from such an easy-to-acquire signal has generated considerable excitement in the digital health and wellness communities, and speech has been referred to as the newest vital sign \cite{vikram2022digital}. The promise is that any neurological, mental health, or physical disturbances that impact the speech production process can be passively detected from patients’ speech patterns. To that end, there has been considerable interest in different academic communities and in industry in developing AI models for diagnosis, prognosis, and tracking of different clinical conditions using only speech (e.g. see \cite{faurholt2016voice, rapcan2010acoustic} for mental health, \cite{luz2021detecting, vasquez2018multimodal} for cognition, \cite{quintas2024automatic, stegmann2023speech, martens2015automated} for nueromuscular disease, \cite{cummins2015review} for emotion recognition, \cite{benway2024artificial, ng2024ssd} for speech sound disorder, etc.).

 \begin{figure}[h!]
  \setlength\belowcaptionskip{-0.8\baselineskip}
  \centering
  \includegraphics[width=0.90\linewidth]{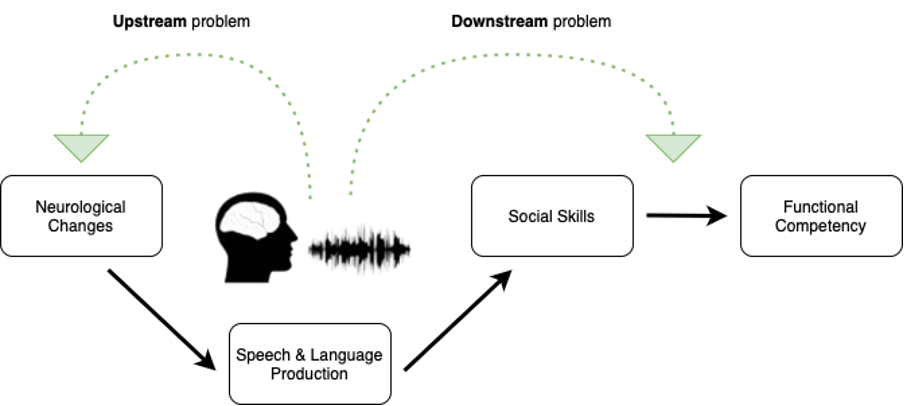}
  \caption{Central role of speech and language in neurology.} 
  \label{fig:system_design}
\end{figure}

 Speech has a dual role in neurology applications (Figure~\ref{fig:system_design}), capturing both upstream information, such as the presence and prognosis of neurological changes, and downstream consequences, like the impact on quality of life and social interactions \cite{voleti2023language}. In Parkinson’s disease, for example, speech AI is used not only to detect pre-clinical symptoms (upstream) but also to assess how dysarthria affects social communication (downstream). Similarly, in schizophrenia, speech AI detects early signs of psychotic episodes and evaluates their effects on social skills and functional capacity \cite{bedi2015automated, voleti2023language}.

In the development of speech-based clinical solutions, supervised AI on diagnostic outcomes has played a central role. Survey articles on dementia \cite{de2020artificial} and Parkinson’s disease \cite{moro2021advances} show that most clinical speech AI approaches involve starting with a labeled dataset of paired speech samples and diagnostic labels, extracting high-dimensional (often clinically uninterpretable) features, and building predictive models based on these labels \cite{deng2013machine}. However, labeled clinical datasets are much smaller and more variable than the large datasets used for consumer-grade automatic speech recognition (ASR) models. For instance, studies on speech-based dementia classification often work with datasets spanning only tens to hundreds of minutes of speech, in stark contrast to the large-scale ASR datasets \cite{shi2023speech, ardila-etal-2020-common, panayotov2015librispeech}.

\begin{figure}[h!]
  \setlength\belowcaptionskip{-0.8\baselineskip}
  \centering
  \includegraphics[width=0.55\linewidth]{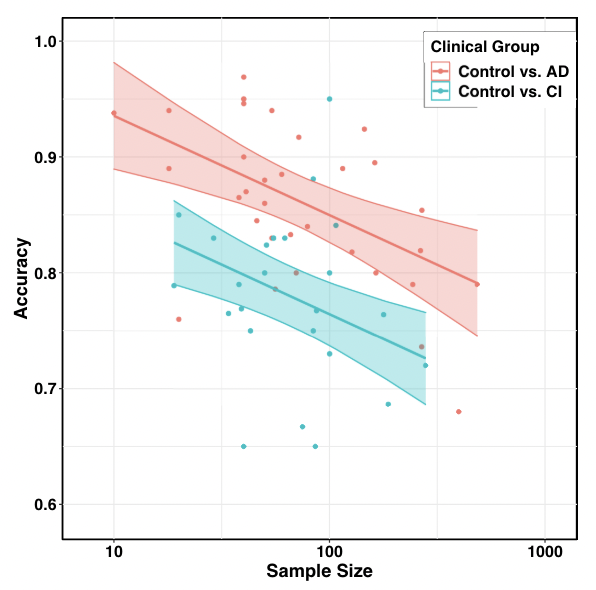}
  \caption{The negative association between reported accuracy and sample size for two types of classification problems across three meta-analyses. AD = Alzheimer’s Disease; CI = Other forms of cognitive impairment.} 
  \label{fig:negative trend}
\end{figure}

Recently, researchers have started focusing on the reliability of the supervised AI approach to predict diagnostic labels. For example, research has shown that AI models for disease prediction in fields like neuroimaging often report inflated accuracies that correlate negatively with sample size, suggesting overfitting and potential issues with generalizability \cite{arbabshirani2017single, vabalas2019machine}. Our own analysis of 59 speech-based classifiers used for detecting Alzheimer’s disease revealed similar concerns \cite{berisha22_interspeech}, with negative correlations between reported accuracy and dataset size (Fig.~\ref{fig:negative trend}), contrary to typical learning curve expectations.  For correctly-trained AI models, as the sample size increases, the accuracy of the model should increase monotonically according to a power law model \cite{viering2022shape}.

We attribute these inconsistencies to several factors, primarily data leakage and variability in model accuracy estimates due to analytic flexibility. Data leakage, where models unintentionally access information from the test set during training, is a significant concern \cite{kapoor2023leakage}. Importantly, data leakage can also occur through repeated use of the same test data over time, even without directly training on it—models may gradually become fine-tuned to the test set through iterative experimentation and evaluation. Additionally, the flexibility available to model developers can lead to different outcomes even when working with the same dataset. For instance, a recent study \cite{coretta2023multidimensional} involving 46 research teams, each given the same dataset and research question, resulted in widely divergent models and performance estimates. This variability, combined with the tendency to publish only high-performing models, contributes to systemic overoptimism in our field \cite{berisha22_interspeech}. In short, this evidence points to systemic overoptimism about how well popular approaches to clinical speech AI model development actually work. 

Motivated by the growing interest in this field and the increasing evidence that data-driven approaches successful in consumer speech AI applications have not yielded translational success in clinical speech AI, we present this tutorial article. We argue that straightforward application of supervised AI tools from other areas is unlikely to yield the same benefits here due to the complexity and diversity of the problem space and data limitations. In light of these constraints, our aim is to provide an interdisciplinary overview tailored for those interested in developing and evaluating clinical speech models.

We begin with a general overview of the human speech production mechanism, serving as a guide for understanding the clinically relevant information available in the speech signal. We include examples from patients with motor disorders (e.g., Parkinson’s disease, ALS), cognitive disorders (e.g., Alzheimer’s disease), and thought disorders (e.g., schizophrenia, bipolar depression).

Next we provide an overview of how clinical speech AI is different from more conventional applications of speech AI. This discussion is focused on the need for more thorough model validation, an understanding of speech and label variability, and the need for custom representations for the very diverse applications of clinical speech AI. 

We then use the technical pipeline for AI models outlined by the Food and Drug Administration \cite{food2019proposed} as a framework to guide our overview of the key aspects of clinical speech AI model development. First, we discuss considerations for selecting appropriate speech elicitation tasks, followed by a discussion on the technical aspects of speech data collection. These sections provide practical guidelines for choosing tasks that maximize the signal of interest and ensuring hardware setups are optimized to collect data of sufficient quality for reliable analysis. 
%We further provide a practical guideline for speech recording setup, covering key factors such as microphone specification, microphone placement, and noise level control, etc. to create a reliable setup. 

Following data collection, we turn our focus to clinically-interpretable speech representations. This section is anchored by discussing the difference between traditional AI ``speech features" and clinically validated ``speech measures." We draw connections to measurement theory and advocate transitioning from feature selection approaches focused on increasing AI model accuracy to identifying validated speech measures connected with underlying clinical constructs of interest. Since this transition is not straightforward, we provide an overview of traditional representations used in clinical speech analysis and highlight ideas for bridging the gap between ``features" and ``validated measures." These interpretable representations serve as the foundation for downstream clinical models that can offer users justifications for their outputs.

A key distinction between clinical speech AI applications and traditional speech AI lies in the need for thorough and continuous validation at multiple levels. In this tutorial, we reference the V3 framework—verification, analytical validation, and clinical validation—proposed by \cite{goldsack2020verification}. We adapt this framework to address the specific challenges of clinical speech AI and provide examples throughout to illustrate how it can be effectively applied in this domain, both during initial model development and continuously post deployment. We conclude the tutorial with a discussion of ethical considerations, focusing on the premature deployment of models, privacy concerns, and issues of fairness and bias in clinical speech AI.

This paper is organized as follows: \begin{itemize}
    \item Section \ref{speech_production} introduces the mechanisms behind human speech production and highlights the clinically-relevant information contained in the speech signal.
    \item Section \ref{comparison} discusses how clinical speech applications differ from mainstream speech AI technologies and why these differences are important.
    \item In Section \ref{clinical_speech_pipeline}, provides an overview of the clinical speech AI development pipeline, which serves as a guide for the technical discussions in Sections \ref{data_acquisition} through \ref{Post_deployment_ClinicalMLModels}.
    \item Section \ref{data_acquisition} focuses on the design of effective speech elicitation tasks and the technical aspects of collecting clinical speech data.
    \item Section \ref{speech_features_speech_measures} discusses the need for a shift from traditional speech features to clinically interpretable and validated speech measures.
    \item Section \ref{Design_ClinicalML_Model} emphasizes key considerations for building AI models capable of predicting clinical outcomes.
    \item The discussion continues in Section \ref{Post_deployment_ClinicalMLModels}, which addresses the importance of post-deployment monitoring to ensure model reliability over time.
    \item Ethical, privacy, and security issues related to clinical speech AI development are examined in Section \ref{Ethics_privacy_security}.
    \item Finally, the paper concludes with a summary in Section \ref{Conclusion}.
\end{itemize}
\section{The human speech production mechanism}\label{speech_production}
To develop interpretable AI models for clinical speech data, an understanding of how humans produce speech is of fundamental importance. Measuring changes in what we say (words, syntax) and how we say them (acoustic features, how speech sounds) allows for detection and classification of differences of interest. While the complete speaking process is mediated by complex neural circuits synergistically activating more than 100 muscles to create the acoustic signal, we can adopt a simplified stage approach to compartmentalize the various components involved in speaking. We can then reference these stages to discuss how various diseases/conditions map to deficits in these stages. Levelt’s (1989) book, ``Speaking: From Intention to Articulation" presents a modular model of all known aspects of speech production. It provides a valuable framework for understanding the various processes at play and how things can go wrong \cite{levelt1993speaking}. Here we use a simplified adaptation of Levelt’s model, focusing on three stages that are relatively well-understood and supported by abundant research: Conceptualization, Formulation, Articulation.

\noindent{\bf Conceptualization:} Before anyone utters a word, they must develop a thought, idea, or message. They also must experience the intent to communicate that thought to someone. Conceptualization occurs within the context of an individual’s perception of self, of time, and place, and an appreciation for the perspective of the receiver of this message (i.e., Theory of Mind). Conceptualization requires adequate functioning of the parts of the brain responsible for judgement, reasoning, memory, emotion, and social motivation \cite{moll2003morals}. 
There are several conditions that interfere primarily with the Conceptualization Stage of speaking, including psychiatric disorders (schizophrenia, bipolar disorder, major depressive disorder), dementias that include personality changes and/or hallucinations (frontotemporal dementia, Lewy body dementia), and the later stages of Alzheimer’s disease. Speech characteristics of impaired conceptualization result from the presence of `negative symptoms' (psychomotor retardation, apathy), `positive symptoms' (mania, hallucinations), and a distorted sense of reality (disorientation, paranoia, psychosis). Speech characteristics of negative symptoms include reduced speech output, imprecise articulation, and a tendency toward monotonicity and low or monoloudness \cite{sapir2007voice}. Positive symptoms are associated with rapid, pressed speech and a lack of coherence in the message being conveyed \cite{harvey1983speech}. Individuals with a distorted sense of reality may speak very quickly or slowly, change topics randomly, or produce incoherent speech. Signal for these speech characteristics can be found in both the acoustic speech signal as well as in analysis of the transcripts of the spoken message.  %(XXX).

\noindent{\bf Formulation:} In this second stage, words are selected and sequenced to best convey the conceptualized message. The words must be sequenced according to the rules of the language being spoken. But the word selection and sequencing also involve abstract decisions about the level of specificity of words to use (`car' versus `convertible'), their emotional valence and intensity (`dislike' versus `despise'), and language devices (sarcasm, humor, double entendre). These decision pair with paralinguistic decisions on how the outflow of the speech message (prosody) will enhance the intended meaning. For example, the sentence, ``I despise riding in convertibles, except at stop lights," might be spoken with an emphasis on `despise' and a longer-than usual pause at the comma to invoke intrigue with the final clause. This string of richly phenotyped mental words is then translated into a score of sensorimotor commands to muscles necessary to produce the sounds of the sequence of words, with timing and emphasis that underscores the message’s meaning.
Health conditions that interfere with the Formulation Stage typically involve damage to the cortical and subcortical language circuits of the brain. A cerebrovascular accident (CVA, or stroke) of the left cerebral hemisphere can cause aphasia, characterized by varying patterns of difficulty finding the words one wishes to speak, difficulty sequencing the words into meaningful sentences, and difficulty understanding what others are saying. In contrast, a CVA in the right cerebral hemisphere can leave language preserved, but one may struggle with perceiving and producing paralinguistic information that signals humor, sarcasm, or emotion. Individuals with early to middle-stage Alzheimer’s disease can also struggle to find words, and the words they do find tend to be less specific \cite{jacobs1995neuropsychological}. As with the Conceptualization Stage, evidence of all these speech characteristics can be found in both the acoustic speech signal as well as in analysis of the transcripts of the spoken message.

\begin{figure}[h!]
  \setlength\belowcaptionskip{-0.8\baselineskip}
  \centering
  \includegraphics[width=0.45\linewidth]{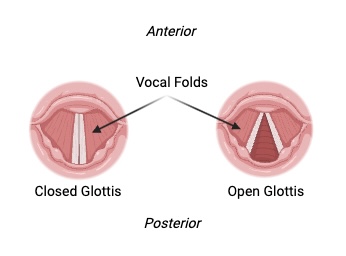}
  \includegraphics[width=0.45\linewidth]{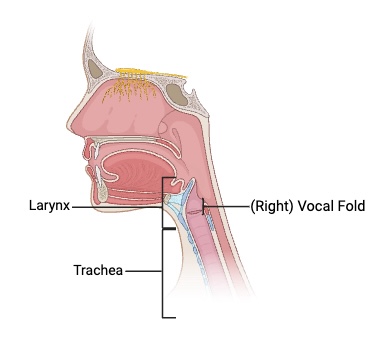}
  \caption{Exhaling air is initially blocked by the closed vocal folds (cords) in larynx. As air pressure increases, the vocal folds open, start to vibrate, and modulate the exhaling air. Pictures were created with BioRender.com.} 
  \label{fig:vocal_fold_larynx}
\end{figure}
 \begin{figure}[h!]
  \setlength\belowcaptionskip{-0.8\baselineskip}
  \centering
  \includegraphics[width=0.65\linewidth]{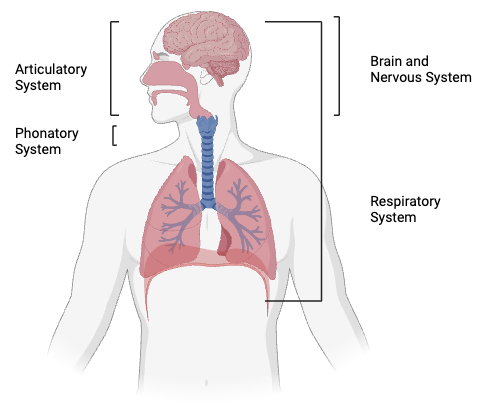}
  \caption{Systems involved in speech production. Picture was created with BioRender.com.} 
  \label{fig:4systems}
\end{figure}
\noindent{\bf Articulation:} After the message has been conceptualized and formulated into the mental sequence of words with an accompanying set of sensorimotor commands, the commands are executed to produce speech. First, one inhales more deeply than at rest and begins to release a column of air that will travel upward through the vocal tract to be shaped into the words we hear. The ascending column of air will first encounter the larynx where it will be impeded by the resistance of the closed vocal folds (cords). As pressure from the air builds up beneath them, the vocal folds blow open and begin to vibrate. This vibration chops the ascending column of air into bursts of audible phonation (voice). 
The faster the vocal folds vibrate, which is finely controlled by laryngeal muscles, the higher the pitch of the voice (fundamental frequency). The loudness of the voice is modulated by adjusting the air pressure beneath the vibrating vocal folds. This requires precise control of the muscles that compress the lungs and those that regulate the firmness of vocal fold closure during vibration. The higher the air pressure and the tighter the vocal fold closure, the greater the amplitude of the vibration, and the louder the voice becomes (decibels). Finally, the clarity of the voice (vocal quality) has to do with physical characteristics of the vibration of the vocal folds. The more symmetrically the two vocal folds move during vibration, and the more completely they meet in the middle of the airway, the clearer the vocal quality sounds (harmonics-to-noise ratio). 
As the column of air moves through the larynx, the portions of the column that encountered the vibrating vocal folds have been set into audible acoustic energy that will ascend to be filtered through the upper vocal tract to produce voiced sounds (e.g., vowels, nasals, liquids and glides). The portions of the air column that were undisturbed as they passed through open, non-vibrating vocal folds ascend to be shaped into unvoiced sounds (e.g., unvoiced stops and fricatives). The filter function of the upper vocal tract is created by moving the articulators (tongue, jaw, lips, soft palate) to create time-varying shapes and sizes of cavities, constrictions, and closures. The muscles of the soft palate and throat work in concert to shunt the acoustic energy through the nasal and/or oral cavities to produce nasal and non-nasal sounds, respectively. 
 \begin{figure}[h!]
  \setlength\belowcaptionskip{-0.8\baselineskip}
  \centering
  \includegraphics[width=0.55\linewidth]{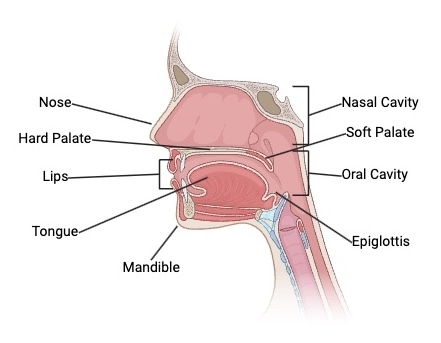}
  \caption{Speech articulators can create time-varying shapes and sizes of cavities, constrictions, and closure to filter phonated sounds passing through. They can also direct the generated sound through the nasal and/or oral cavities.  Picture was created with BioRender.com.} 
  \label{fig:speech_articulators}
\end{figure}
Different resonant cavities created by movements of the tongue, jaw, and lips produce formants (regions of enhanced energy in the frequency spectrum), whose patterns are generated by the vocal tract filter acting upon the source phonation and distinguish among vowel sounds. Constrictions created by close approximations between the lips, tongue, and/or hard palate generate turbulent noise as the air rushes through to produce fricative sounds like `s.’  Closures of the vocal tract allow air pressure to build up and burst through to create stops like ‘p.’ Clear, highly intelligible speech is the result of maximally distinctive resonant cavities, narrow and precise constrictions, and firm and precise closures, precisely timed across systems to produce the sounds of the spoken message.
Any condition that interferes with this highly synchronized movement among the respiratory, phonatory, articulatory, and resonatory systems will be evident in the acoustic signal of speech. Focal conditions have specific effects on speech. For example, a paralyzed vocal fold will result in breathy phonation, without impact on other speech subsystems. Conditions like amyotrophic lateral sclerosis impact multiple targets in the brain and spinal cord, causing spastic and/or flaccid paralysis of muscles of the limbs, trunk, head, and neck. This results in reduced respiratory support, reduced articulatory precision, hypernasality, strained vocal quality, and overall slowed speech. The constellation of speech symptoms for any given condition derives from the location and extent of nervous system damage \cite{duffy2012motor}. Further, conditions that impact the structure and function of the speech organs also will manifest in the speech signal.  Cleft palate, laryngeal cancer, chronic obstructive pulmonary disease, and many other conditions, impact speech in predictable ways \cite{pereira2023differences, orlikoff1996dysphonia, binazzi2011dyspnea}.

\section{Comparing clinical speech applications to other speech applications}\label{comparison}

 \begin{figure}[h!]
  \setlength\belowcaptionskip{-0.8\baselineskip}
  \centering
  \includegraphics[width=0.65\linewidth]{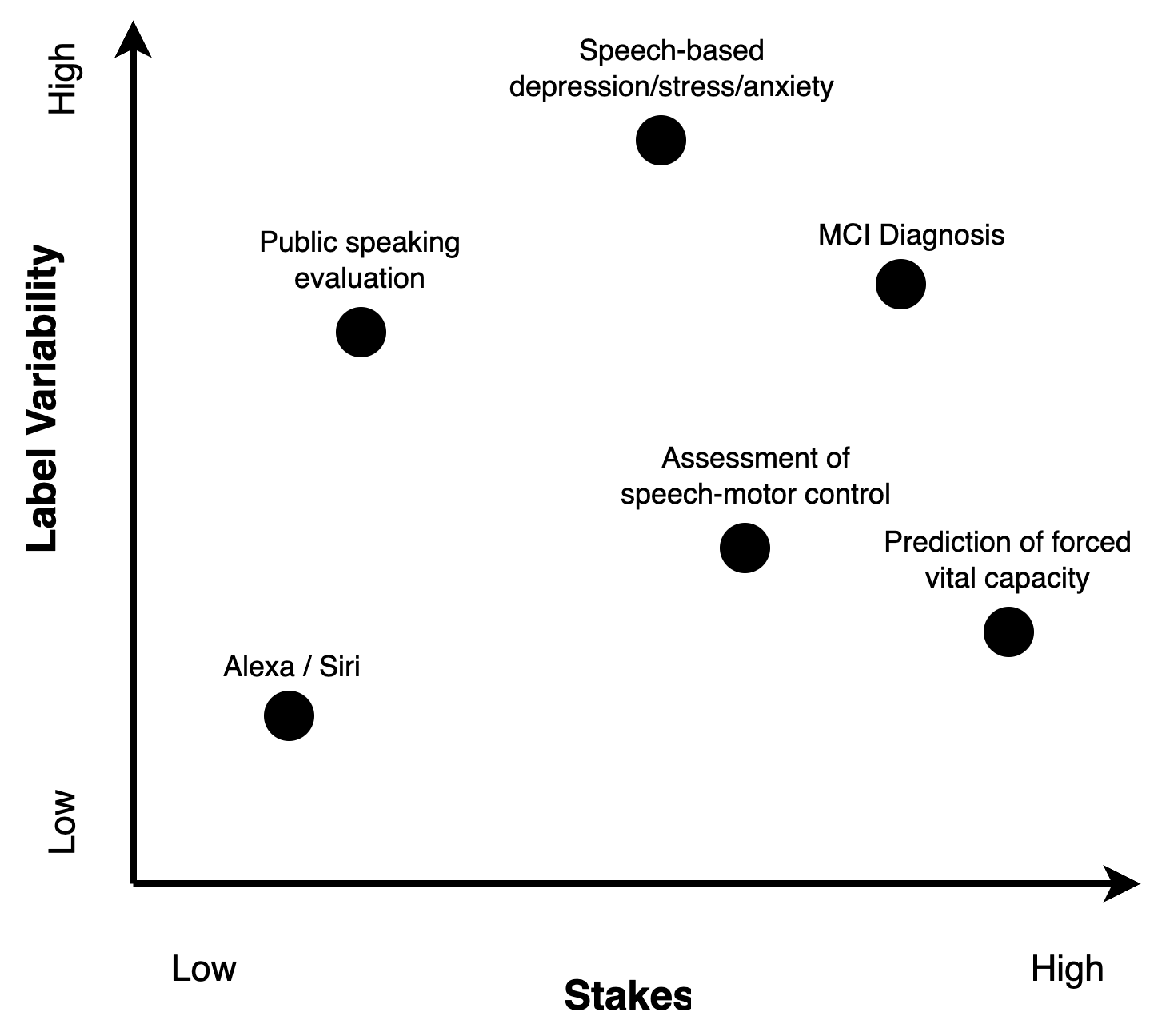}
  \caption{Differences between traditional applications of speech AI and clinical speech AI} 
  \label{fig:stake}
\end{figure}

In this section we highlight the important differences between traditional applications of speech AI and clinical applications of speech AI and explain that straightforward application of techniques that have worked well in traditional domains are often not appropriate for clinical applications. In the two-dimensional plot in Fig. \ref{fig:stake} we depict important differences between traditional applications of speech AI and clinical speech AI. The $x$-axis on the plot represents the stakes of the application. In this context ``stakes" refers to the level of consequence or impact that outcomes of speech AI applications have. The $y$-axis represents the amount of variability in the ground-truth labels used to train the supervised learning algorithms. For example, transcripts for speech recognition are reliable; however, there is considerable variability in diagnosing mild cognitive impairment or depression \cite{folsom2006diagnostic, weinstein2022diagnostic}. 
On this two-dimensional plot, we plot certain applications of speech AI and refer to them in the description below.

One of the primary differentiators is the elevated stakes associated with clinical speech AI, where the outcomes have a direct bearing on human health. For example, a mis-recognition in ASR might result in a trivial transcription error (e.g. a misunderstanding by Siri or Alexa), whereas inaccuracies in clinical speech AI could lead to misdiagnosis or delayed treatment of severe health conditions such as Alzheimer’s or Parkinson’s disease. Adding to the model building challenge is the high diagnostic label variability associated with these conditions, especially prominent in the early stages of neurological diseases. Recent meta-analyses reveal that the misdiagnosis rates in Alzheimer’s disease exhibit a wide range with sensitivities from 71\% to 87\%, and specificities from 44\% to 71\%. Similarly, the diagnosis of Parkinson’s disease in the initial five years is reported to be inaccurate nearly half the time  \cite{beach2018importance}. This contrasts starkly with traditional speech AI like speaker recognition, which has a clear ground truth. Similarly, the ground truth in ASR has relatively low inter-transcriber variability \cite{xiong2017toward}. The  variability in diagnostic labels presents an added layer of complexity in clinical speech AI. This variability could be attributed to the evolving nature of diseases, subjective interpretation by different clinicians, or the presence of comorbid conditions that might obscure the primary diagnosis. It necessitates more thorough modeling and validation approaches, as we will highlight throughout this tutorial. 
%In Section xxx, we provide an overview of several of these methods.

For many traditional applications of speech AI, models are evaluated using simple measures of accuracy (e.g. word error rate in ASR; equal error rate in speaker verification). In contrast, clinical models require more comprehensive validation. Beyond measures of accuracy, validation in clinical settings requires a broader set of criteria to ensure that the models are reliable, robust, and clinically viable. This involves conducting rigorous testing across diverse cohorts, capturing real-world deployment feedback, and undergoing regulatory scrutiny to ascertain the safety and effectiveness of these models. A starting point for this analysis is the Verification, Analytical validation, Clinical Validation (V3) framework recently proposed for digital health. In section \ref{speech_features_speech_measures}-\ref{Design_ClinicalML_Model}, we provide an overview of how this validation can be conducted for speech-based digital health tools to ensure that developed models can perform well under various conditions and are capable of handling the inherent uncertainties associated with clinical data.

Traditional speech AI often relies on standard input representations like Mel Frequency Cepstral Coefficients (MFCCs), the mel-spectrum, other open-source feature sets \cite{schuller16_interspeech}, or deep-learning-based speech representations \cite{mohamed2022self}. However, the diversity in clinical applications calls for more bespoke, individually validated representations. In contrast to non-clinical applications where speech data is available at large scale, the scarcity of large-scale clinical data requires reduction of sample complexity; this can be achieved via simpler, interpretable, and individually-validated features. Moreover, the high-dimensional nature of traditional input representations often leads to models that lack interpretability — a critical aspect in clinical scenarios for ensuring reasonable and reliable model outputs. 
There is great utility in developing acoustic markers of the underlying speech symptoms. The ground truth for these models can be attained directly from the audio and/or transcript, making it much easier to obtain than clinical diagnostic labels. 
%For example, breathy voice is a characteristic feature of Parkinson’s disease; however, there are no acoustic markers of breathy voice. 
For example, breathy voice is a characteristic symptom of Parkinson’s disease. Feature parameters that measure breathiness in speech, such as cepstral peak prominence \cite{murton2020cepstral}, can be the useful markers of disease progression. Furthermore, in the case of dysarthria secondary to different neurological conditions, there are existing protocols for assessment of perceptual dimensions with well-characterized inter and intra-rater variability \cite{bunton2007listener)}. 
Referring back to Fig. \ref{fig:stake}, we note that these problems of quantifying interpretable acoustic markers have lower stakes as they do not produce direct diagnostic decision and have lower label variability as the perceptual constructs (e.g. breathy voice) are narrower and better defined when compared to the broad diagnostic labels. Relatedly, data from healthy individuals, which is much easier to obtain, can be used to augment smaller clinical databases for development of these features and for developing normative ranges. In Section \ref{speech_features_speech_measures}, we highlight several approaches for developing interpretable representations for clinical speech applications and provide examples of how healthy corpora can be used to develop acoustic and linguistic measures of perceptual constructs of interest.

%The discussion above highlights important considerations for algorithm developers, before they begin designing models. Oftentimes, developers use existing databases consisting of paired speech samples and diagnostic labels for building models. In many cases, the variability of the labels in these databases is unknown, making it difficult to account for uncertainty. To compound the issue, small sample size of most databases make overfitting easier and increase variability, contributing to the overoptimism highlighted in the introduction. For small sample sizes (e.g in the tens) and databases with unknown label uncertainty, algorithm developers should forgo supervised learning on broad diagnostic labels for more manageable problems. 

\section{The clinical speech pipeline}\label{clinical_speech_pipeline}
Clinical AI models are typically trained via supervised learning to analyze a subject’s speech data, extract the clinically-relevant information, predict the clinical labels (e.g. severity score, intelligibility score, subtype of a speech disorder, etc.), and provide actionable insight to clinicians. 
 \begin{figure}[t!]
  \setlength\belowcaptionskip{-0.8\baselineskip}
  \centering
  \includegraphics[width=0.75\linewidth]{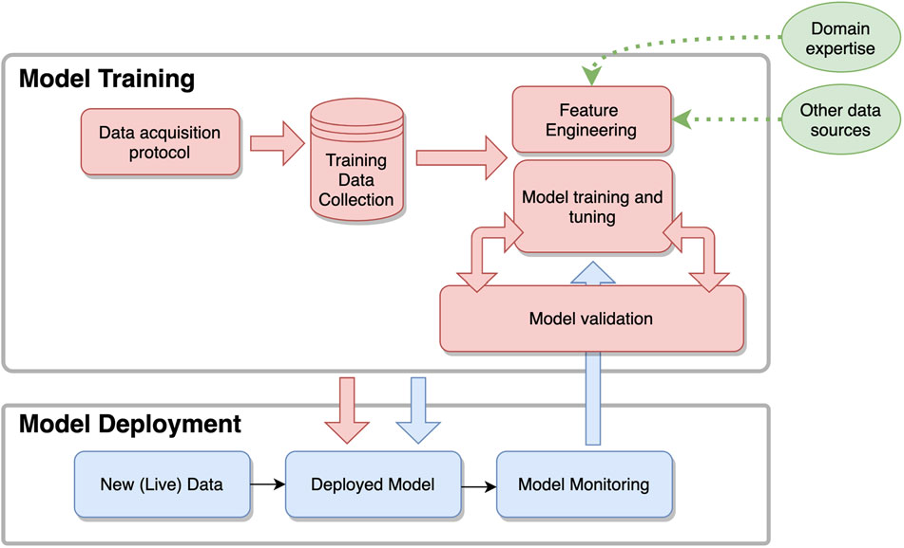}
  \caption{Technical pipeline of clinical AI model training and deployment.} 
  \label{fig:model_training}
\end{figure}
As described by the FDA \cite{us2022clinical}, the technical pipeline for training and deployment of clinical AI models as AI-based software-as-a-medical device (SaMD) tools is illustrated in Figure \ref{fig:model_training}. 
The pipeline starts with the design of  data collection protocol. It defines how the clinical population of interest is recruited, how speech elicitation stimuli are designed, and how data is collected. During data collection, participant speech is captured by speech recording devices in the form of acoustic signals. Clinical labels associated with the participant, such as symptoms of disease, clinical scores, and other clinically meaningful labels, are provided by experts and associated with the collected speech. In AI parlance, the speech signals and clinical labels become the ``data’’ and ``labels’’ stored in a structured database. Feature engineering plays an important role in extracting useful information from the speech data to facilitate clinical speech AI. Domain knowledge in speech signal processing, speech production, clinical speech science, etc., can be incorporated into the process. With a sufficient amount of features and clinical labels, statistical models are trained to identify the relationship between the features and the labels. Validation of the trained models is performed using a held-out set of test data to determine the feature and model to be deployed. Following deployment, real-world model performance is continuously monitored after deployment. With speech data newly acquired during deployment, the model can be iteratively updated and redeployed. 

Errors in each part of the model development process can negatively impact the efficacy of the analytic system. For example, if speech data is collected inappropriately (e.g. insufficient consideration in selection of speech elicitation tasks, improper implementation of recording speech, etc.), the extracted features may not carry sufficient clinically-relevant information, leading to failure of statistical modeling and model deployment. If methods of feature extraction do not consider clinical knowledge, or the statistical models are overly complex, clinicians may lack necessary evidence to interpret the model output. If models are trained on data sets that are not representative of the deployment scenario, model performance could be impacted. These factors could limit the adoption of the models in clinical settings. 
Given the spectrum of factors that influence %could impede 
model development in clinical speech AI, potential approaches that steer the model development towards better interpretability and reliability are discussed in the following sections. 

In Section \ref{data_acquisition}, we will discuss the relationship between degrees of freedom in speech content and the computational load asserted to the speakers. The discussion provides the researchers with a mental model for determining the speech elicitation tasks that magnify speech changes for a given clinical condition. Then the section discusses important considerations regarding  data collection. We provide practical guidelines for speech recording setups for real-world settings, with the goal of obtaining speech data reliably and in a way that captures underlying symptoms associated with a condition. Finally, the section also considers verification of the recording setup to ensure that acquired data is of sufficiently high quality for the application of interest.

Section \ref{speech_features_speech_measures} %focuses on feature engineering for clinical AI model development.
proposes a shift from using speech features to utilizing {\em speech measures} as an important step to develop clinically interpretable and generalizable clinical speech models.  
We draw a distinction between conventional high-dimensional \emph{speech features} used by the technical speech community, and interpretable \emph{speech measures} which are required for validated clinical/scientific studies of speech. We highlight the limitations of high-dimensional uninterpretable speech features, and describe the benefits of developing speech measures linked to constructs of clinical interest during feature engineering. 
% The goal is to promote a paradigm shift from using speech features to utilizing speech measures as an important step to develop clinically interpretable and generalizable clinical speech models. 
The section further discusses the validation of speech measures for clinical applications, as informed by the V3 validation framework \cite{goldsack2020verification}. We present examples of speech measures validated in accordance with this framework across various studies. We further highlight the drawback of relying on invalidated features during clinical model development.
The Section will then discuss technical details of speech measure design. These strategies range from knowledge-driven to data-driven approaches. While the knowledge-driven approach can be readily adopted from clinical and speech science research, our further emphasis will be on introducing the data-driven approach to derive valid and reliable speech measures. This section demonstrates new opportunities to the clinical community on the integration of AI and clinical knowledge in speech analyses. The process of designing novel speech measures needs to be coupled with rigorous analytical validation. We will demonstrate existing works that have performed analytical validation, which will explain the importance and implication of doing so. 
%The section further describes strategies for strengthening the reliability of a speech measures during developments.

The Section \ref{Design_ClinicalML_Model} then shifts the focus on the design of clinical AI model, which predicts clinical labels based on the input speech features. We discuss the most common issues that impede the development of generalizable and explainable clinical AI model. We then provide strategies to mitigate these issues through the use of validated and interpretable lower-dimensional speech measures as inputs, and the adoption of AI models that intrinsically offer explanability. Similar to speech measures design, as discussed in Section \ref{speech_features_speech_measures}, we will introduce the clinical validation framework for the developed model. The process not only concerns the reliability of the predicted clinical label, it further  validates whether the design of the model (from model input to model output) answers the clinical question of interest. We will illustrate examples and counter-examples to emphasize the importance of clinical validation before deploying the clinical AI model in real-world settings.

Section \ref{Post_deployment_ClinicalMLModels} focuses on post-deployment monitoring of developed models, an important component to ensure the ongoing efficacy and safety of AI-based clinical speech models. This involves continuous collection and analysis of real-world data to assess model performance, detect potential drifts in data distribution, and identify any unforeseen issues. 
%Feedback loops are established to iteratively update the models, incorporating new data and insights to enhance robustness and adaptability. Additionally, regular audits and compliance checks are conducted to meet regulatory standards and maintain trust among clinicians and patients. 

While the above sections mainly discusses the technical aspects of clinical AI models, Section \ref{Ethics_privacy_security} will shift the focus to responsible development of clinical speech AI. We will discuss potential model bias induced  by demographic factors in the datasets and human-factors during model development. The section will further discuss potential hazards of relying on ``open-sourced models" developed for other applications. We will further address the responsible development from the user perspectives, where models could be potentially misused in a way that makes false claims about some person's health condition, violating one's privacy. We emphasize the importance of regulations mandated by the governing bodies.

\section{Speech data collection}\label{data_acquisition}

This section covers the key aspects of speech data collection, starting with the design of speech elicitation tasks that align with clinical objectives and patient populations. It then covers the technical requirements for ensuring high-quality  data collection, including the verification of hardware to guarantee accurate and consistent speech recordings. 

\subsection{Speech elicitation task design}

Speech elicitation tasks are constructed prior to speech  data collection. Certain forms of elicitation tasks allow symptoms of a given clinical condition to be amplified. For example, %adding Caterpillar task here...
diadochokinetic (DDK) tasks are used to assess the speed and coordination of an individual's speech muscles by having them rapidly repeat simple syllables like
/p\textipa{@}/, /t\textipa{@}/, and /k\textipa{@}/. 
Maximum phonation time (MPT) is used to assess the glottic efficiency by instructing a speaker to sustain a vowel sound produced on one deep breath. 
Instructing speakers to raise and lower their pitch level as much as possible allows analyses of the phonation mechanism using fundamental frequency range (FFR), or others statistics of the F0 in the speaker's voice \cite{reich1989factors, wit1993maximum, ordin2017cross}.
These tasks are considered `maximum performance tasks' because they challenge the speaker's capability to perform a particular speech task at their highest possible speed and precision. The examples above are particularly effective for speakers with voice and motor speech disorders,  where the speech deficits are mainly contributed by the physiological impairment and musculatory control of speech production organs.
In other conditions where speech production mechanism is intact but the elicited speech is affected by problems in language, memory, or cognition, other speech elicitation tasks can also be optimized to amplify the cardinal symptoms. 
% Similarly, other speech elicitation tasks can be optimized to amplify cardinal symptoms associated with other conditions. 
A structured interview, which comprises a pre-defined set of questions concerning the anxiety level in a speaker, 
can be utilized to assess depression patients \cite{shear2001reliability}.
Speech pauses, speaking rate, and length of speech in the patients’ responses are associated with the severity of depression \cite{stassen1991speech}; 
For dementia patients whose cognitive function is impaired, picture description task, such as Cookie theft task, can be used to evaluate cognitive-linguistic function \cite{cummings2019describing}. The manner in which patients describe the picture can summarize their language proficiency, reflecting the severity of the disease and revealing how their language ability is affected. 
% Language use in different types of Aphasic patients can be analyzed by the lexical semantics in speech [Ref].
Another example is children with developmental speech sound disorders who persistently produce certain phonemes incorrectly. To identify the type of speech errors, a picture naming or multisyllabic word repetition task can be adopted, which fix the speech content and guide the children in producing the target phonemes of interest \cite{goldman1969goldman, cheung2006hong, benway2020differences}.
 
 \begin{figure}[h!]
  \setlength\belowcaptionskip{-0.8\baselineskip}
  \centering
  \includegraphics[width=0.95\linewidth]{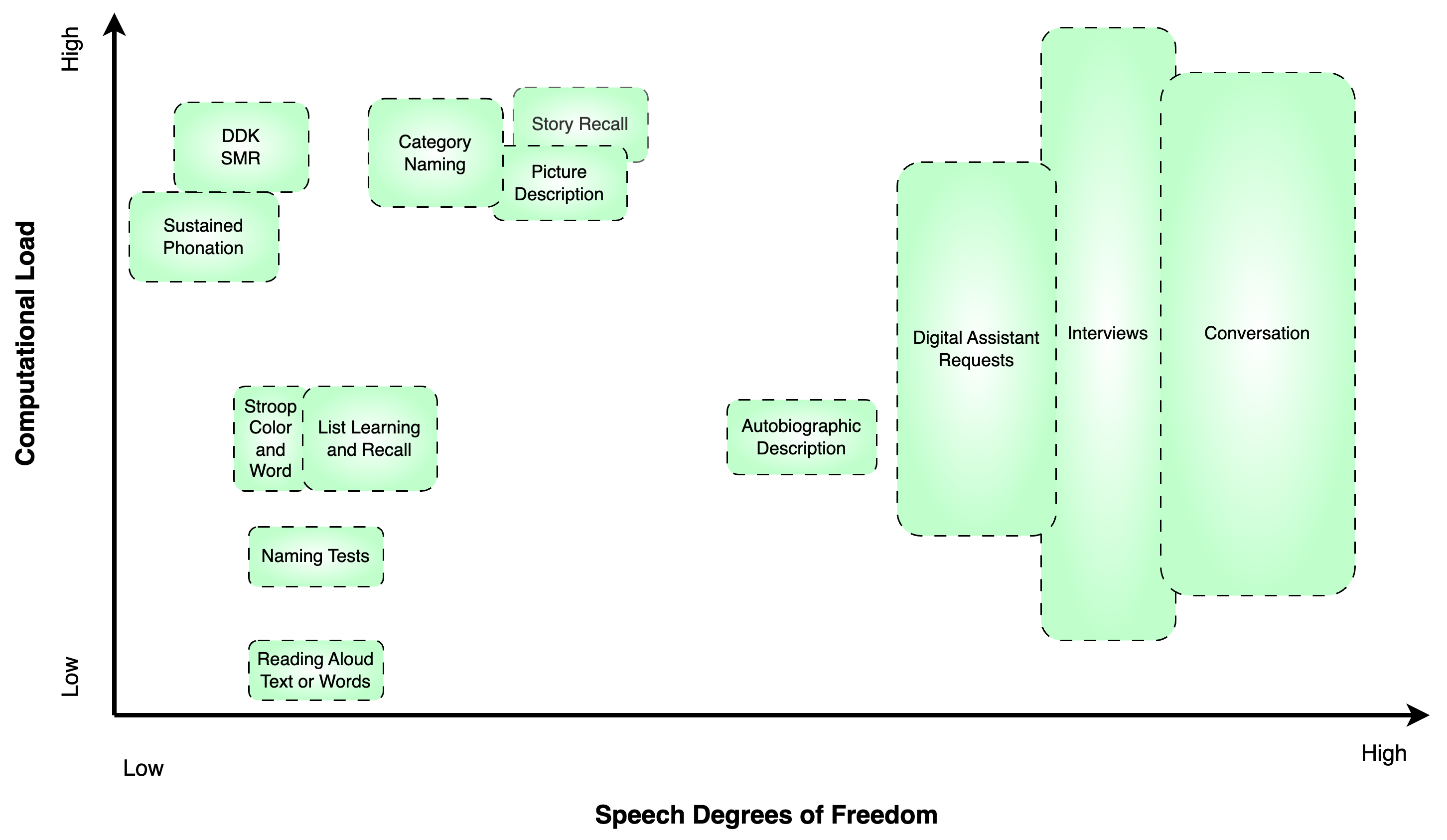}
  \caption{Categorization of speech elicitation tasks} 
  \label{fig:speech_elicitation_tasks}
\end{figure}

%Adding Julie's part
Several variables must be considered for the successful collection and analysis of speech data. When using existing speech databases for model development, these considerations also apply to understand the appropriate use cases for the existing data:
\begin{itemize}
    \item What is the goal of the model development (e.g., classification, diagnosis, differential diagnosis, change (progression/remission/improvement), at a group level or individual, cross-sectional or longitudinal)? The answer to this question estimates the level of sensitivity and specificity the modelling will require to be successful, and how much speech will need to be collected and at what intervals.
    \item What aspects of speech production (Conceptualization, Formulation, Articulation) are most likely impacted by the disease or condition to be analyzed and what is their severity level? These deficit patterns delimit the type of pressure the speech task should place on the participant to enhance the signal. For example, if the articulation aspect of speech production is primarily impacted by the disease or condition, and these deficits manifest only mildly in the study population, maximum performance speech tasks can be selected to amplify articulation deficits.
    \item Are there participant-related limitations that must be considered in task design (e.g., vision, hearing, cognition, reading ability, time to fatigue)?
    \item What are the language and cultural considerations of the participant population (language(s), native versus non-native speakers, dialect, accent; rhythm and phonology, and semantics and syntax of the language(s) collected)? 
    \item How will the spoken responses be processed (ASR, digitized acoustic signal) and what type of measures will be extracted (NLP, spectral-temporal measures)?    
\end{itemize}

\begin{table}[th!]
\caption{Prospective collection of speech samples to classify between mild cognitive impairment and no cognitive impairment in a genetically at-risk population of 50–80-year-old men and women in a health care system in New England major cities.}
    \resizebox{\textwidth}{!}{

\begin{tabular}{l|l}
\textbf{GOAL}                                                                          & Classification/Cross-Sectional                                                                                                                  \\ \hline
\textbf{DEFICITS/SEVERITY}                                                             & \begin{tabular}[c]{@{}l@{}}Formulation: memory, language, executive function, \\ / Unimpaired to mildly impaired \end{tabular} \\ \hline
\textbf{\begin{tabular}[c]{@{}l@{}}PARTICIPANT  \\ LIMITATIONS\end{tabular}}           & Must be able to read English text at fifth grade level                                                                                                                                              \\ \hline
\textbf{LANGUAGE/CULTURE}                                                              & Native English speakers                                                                                                                         \\ \hline
\textbf{\begin{tabular}[c]{@{}l@{}}SPOKEN RESPONSE \\ PROCESSING/METRICS\end{tabular}} & \begin{tabular}[c]{@{}l@{}}Automatic Speech Recognition and Large Language Models, \\ / Natural Language Processing metrics \end{tabular}                                                                                        
\end{tabular}
}
\label{tab:example1}
\end{table}

These considerations can guide to trim the search space for the suitable speech elicitation task. See the example in Table \ref{tab:example1}, which considers speech elicitation task design for the development of a tool to classify between cognitively intact individuals and those with mild cognitive impairment based on speech. Development of the tool calls for speech elicitation tasks that place pressure on cognitive functioning (memory and language formulation) to amplify subtle differences between the groups’ performance metrics. It is useful to consider potential speech elicitation tasks in terms of the spoken response variability and how difficult the task is to complete. 

Figure \ref{fig:speech_elicitation_tasks}  portrays examples of speech elicitation tasks plotted relative to the degrees of freedom of the spoken responses ($x$-axis) and the degree of computational and/or sensorimotor pressure inherent in completing the task ($y$-axis). For example, reading aloud a list of words has a low number of degrees of freedom because each word has exactly one correct spoken response. Further, reading a word places little pressure on the cognitive-language system and does not require higher order executive function or memory. Describing what’s happening in a picture requires a synthesis of the objects and actions depicted in the picture, and inferences about their temporal and spatial relations, thereby making it a more computationally challenging speech elicitation task than word reading. But because the picture constrains the content of the response, degrees of freedom are lower than for tasks with unconstrained topics, such as an open conversation. The reduction in degrees of freedom can provide benefit and can serve as a drawback. For example, a cross-sectional study design may benefit from control over the content of speech as an open conversation unconstrained by topic could add noise to the data, obscuring any small group differences. However, the reduced freedom in describing a picture may limit the richness and spontaneity of language use, which could be a drawback in studies aiming to capture more naturalistic speech patterns. 

In the MCI example posed above, we would expect automatic speech recognition (ASR) to be relatively accurate in a conversational task for native speakers of English in this age range. ASR would be even more accurate in tasks with lower degrees of freedom, in which ASR model pre-training could benefit accuracy. But because the intended analysis level is natural language processing on connected speech, a story recall task (immediate and delayed following intervening distractor tasks) would be a reasonable option for placing pressure on the cognitive system, while keeping degrees of freedom relatively constrained. The task is difficult because the story’s contents need to be understood, remembered, retrieved, and retold. Adding a delayed condition places further pressure on the system to induce response errors in the context of mildly impaired cognition. This also allows for a comparison of performance in the immediate-versus-delayed condition within individuals, providing a metric of performance stability.

\begin{table}[th!]
\caption{Prospective collection of speech samples to identify impact of intervention on disease progression in chronic obstructive pulmonary disease in women between the ages of 60-75 years old.}
    \resizebox{\textwidth}{!}{

\begin{tabular}{l|l}
\textbf{GOAL}                                                                          & Longitudinal/within-subject                                                                                                                  \\ \hline
\textbf{DEFICITS/SEVERITY}                                                             & \begin{tabular}[c]{@{}l@{}}Pulmonary function impacting Articulation Phase \\ / Mild to severe \end{tabular} \\ \hline
\textbf{\begin{tabular}[c]{@{}l@{}}PARTICIPANT  \\ LIMITATIONS\end{tabular}}           & Must be able to follow verbal instructions in native language                                                                                                                                          \\ \hline
\textbf{LANGUAGE/CULTURE}                                                              & Can perform task in native language                                                                                                                      \\ \hline
\textbf{\begin{tabular}[c]{@{}l@{}}SPOKEN RESPONSE \\ PROCESSING/METRICS\end{tabular}} & Temporal measures extracted from the acoustic signal                                                                                    
\end{tabular}
}
\label{tab:example-2}

\end{table}

Different elicitation tasks also impose varying levels of computational load on the human brain, as shown in the $y$-axis of Figure \ref{fig:speech_elicitation_tasks}.  Tasks like picture naming and passage reading induce minimal computational load, as the spoken content is predefined by clinicians and requires little effort from the speaker. In contrast, more complex tasks such as picture description, interviews, conversations, and story recall necessitate greater coordination of memory, attention, and organization, thereby imposing a higher cognitive load on the speakers. %computational load

An example of a high computational load task is illustrated in Table \ref{tab:example-2}, in which participants will provide speech samples intermittently over time to track evidence of chronic obstructive pulmonary disease (COPD) progression in the context of an intervention. This means that the speech elicitation task must be feasibly executed reliably across a large severity range. Due to the variability in the patient population, reducing degrees of freedom in the speech elicitation reduces possible confounds. However, a desirable elicitation task also taxes the respiratory system (high computational load). A maximum performance tasks that put pressure on the respiratory system include taking a deep breath and sustaining a phonation for as long as possible or counting as high as possible or saying the alphabet on a single breath. In the context of significant pulmonary function deficits, reading a passage aloud also would be considered a challenging task in which speakers need to pause increasingly frequently to take a breath while reading the passage aloud. 

% As shown in Figure \ref{fig:speech_elicitation_tasks}, speech elicitation tasks can be categorized along two dimensions, namely the degrees of freedom in speech
% %the content structure 
% and the cognitive computational load required to produce the speech. Structured speech tasks, such as picture naming, DDK tasks, and reading tasks, impose certain constraints on the spoken content. They are less variable than spontaneous speech and have lower degrees of freedom. 
% %can be designed as maximum performance tasks.
% The structured elicitation tasks are convenient for administration, give the experimenter design flexibility, make automated transcription less challenging, and generally have improved repeatability, facilitating longitudinal comparisons \cite{wolk1998phonological}. This also has important implications regarding study replicability; results from structured tasks are more likely to replicate across studies owing to the reduced variability. Unstructured tasks, such as interviews and spontaneous conversation, have little constraint in the spoken content. Speakers are allowed to speak as in daily-life communication and are ecologically more valid as measures of communicative ability. The increased degrees of freedom results in more variable speech output. As a result, measuring important speech parameters, such as intelligibility, fluency, and responsiveness, requires analysis over longer windows. 

Speech production involves the precise integration of language, memory, cognition, and sensorimotor functions. Researchers should review the existing literature to identify the stages where the speech production of clinical populations diverges from that of healthy populations. Selecting appropriate elicitation tasks is crucial to elicit speech that carries clinically relevant information and to examine the upper limits of the subjects’ performance. For example, patients with cleft lip or with oral cancer have damaged sensorimotor functions, while their language skills, memory and cognition function may be intact. Given the relevant speech problems are manifested in articulation, structured speech tasks that can control phonetic content of speech, such as isolated pseudo-work tasks, and sentence reading tasks, can be used for  data collection \cite{sell2005issues, woisard2021c2si}. In patients with voice-related disease that also suffer from sensorimotor issues, structured maximum performance task such as maximum phonation task are likely to be more useful \cite{omori2011diagnosis, karlsen2020acoustic}. Patients diagnosed with amyotrophic lateral sclerosis (ALS) typically encounter difficulties in transducing neuromuscular commands into speech movements, while they also have significant impairment in respiratory function. Structured speech tasks that are focused on muscle coordination in speech production, such as Diadochokinetic test and single word intelligibility test can be adopted to understand the maximum repetition and phonetic contrastivity in speech, respectively \cite{mulligan1994intelligibility}. Lengthy and unstructured tasks, such as conversation, can be used to analyse the respiratory conditions via measuring the loudness, breath and disturbance in speech \cite{ball2004communication}. When the target population includes patients with higher-level cognitive or perceptual issues, such as Alzheimer’s disease or aphasia, speech tasks imposing cognitive loads—like picture description and storytelling —are useful in highlighting deficits in speech production. Cognitive deficiencies result in less organized, repetitive, and less detailed speech descriptions\cite{usita1998narrative, leyton2014verbal}. 
On the other hand, patients with depression or schizophrenia face challenges in the conceptualization of speech production. Designing elicitation tasks for these patients is complicated by fluctuating conditions, such as a mix of positive and negative symptoms and nonspecific symptoms attributable to various etiologies. Lengthy, unstructured tasks like interviews and conversations can increase the likelihood of obtaining useful data for analyzing depression-related speech. Cognitively demanding tasks may also provoke speech in patients with schizophrenia that is distinct from that of healthy individuals\cite{cohen2014speech}.

The diagram in Fig. \ref{fig:speech_elicitation_tasks} provides a mental model for aligning the properties of speech elicitation tasks with the symptoms of a condition, however validation experimentation is required to evaluate the effectiveness of a given elicitation task for a given condition. Multiple elicitation tasks can be considered for a specific disease to evaluate which leads to the highest accuracy or largest effect size. For example, in assessing language abilities post-stroke, speakers produced more words and structurally complex, lexically diverse, and syntactically accurate speech in storytelling tasks compared to picture description \cite{schnur2024differences}. This implies that storytelling better amplifies language difficulties in stroke compared to picture description. Word retrieval ability in aphasic patients can be assessed through both picture naming and conversational tasks. Experimental results indicated that patients showed better lexical retrieval and self-correction of errors in conversations than in single-word naming tasks \cite{mayer2003functional}. These studies indicate that researchers should consider the design of the elicitation task as an important part of the development pipeline. Adopting multiple elicitation tasks to compare the analytics results obtained across different tasks, as is common with analysis of different feature representations, can lead to differences in performance.
% [INSERT A PARAGRAPH HERE ON THE RELATIONSHIP BETWEEN FEATURES AND ELICITATIONS]

Further complicating the analyses is the relationship between elicitation tasks and speech features when it comes to interpretation. For instance, the variability of F0 can be measured in reading passages or in sustained vowel tasks \cite{tykalova2021effect, lee2012variability}. While the former is focused on vocal coordination in natural conversation setting, the latter is focused on vocal performance in a controlled setting. The pitch variability derived from the two elicitation tasks would be different and not directly comparable. 
In \cite{abbiati2023speech}, spatiotemporal index (STI) is measured on English-speaking young adults using four different sentences varied by length and complexity. Results showed that the sentence type had significant effect on the measured STI values. Although these sentences all belong to reading task, the variability of design in sentences affect the speech feature, meaning the differences in sentences could result in different interpretations of results. 
Understanding the differences between elicitation tasks, the variability within each task, and the relationship between tasks and features is important for clinical model development.

\subsection{Hardware verification for speech  data collection}
% While a well-designed speech elicitation task ideally provides acoustic and/or linguistic information pertinent to the clinical condition of the speakers, 
The protocol for speech  data collection is a critical step to ensure acoustic and/or linguistic information pertinent to the clinical condition of the speakers can be accurately measured from the digital speech signal. 
There is now significant flexibility in selection of devices (e.g. smartphones vs. professional recorder) and platforms (face-to-face vs. remote session) for speech  data collection.  However, such convenience comes at the cost of potential contamination of the acquired data by unwanted effects, which can negatively impact subsequent analysis.
For instance, given the rising popularity of telepractice, Tran et al. investigated how speech compression algorithms employed in data transmission could impact acoustic measures used in dysarthric speech assessment \cite{tran2022investigating}. Analyses of speech data collected at different sampling frequencies, bit rates, and coding formats showed that articulatory-based measures such as goodness of pronunciation (GOP) were notably affected by the different settings of  data collection. Similarly, a study by Ge et al. compared the differences of speech data collected by smartphones and laptops \cite{ge21b_interspeech}. This study showed that the measured second and higher-order formant frequencies, spectral moments, and voice quality measures varied drastically between devices. In the same way,  room acoustics, epsecially reverberations can influence acoutic features~\cite{JHMDAGA:2019}.
These discrepancies could lead to  misinterpretation of segmental and voice quality contrasts in speech and demonstrate that different  data collection settings can lead to divergent results. 
%In the clinical context, the inconsistency could increase the risk of providing inaccurate information to the clinicians. 
Therefore, researchers must exercise caution in speech  data collection to minimize the influence of confounding factors.
% Researchers should be cautious in speech  data collection to ensure the data is least affected by irrelevant factors, while the acoustic measures can truly convey the clinical conditions of the speakers. 

% \noindent {\bf Verification framework in speech  data collection:} 
In biometric monitoring technologies for digital health, a framework was recently proposed to verify the process where sensor-generated signals are transformed into data \cite{goldsack2020verification}. A component  of the validation framework, known as hardware verification, aims to validate if  the analog speech signal is accurately captured, and if post-processing firmware produces appropriate outputs for the analysis.
The hardware verification can be divided into intra-sensor and inter-sensor comparisons. Given the ground-truth data, the intra-sensor comparison validates the consistency of the data acquired by the same sensor overtime, whereas the inter-sensor comparison validates the consistency of the data captured across different sensors. 
For instance, when the clinical task aims to study the vowel space of a disordered speaker, the extracted F1 and F2 values from speech signals should be consistent across different recording devices, within acceptable margins, and should align with ground truth values annotated by human experts (inter-sensor comparison). Additionally, measurements of F1 and F2 taken over time using the same recording device should remain stable (intra-sensor comparison). These comparisons ensure that data collection produces reliable, reproducible results and that the data remains accurate, regardless of the recording equipment used.

To further illustrate these concepts, we present examples of both inter-sensor and intra-sensor comparisons in the following sections. These examples demonstrate how the verification framework can be applied to ensure the accuracy and consistency of clinical speech data, ultimately supporting the validity of the analysis and interpretations derived from such data.

% Using step counter as an example, if the amount of total steps in a walk is known, a person could carry the counter to carry out the same walks multiple times. Measure can be made to verify if the step counts recorded by the counter are accurate and consistent across different walks and different counters.  
% In clinical speech AI, sensors such as speech recorders should acquire clinical speech data that carries consistent and accurate clinical information. 
% For instsance, when the clinical task aims to study the vowel space of disordered speakers, for a specific speaker, the F1 and F2 extracted from speech signals acquired from different recorders should be consistent, and match the ground truth values annotated by humans. Rapid measurements of F1 and F2 from speech signals acquired from the same recorder should also be consistent. Examples of inter-sensor and intra-sensor comparison for the verification will be illustrated in the followings.

\noindent {\bf Examples of inter-sensor comparison:} 
% Fahed et al. demonstrated the protocol of verifying devices in clinical speech acquisition \cite{fahed2022comparison}. In the telepractice setting, different speech recording devices such as smartphones, tablets, and professional high-quality studio microphones, were used to collect the speech data of Huntington’s disease (HD). 
Considering the importance of equivalence between speech features derived from data collected by mobile devices and professional standard microphones, Fahed et al. compared the clinical data (patients with Hungtington's disease) collected by devices such as smartphones, tablets, and professional high-quality studio microphones in a telepractice setting \cite{fahed2022comparison}. 
Acoustic measures related to voice quality, such as fundamental frequency (F0), harmonic-to-noise ratio (HNR), jitter, and shimmer were computed from the speech collected from sustained vowel tasks. 
The analysis revealed significant differences in values of harmonic-to-noise (HNR) ratio and statistics of shimmer, while considerable agreement between mobile devices and professional microphones was found in features of F0 and statistics of jitter. 
The hardware verification in the study suggests caution should be taken when using HNR and shimmer in clinical speech AI if the speech data is to be acquired with different mobile devices post deployment. 
% Nevertheless, for each feature, values extracted from mobile devices and professional microphone were shown to be significantly correlated. The analysis further showed that the voice related features could be reliably measured in particular sustained vowels. Acoustic measures such as HNR should be measured cautiously as they could vary notably between devices. 
In \cite{szabo2001voice}, Szabo et al. validated the use of different voice accumulators in continuous monitoring of F0, phonation time, and sound pressure level of speech. The voice accumulator is a device that provides objective documentation of a speaker’s voice use, but there is a large variety in the implementation of acoustic measures. 
Consistency of measurement performed by different models of devices at different times is desired. F0 measured by trained human annotators using professional signal-processing software served as the ground truth for comparison with the measures made by hardware devices. The results showed that the F0 and phonation time measured by different accumulators correlated well with the measures performed by human annotators, but there was a large inter-subject variability when using different voice accumulator models. 
The results suggested that while the accumulators were shown to produce reliable measures of F0 and phonation time, a careful physical setup (placement, firm attachment of the devices) is necessary to further ensure the reliability of the measurements.

\noindent {\bf Examples of intra-sensor verification:} 
Vocal intensity and F0 are important parameters to evaluate the outcomes of phonosurgery. These two-dimensional parameters can be represented by voice range profile (VRP). 
The dual-microphone setup which is commonly employed to measure VRP was verified in \cite{printz2018test}. 
To perform the intra-sensor verification, a test-retest experiment was conducted in which healthy control participants repeated the same voice examination tasks within 6-37 days. The F0, semitone range, sound pressure level, and voice range profile were extracted from the speech collected during these repeated tests. The analysis showed that the majority of acoustic parameters extracted from the test and retest achieved satisfactory Pearson correlation coefficients of over 0.7 on average, suggesting that the dual-microphone setup could be reliably used to acquire speech data for analysis. 

% In speech  data collection, cautions must be taken to ensure speech data captured by hardware and post-processed by firmwares does faithfully preserve representation of clinical speech. The validation confirms the usability of data for model development. When different research teams follow the verification protocol, comparisons between studies become reliable and meaningful.

\subsection{Guidelines for Speech Recording Setup} 
The goal of speech  data collection is to convert the sound pressure of natural speech into an electrical signal that preserves the same characteristics as in the natural speech. The quality of acquired data is not only affected by hardware and firmware of the sensing device, but the physical placement as well. For example, speakers with vocal fatigue may experience difficulties in producing speech with a volume on par with healthy controls. If the distance between the speech recorder and the speaker’s mouth is not adjusted accordingly for the fatigued speech, and collected speech signal could have a low signal-to-noise ratio (SNR) as the speech can be impacted by the surrounding background noise during the recording session. This may limit the useful clinical information available in the speech signal. 

One set of guidelines for speech  data collection for patients with movement disorder were suggested in \cite{rusz2021guidelines} to improve consistency of data quality for clinical assessments. The paper discusses three parts of the  data collection pipeline, namely the recording environment, the recording process, and the acoustic output. The challenges of recording speech from participants with movement disorders is that multiple symptoms such as tremors, dystonic postures, cognitive problems, and fatigue, could alter the loudness of speech. To control the recording environment, it is desired that the noise level is at least 10 dB less than the quietest speech utterance recorded in the session, which could be achieved by adjusting the mouth-to-recorder distance, and reducing  environmental noise with clear instructions. Regarding the selection for speech  data collection, a head-mounted microphone is preferred as it ensures that the distance between the microphone and the mouth is consistent. The frequency response of the microphone should be ideally consistent within the frequency range of human speech. Firmware post-processing of speech data that could alter original characteristics of speech signals, such as speech compression, equalizer, gain control, and noise cancellation, should not be used during acquisition. 

Another tutorial by Svec et al. \cite{svec-ajslp}  focused on microphone selection for voice production studies. As before, the authors suggested that the microphone should cover the complete spectrum of human speech with a flat frequency response. This can be achieved by electret and condenser microphones. Microphones that are colorized for enhanced perception of sound should be avoided. In terms of the directionality, it is preferred that the microphone be positioned in cardioid as the configuration does well to capture speech signals coming from the front. Audio signals coming from other directions, such as background noise, are suppressed in the process. Meanwhile, the dynamic range of microphones is desired to cover the sound level of the loudest phonation and ensures the noise level is minimally 15 dB lower than the voice level. A bit resolution of audio recordings that is higher than consumer-grade, i.e. 16 bit, is recommended. While the recommendations from \cite{rusz2021guidelines} and \cite{svec-ajslp} are useful when scientists have vast control over the  data collection process, this is often not the case.

 \begin{figure}[h!]
  \setlength\belowcaptionskip{-0.8\baselineskip}
  \centering
  \includegraphics[width=0.99\linewidth]{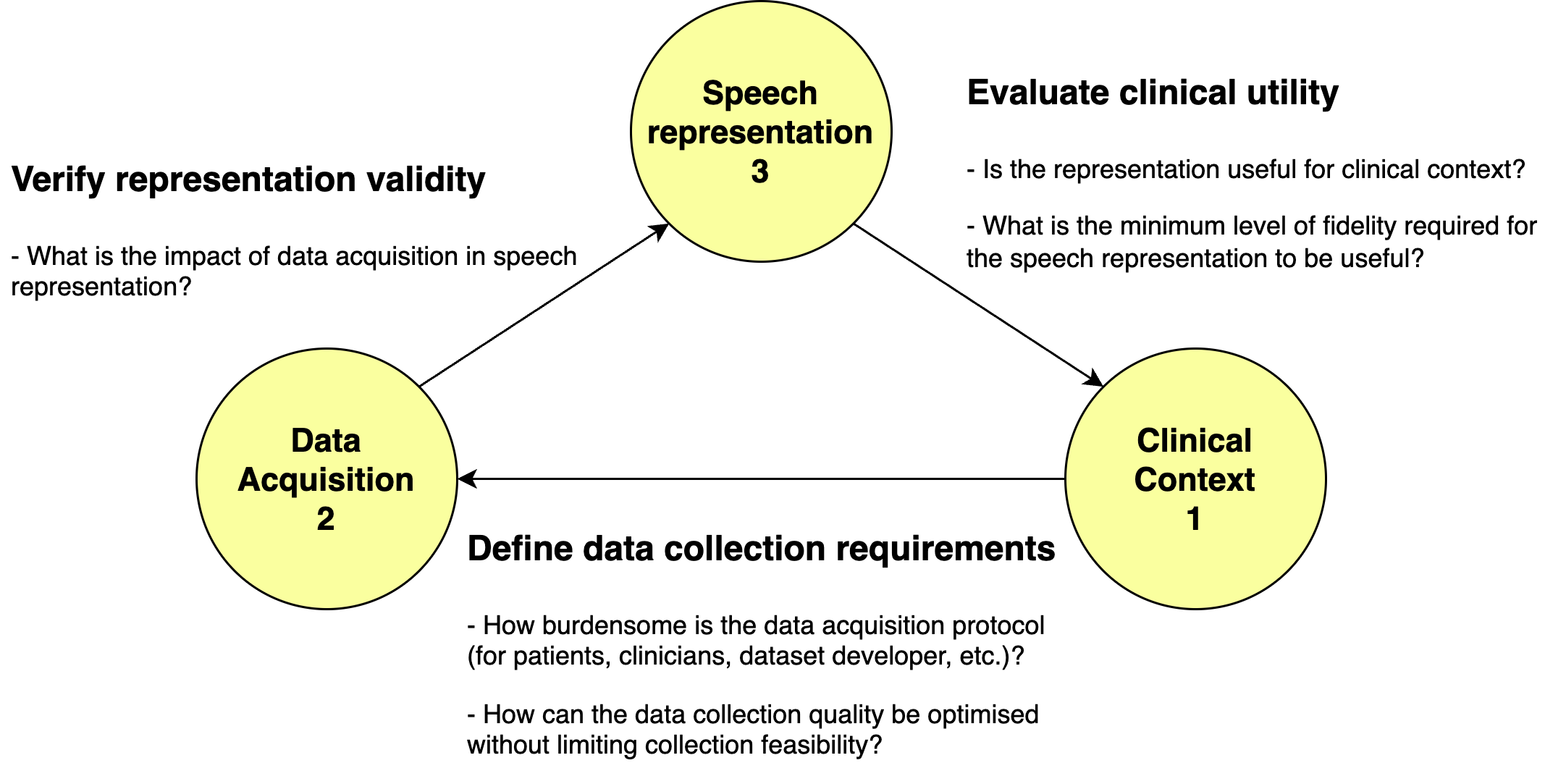}
  \caption{High-level diagram of considerations for speech  data collection.} 
  \label{fig:3-requirements}
\end{figure}

In real-world settings, there is often a trade-off between data quality and data collection feasibility. For example, consider the design of a tool for remote assessment of speech in patients with ALS. Attempting to control the remote data collection protocol per the suggestions above can complicate data collection, making it prohibitively burdensome for patients. This can affect data collection feasibility and hinder model development. In Fig. \ref{fig:3-requirements}, we provide a high-level diagram for thinking about  data collection. For a given clinical context, 
it becomes important to balance the minimization of environmental and human variability during  data collection with the feasibility of data collection. The clinical context defines the  data collection requirements. For example, for real-world data collection studies, it may be required that patients use their own devices. For in-clinic studies, while the patients may use the same device (e.g. an iPad), there will be variability in recording environments across different clinics and times of day. To  balance feasibility and data quality, researchers can measure parameters during  data collection to control noise and provide feedback to speakers for adjustments. For example, a camera can measure the distance and alignment between the speaker and the microphone, and a sound level meter can monitor speech intensity, prompting speakers to adjust their volume or to decrease the distance to the microphone. The desired clinical context in which the clinical AI model will be eventually deployed defines the collection criteria and acceptable mitigation strategies to reduce variability. Nevertheless, there will always be some minimum level of data variability during speech acquisition. 

Next, it’s important to understand how this variability affects speech representation. Variability in  data collection can lead to differences in feature representations. For instance, speech signals can be affected by reverberation, environmental noise, microphone placement, and background sounds, leading to inaccurate measures such as F0 and formant frequencies \cite{pan2000effects, dineley23_interspeech}. Measures of voice, such as pitch, jitter, and shimmer are highly sensitive to external factors. Even minor changes in the input signal can significantly affect these features. On the other hand, measures of language, such as word embeddings, are less sensitive to background noise since modern ASR engines are designed to handle small variations in input signal quality, making linguistic features more robust to environmental noise. Feature-level verification can quantify the impact of  data collection variability on feature measurement error. 

Finally, this error impacts the analytic system’s ability to detect clinically relevant speech changes, potentially reducing their clinical utility for the given clinical context. Feature evaluation criteria (e.g. repeatability, statistical tests, information gain, etc.) 
%(e.g. information gain [xxx]) 
can be used to evaluate the impact of feature noise — caused by variability in  data collection protocols — on the ability to develop a model for the specific clinical context. In addition to evaluating the impact on individual features, external meta features associated with variability (e.g. microphone type, background noise level, distance to microphone) can be collected during acquisition. These meta features can be used both to monitor the acquisition process and as supplementary data for statistical modeling. When trained systems process speech measures for clinical model development, they can use this additional information to mitigate noise effects and improve clinical models.

\section{Speech representations for clinical speech AI}\label{speech_features_speech_measures}

% \subsubsection{Speech features vs. speech measures:}
% If the acquisition of speech data is performed appropriately, i.e. satisfying the steps as discussed in Section \ref{data_acquisition}, clinical speech data acquired from disordered speakers is expected to carry information about the presence, severity, and prognosis of neurological problems. 

The speech, neuroscience, and clinical science communities have long studied the links between health conditions and speech. By focusing on the physiological, neurological, and psychological aspects of speech, meaningful attributes have been derived through acoustic and linguistic analysis, providing a deep understanding of how speech characteristics relate to clinical conditions.

For example, speakers with Parkinson’s disease tend to produce monotonic speech \cite{kim1994monotony, skodda2008speech}. In speakers with dysarthria, speech intelligibility is associated with reductions in vowel space area (VSA), which serves as a proxy for the kinematic displacements of articulators during speech production \cite{turner1995influence, weismer2001acoustic, berisha2014characterizing}. Similarly, speakers with Alzheimer’s disease (AD) exhibit reduced fluency, difficulties in word finding, and an increased number of grammatical mistakes in their speech \cite{mueller2018connected}. These clinical findings demonstrate that certain speech \emph{constructs} are consistently important across a variety of conditions, motivating the development of a broad collection of speech {\em measures} to operationalize these constructs.

\begin{figure}[h!]
  \setlength\belowcaptionskip{-0.8\baselineskip}
  \centering
  \includegraphics[width=0.95\linewidth]{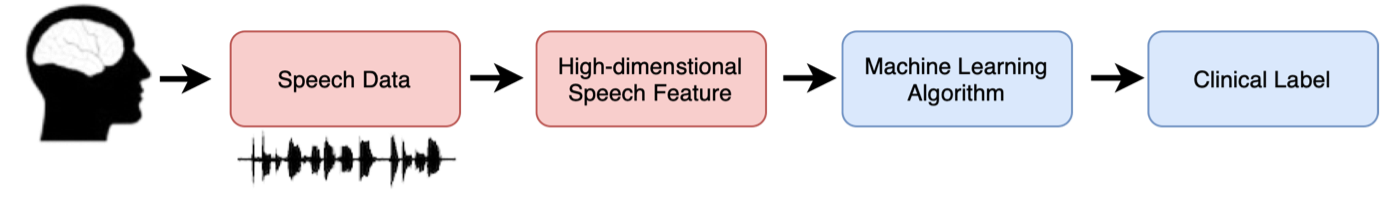}
  \caption{Traditional pipeline of AI model training for clinical speech AI} 
  \label{fig:trad_pipeline}
\end{figure}

In the current AI-driven clinical speech AI landscape, model developers have largely adopted feature representations that are not directly linked to specific clinical constructs. The standard pipeline for training clinical AI models based on speech features is shown in Figure \ref{fig:trad_pipeline}. This pipeline typically uses high-dimensional features as inputs to AI models trained on clinical labels. These features, often borrowed from other speech applications, are not inherently aligned with the theoretical foundations of clinical speech constructs. They lack individual validation, vary in their implementation, and are primarily selected to optimize model accuracy. These representations originate from non-clinical applications like ASR and speaker verification (SV), where features are evaluated based on their ability to reduce error rates, with little consideration for their interpretability in clinical contexts.
However, feature interpretability is critical for adoption in real-world clinical settings \cite{vellido2020importance}. A report issued by the U.S. Food and Drug Administration (FDA) highlighted the risk of automation bias, where humans may over-rely on suggestions from automated systems \cite{us2022clinical}. This bias is more likely to occur when the system provides a single output, rather than a range of options or supporting information. This is particularly concerning when highly complex features are used, as clinicians must rely solely on model outputs to make decisions without understanding the underlying features. In real-world settings, this creates risks, especially when the system underperforms and backtracking the source of the problem becomes difficult.

To develop more interpretable and reliable speech models, we posit that developers should move away from high-dimensional features and instead focus on interpretable {\em speech measures} that are linked to clinically relevant constructs. The relationship between these speech constructs and their corresponding acoustics is often complex and not easily quantifiable. Therefore, especially in high-stakes clinical applications, it is critical to operationalize abstract constructs into objective measures that can be scrutinized for validity and reliability. %for example, using cepstral peak prominence (CPP) to measure voice quality, phone log-likelihood ratio to measure articulatory precision, etc.

\begin{figure}[t!]
  \setlength\belowcaptionskip{-0.8\baselineskip}
  \centering
  \includegraphics[width=0.95\linewidth]{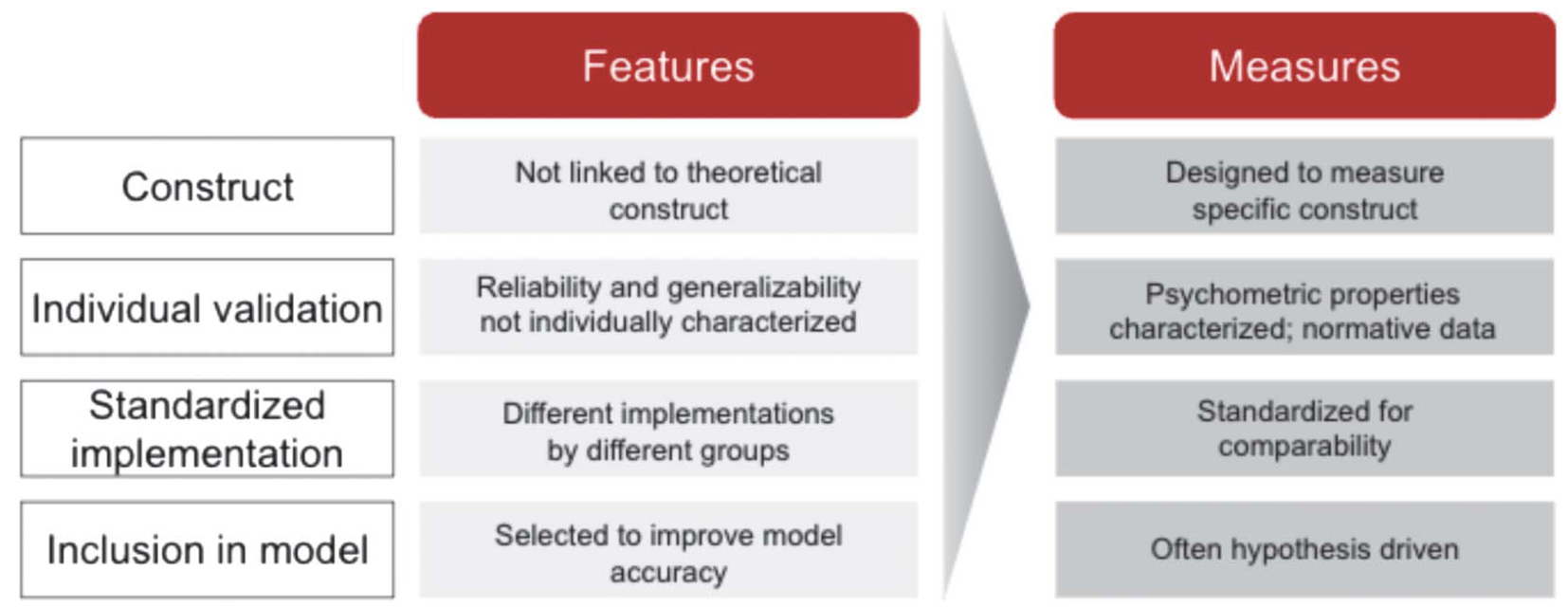}
  \caption{A comparative overview between speech features and speech measure \cite{liss2024operationalizing}.} 
  \label{fig:features_measures}
\end{figure}

The proposed transition from conventional speech features to interpretable speech measures is illustrated in Figure \ref{fig:features_measures}. Speech features used to date, typyically  have been based on short-time analysis techniques such as the short-time Fourier transform, MFCCs, and the mel-filterbank, each of which provides a different description of energy distribution across frequencies over time (Fig. \ref{fig:trad_pipeline}). These features have been used to assess a variety of disorders, including depression \cite{cummins2015review}, Parkinson’s disease \cite{vasquez2018multimodal}, oral cancer \cite{halpern20_interspeech}, and hearing impairment \cite{moller1999analysis}. More recently, deep learning models have extracted features to assess diseases such as Parkinson’s, oral cancer, stuttering, and cognitive impairment \cite{quintas20_interspeech, bayerl22b_interspeech, braun23_interspeech}. These neural network embeddings are trained to capture implicit and abstract information from speech signals, differentiating between healthy and disordered speech.

While attempts have been made to interpret these embeddings through visualization techniques like principal component analysis (PCA) or $t$-distributed stochastic neighbor embedding ($t$-SNE) \cite{van2008visualizing}, these are dimensionality reduction technques that cannot be used to infer clinical relevance. Clustering embeddings with additional labels, such as gender or language, has been used to explain how they encapsulate diagnostic information, but these approaches do not always align with the goal of linking features to clinically validated constructs.

Although model performance, statistical analyses, or visualizations may partly indicate the efficacy of speech features, these methods often misalign with the clinical approach to analyzing speech. They are not specifically designed to link to well-defined theoretical constructs. For instance, a clinical practitioner may struggle to interpret how the 13th MFCC coefficient relates to a clinical condition. The methods for extracting these features also vary \cite{zheng2001comparison}, and without individual validation, the reliability and generalizability of these features have not been established. As a result, when researchers attempt to replicate published work, there is a high likelihood of producing inconsistent performance and analytical results.

% On the other hand, a speech measure aims to provide an objective and interpretable measure of speech constructs that is closely tied to clinical assessment and diagnosis.
In contrast, speech measures are operationally defined to represent theoretical constructs directly related to the disease of interest and can be derived using automated methods. For example, speech intelligibility can be measured using ASR systems that quantify word error rates as a proxy for clarity \cite{meyer15_interspeech, schuster2006evaluation}. Acoustic features such as vowel space area (VSA) reduction, which relates to articulatory precision, can be automatically extracted to assess speech motor control in conditions such as dysarthria \cite{kim2011vowel, kim2011acoustic, lansford2014dysarthria}. Similarly, prosodic features like pitch variability can be measured through algorithms analyzing fundamental frequency contours, aiding in the quantification of monotonicity in speech for conditions such as Parkinson’s disease \cite{kim1994monotony}. These measures align more closely with clinical constructs, offering greater consistency and scalability in clinical speech AI applications. To ensure these measures accurately reflect clinical constructs, ground truth data can be established using existing clinical instruments to establish the validity and reliability of measures. For instance, in dysarthric speech, perceptual evaluation of key clinical features has a well-documented history; for example \cite{darley1969differential}.
% Other speech characteristics can also be measured by auditory-perceptual evaluation. In \cite{darley1969differential}, dysarthric speech was rated by human listeners using 38 perceptual dimensions of speech. Each dimension is rated by a 7-point scale ranging from ``typical’’ to ``severely atypical’’. 

 \begin{figure}[h!]
  \setlength\belowcaptionskip{-0.8\baselineskip}
  \centering
  \includegraphics[width=0.75\linewidth]{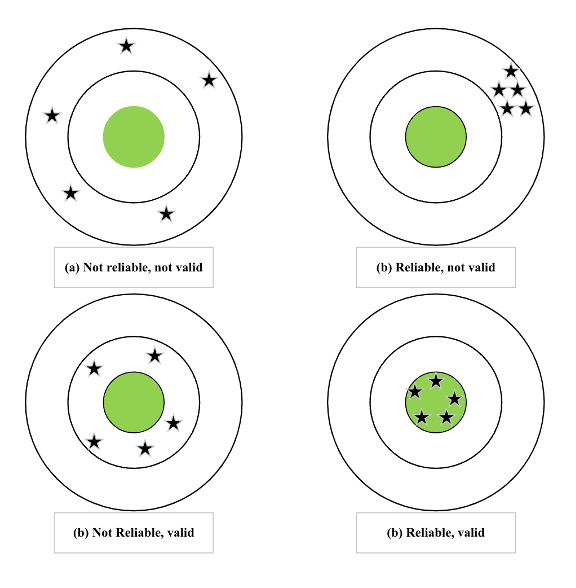}
  \caption{Conceptual relationship between reliability and validity of a measure. Scattered distribution of data point indicates a low reliability and high variability of measures. The radius of the circle refers to the degree of validity. A small radius indicates high validity of the measure. Measures falling in the green zone is ideal as they have both good reliability and validity.} 
  \label{fig:reliabitliy-validity}
\end{figure}

\noindent {\bf Valid and Reliable Measures:} A valid speech measure should achieve both high reliability and validity (Figure \ref{fig:reliabitliy-validity}). Failure to meet either criteria can result in unreliable or meaningless measures, inaccurate statistical test results, and potentially misleading clinical decisions. Reliability refers to the consistency of a measure \cite{allen2001introduction}. In clinical research, two types of reliability are commonly considered: test-retest reliability internal consistency.

Studies have shown that some commonly used speech measures lack reliability or validity. For example, in \cite{stegmann2020repeatability}, the repeatability of popular speech and language features extracted from open-source toolkits, such as openSMILE \cite{eyben2010opensmile} and Praat \cite{boersma2018praat}, was evaluated on several types of disordered speech. The intraclass correlation coefficient (ICC) and within-subjects coefficient of variation (WSCV) revealed that most of the features tested did not meet the acceptable thresholds for clinical decision-making \cite{fleiss2011design, portney2009foundations}.

Similarly, in \cite{iter2018automatic}, a computational approach was proposed to assess the speech of individuals with schizophrenia by measuring incoherence and tangentiality. While the proposed metrics showed significant statistical differences between the schizophrenia group and healthy controls, as well as strong classification performance, Hitczenko et al. \cite{hitczenko2021automated} demonstrated that the tangentiality measure did not show significant differences between the groups when tested on a different dataset. Furthermore, the coherence measure was found to correlate more strongly with sentence length and speaker socio-demographics rather than clinically validated markers of thought disorder. These examples highlight the importance of rigorously evaluating both the reliability and validity of measures in clinical speech analytic research.

Validating measures in terms of reliability and validity is essential and is a routine practice in clinical research. Ensuring reliability and validity makes measures clinically meaningful, closely tied to the construct of interest, and repeatable, enabling consistent comparisons between studies conducted by different research teams.

\begin{itemize}
	\item Test-retest reliability assesses whether a construct is measured consistently over time. For instance, in \cite{park2019test}, test-retest reliability was used to validate acoustic, aerodynamic, and perceptual measures in voice assessments. Measures obtained over five consecutive days showed moderate to high intraclass correlation coefficients (ICC=0.64-0.99), whereas subglottal pressure and perceptual voice evaluation had low ICCs, indicating poor reliability.
	\item Internal consistency assesses whether different items (e.g., speech elicitation tasks) in the same test yield consistent results. This can be evaluated using the split-halves method, where the set of items is divided into two parts, and the correlation between the two reflects internal consistency. For example, in \cite{dos2016protocol}, Cronbach’s alpha was used to validate a clinical assessment tool for dysarthric speech, comparing human-rated intelligibility scores from sentence-reading and single-word tasks. 
\end{itemize}

The examples above can be generalized to validation of individual speech measures. To facilitate reliability analyses, the protocol for speech data collection (as discussed in Section \ref{data_acquisition}) should ensure that relevant data are collected. For example, speech representations can be collected over consecutive days to evaluate test-retest reliability; and they can be evaluated over different elicitation tasks and compared to evaluate for internal consistency. 

%A valid measure must also satisfy the criteria of validity, meaning it should measure what it claims to measure. 
Validity of individual features can be assessed based on their relationship with existing, validated measures of the construct of interest (convergent validity). In \cite{rowe2021validation}, acoustic-based speech measures—such as the duration ratio of different syllable sequences, variability in syllable duration, and F2 slope—were compared with validated perceptual ratings of coordination, consistency, and speed in ALS speech. The Pearson and Spearman correlation coefficients confirmed the validity of these speech measures for profiling articulatory deficits in motor speech disorders. Similarly, in \cite{yawer2023reliability}, Yawer et al. evaluated the external validity of a publicly available speech AI tool designed to assess stress. The tool’s stress measures were compared with the well-established Perceived Stress Scale \cite{reis2010perceived}. The poor correlation between the two measures suggested that the widely available tool lacked sufficient validity for clinical use.

Validity analyses require not only comparisons of speech measures with existing clinical scales but also consideration of the quality of these scales themselves. Many existing clinical measures, particularly in speech, rely heavily on human perceptual ratings, which are inherently subjective and prone to variability \cite{bunton2007listener}. These perceptual scales can differ based on rater experience, task complexity, and environmental factors, contributing to variability in the `ground truth' labels. As a result, it is important to account for this uncertainty when evaluating the validity of automated measures, as even a high correlation with these perceptual scales may not fully capture the true clinical construct. Further, improving the reliability of the clinical scales used as references through standardization or advanced rater training could enhance the robustness of validity evaluations.

Following reliability and validity analyses, the next step is to compare deviations from normative distributions in clinical populations. Normative data, typically collected from healthy control groups, serve as a baseline for identifying abnormal patterns in clinical populations. Statistical tests can then be applied to evaluate the efficacy of the derived measures. For example, in developing a measure of breathiness in speakers with Parkinson’s disease \cite{darling2020impact, darling2022longitudinal}, measures such as speech rate, oral pressure, inspiratory duration, and rib cage volume were recorded based on clinical hypotheses \cite{solomon1993speech}. Statistical comparisons between the clinical group and the normative baseline revealed significant differences in speech rate, oral pressures, and rib cage volume, supporting their validity as discriminative clinical measures for Parkinson's Disease. However, inspiratory duration did not show significant differences, suggesting it may not provide any discriminative value as a feature in this population.

When does a measure become valid and reliable? Rather than crossing a clear-cut threshold, the process of establishing validity and reliability is better understood as a continuum. A measure gradually gains credibility as evidence accumulates over time. This process starts with demonstrating that the measure consistently reflects the clinical construct it aims to assess. As the evidence grows, statistical analyses show increasingly strong and significant relationships with existing, validated measures. The measure’s reliability is further strengthened when it consistently performs across different contexts, populations, and conditions. Real-world validation, where the measure is tested in practical clinical settings, adds another layer of confidence. Acceptance by the clinical and research communities follows as the body of evidence supporting the measure's validity and reliability becomes robust enough to inspire trust. Rather than a fixed point, validity is a continuous process of gathering and demonstrating evidence.

  \begin{figure}[h!]
  \setlength\belowcaptionskip{-0.8\baselineskip}
  \centering
  \includegraphics[width=\linewidth]{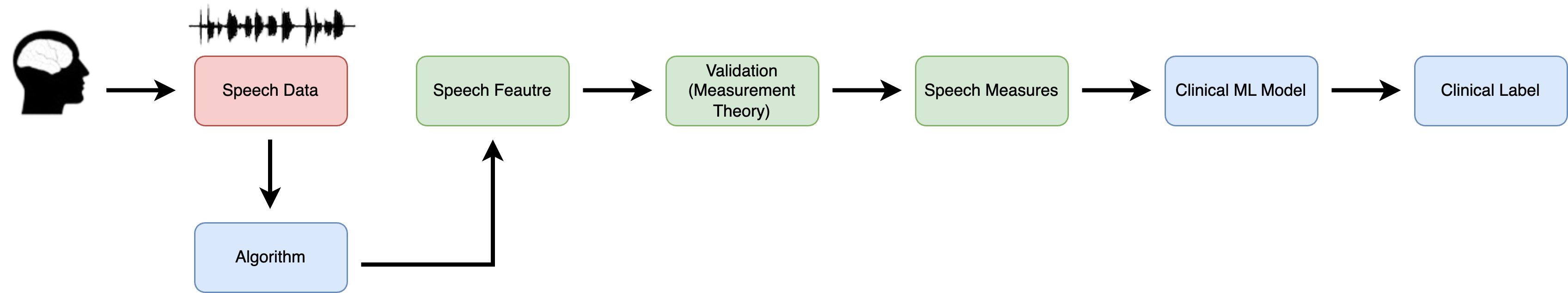}
  \caption{Proposed clinical speech model development using the concept of speech measure.} 
  \label{fig:speech_measure_pipeline}
\end{figure}

 % \section{Moving from speech features to speech measures}\label{Technical_spechfeatures_to_speechmeasures}

\noindent{\bf A pipeline for clinical speech AI:} The traditional pipeline for speech AI is inadequate for clinical applications. To overcome these limitations, algorithm designers should adopt the concept of validated speech measures rather than relying solely on traditional speech features. The development and validation of speech measures should become a standard part of the clinical speech AI pipeline. Figure \ref{fig:speech_measure_pipeline} illustrates the process of integrating speech measures into AI  model development for clinical speech AI. After speech data is collected, computational methods are used to extract parameters. These parameters must undergo reliability and validity testing, in line with measurement theory, before being considered valid ``speech measures" for clinical use. In this pipeline, clinical AI models—unlike general AI models (shown in Figure \ref{fig:trad_pipeline})—are characterized by simplicity, interpretability, and the use of validated speech measures as input features. These models should provide clear insights into how the input features drive decision-making, ensuring that clinicians and developers can confirm the system is effectively utilizing validated speech measures.

A key component in the new pipeline is the development of the representations. In the following sections, we will introduce various approaches to deriving speech measures. Similar to hardware verification in speech  data collection, extracted speech features must undergo analytical validation for reliability and validity before being considered valid measures. This validation process should involve both algorithm designers and clinical experts to ensure each measure is clinically interpretable and comparable to those obtained through gold-standard methods, per the previous section.  %clinical speech AI.  

\subsection{Knowledge-driven speech measure design} 
To design a set of measures for assessing a specific disease, the most straightforward approach is to select measures that have been validated by existing studies. 
We term these measures ``knowledge-driven" as they can be obtained through computational analysis of the speech data. 
A review by Voleti et al. summarized a series of measures derived from the acoustic and language aspect of speech to assess cognitive and thought disorders \cite{voleti2019review}. In acoustics, prosodic measures such as duration of voiced segments, duration of silent segments, loudness, periodicity, and non-verbal cues such as interruptions, interjections, natural turns, reflect irregularities in rhythm and timing of speech \cite{roark2011spoken, konig2015automatic}; Articulatory measures such as formant frequencies, vowel space area, and formant trajectories over time describe atypical movements of speech articulator \cite{horwitz2016relation}. Vocal quality measures, such as jitter, shimmer and harmonic-to-noise ratio (HNR), describe how air flows through the lung and glottis. These acoustic-based measures have clinical meaning, and have been shown to be effective in aiding assesmment of different clinical conditions, from dysarthria to thought disorders.

Many conditions, such as cognitive and thought disorders, are associated with atypical language use. As a result, speech measures can be derived from text transcriptions using natural language processing (NLP) techniques. From a language perspective, speech intelligibility can be assessed by analyzing automatic speech recognition results, which highlight incorrect word production, unintended insertions, and deletions. Lexical diversity, which reflects unique vocabulary usage, can be calculated by analyzing the number of unique words relative to the total word count in the speech sample \cite{brunet1978vocabulaire}. Additionally, the lexical and syntactic complexity of spoken sentences can be analyzed using parse trees to derive grammatical structure, with language complexity measured by statistics related to the branches of these trees \cite{yngve1960model, berg2011structure}. These language-based measures have proven effective in assessing disorders related to language and cognition.

The measures summarized above have been applied in the evaluation of various disorders. For example, measures such as F0, formant frequencies, voice quality, and speech rate have been used to analyze conditions like autism spectrum disorder in children \cite{lee2023knowledge}, Parkinson’s disease \cite{rusz2011quantitative, harel2004acoustic, tsanas2013objective}, and Huntington’s disease \cite{romana2020classification, perez2018classification}. Additionally, measures of language diversity have been used to assess stuttering in children \cite{charest2020properties} and narrative discourse in speakers with aphasia \cite{fergadiotis2013measuring}.

Knowledge-driven measures, particularly those validated in clinical studies, offer an initial baseline for clinical AI models. They are especially valuable when dealing with new clinical applications or rare conditions with limited data. In such cases, starting with measures that are already known to change with the condition provides a reliable baseline. These validated measures can be directly incorporated into AI models, offering a level of robustness and interpretability. However, one tradeoff is that knowledge-driven measures focus on pre-established patterns and may not uncover new or previously-unknown relationships between speech characteristics and the disease.

\begin{figure}[ht!]
  \setlength\belowcaptionskip{-0.8\baselineskip}
  \centering
  \includegraphics[width=\linewidth]{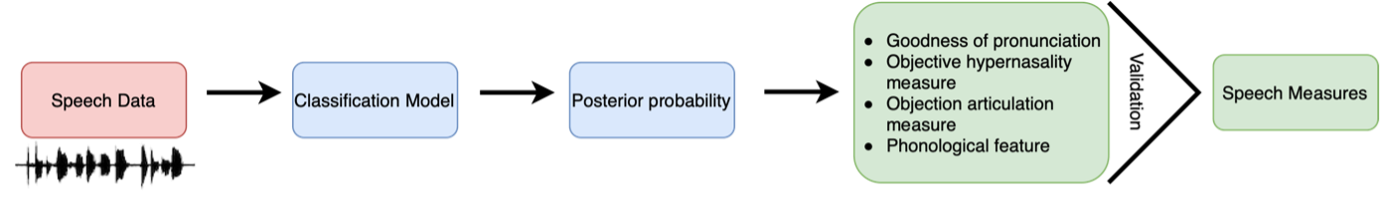}
  \caption{Using acoustic models to derive speech measures.} 
  \label{fig:Data_driven}
\end{figure}

% \subsection{Data-driven approach to speech measure design}
\subsection{Acoustic modeling of speech to derive speech measures}\label{acoustic_model}

While clinical speech samples are limited across many conditions, speech data from the general population is abundant \cite{panayotov2015librispeech, ardila-etal-2020-common, NAGRANI2020101027, chen21o_interspeech}. This creates an opportunity to leverage general population data to train models that capture representations of healthy speech, which can then be applied to assess various aspects of clinical speech.

One example of this approach is acoustic modeling, a core component of more conventional ASR systems. These models are typically trained on speech data from the general population to recognize linguistic units from short segments of speech. With the right training setup, acoustic models can capture important articulatory information. It is well known that clinical speech conditions often manifest at the segmental level—such as distorted sounds, persistent mispronunciations, and misarticulations, cf. \cite{bayerl22b_interspeech, changawala24_interspeech}.

Ideally, a poorly produced phone would be assigned a low probability by the acoustic model. As a result, the model’s probability outputs for linguistic units can reflect the quality of speech production. With proper validation, features derived from these probability outputs can serve as valuable measures for assessing clinical speech. However, it is important to note that ASR systems are not equally reliable across all languages. In particular, for analyses conducted outside the English-speaking domain~\cite{schubert24_interspeech,KM2021}, acoustic errors may still occur even with flawless pronunciation, adding another layer of complexity to the analysis. Therefore, caution must be exercised when interpreting results in languages where the ASR system may not be as robust.

Acoustic model probabilities play a critical role in the goodness of pronunciation (GOP) score,  first introduced in \cite{witt2000phone}. The GOP utilizes the posterior probability of phone recognition to quantify the discrepancy between a speech sound’s actual and expected realization. This measure has been widely used in computer-assisted pronunciation training (CAPT) \cite{witt2000phone, hu2015improved}. Additionally, GOP measures computed from pathological speech have been shown to significantly correlate with clinical assessments of neurological and anatomical disorders \cite{fontan2015predicting}.

By using domain-specific labels in acoustic modeling, posterior probability outputs can offer different clinical interpretations, capturing the quality of speech production from various perspectives. For example, the place and manner of articulation (also known as phonological features) describe where in the vocal tract speech sounds are produced and how the airflow is affected during speech. Speech sound units can be grouped according to these phonological features, which are well understood by clinicians and form part of speech science curricula.

In \cite{jiao2017interpretable}, a recurrent neural network (RNN) was trained using MFCCs and phonological features to classify speech into 16 place and manner of articulation categories. A metric was then computed from the classifier outputs to measure how these classifications differed from those in healthy speakers, allowing for an assessment of articulation abilities in speakers with dysarthria. While the classification model performed well, the derived measures were compared with perceptual ratings of individual articulation items. Results showed that 8 out of the 16 classes had significant and substantial correlations with the perceptual ratings, confirming that certain articulatory measures were reliably linked to the clinical construct.

\begin{figure}[h!]
  \setlength\belowcaptionskip{-0.85\baselineskip}
  \centering
  \includegraphics[width=0.95\linewidth]{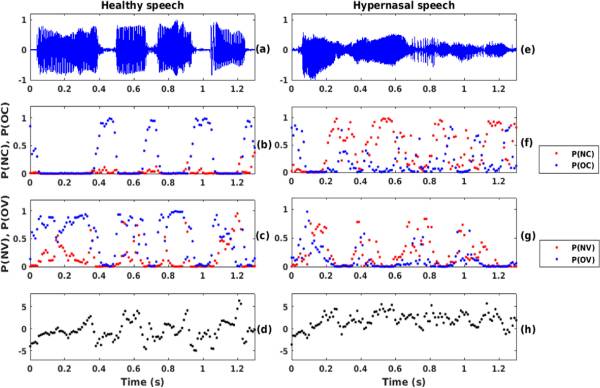}
  \caption{An example algorithm that uses posterior outputs from a pre-trained acoustic model to derive an objective hypernasality measure (OHM) for speech from children with cleft-palate \cite{mathad2021deep}. (a) and (e) illustrate the waveform healthy and hypernasal speech; (b) and (f) illustrate the frame-wise posterior probability of nasal consonants (NC) and oral consonants (OC); (c) and (g) show the frame-wise posterior probability of nasal vowels (NV) and oral vowels (OV); The frame-wise OHM measures are depicted in (d) and (h).} 
  \label{fig:mathad_2021}
\end{figure}

In \cite{mathad2021deep}, the authors propose a new algorithm for assessing hypernasality in cleft speech based on a model trained only on speech from the general population. First, English consonants and vowels from a corpus of speech from the general population were categorized into four categories: nasal consonants (NC), nasal vowels (NV), oral consonants (OC), and oral vowels (OV). An acoustic model was trained on speech from the general population using MFCC features as input to classify each speech frame into one of these four categories. After training, the model was  applied to clinical speech samples and the posterior probabilities from the model were then used to derive an objective hypernasality measure (OHM), which detects nasality in both consonants and vowels.

Figure \ref{fig:mathad_2021} illustrates the utterance ``buy baby a bib" produced by both a healthy speaker and a speaker with cleft palate (CP). The sentence does not contain any nasal consonants, meaning no nasal cues are expected in typical speech production. However, the classifier identified nasality in both consonant and vowel segments of the CP speaker's speech, highlighted in red in Figures \ref{fig:mathad_2021}(f) and \ref{fig:mathad_2021}(g). The OHM values for the CP speaker (Figure \ref{fig:mathad_2021}(h)) were significantly higher than those for the control speaker (Figure \ref{fig:mathad_2021}(d)), indicating hypernasality in CP speech.

To validate the proposed OHM, both internal and external reliability were tested. Internal reliability was assessed using a split-half strategy, where the OHMs computed from half of each speaker’s utterances were compared with the OHMs from the other half. An internal reliability score of 0.9 indicated consistency in OHM across different speech content. External validity, which tests the generalizability of the measure to speech data from different studies, was evaluated using an external dataset that was not involved in model training. An OHM-rater reliability of 0.7 suggested that the OHM had reliability on the same order as the inter-rater variability.

\begin{figure}[h!]
  \setlength\belowcaptionskip{-0.8\baselineskip}
  \centering
  \includegraphics[width=0.80\linewidth]{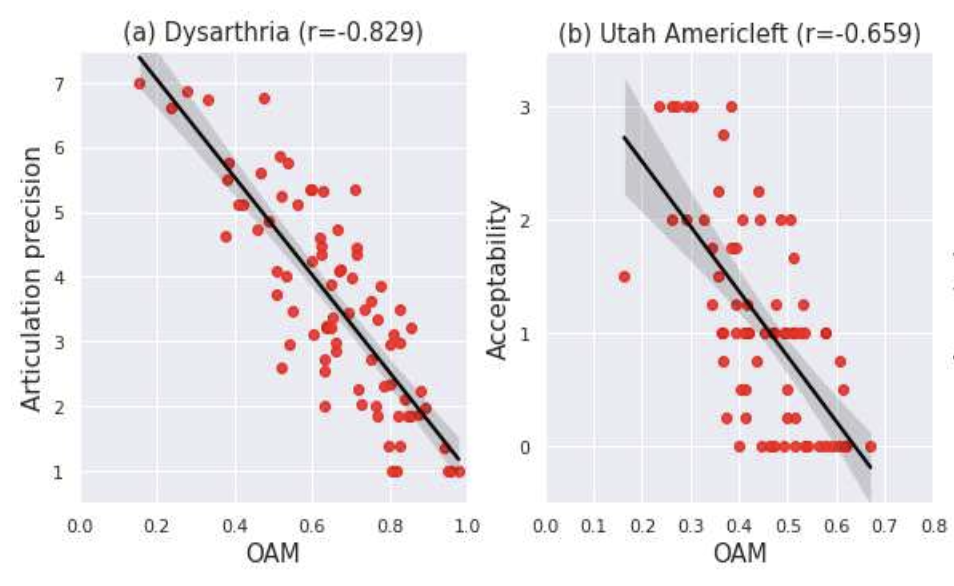}
  \caption{Scatter plots of speaker-level OAM vs. a) perceptual ratings for dysarthric speech; b) acceptability rating of cleft palate speech.} 
  \label{fig:mathad_2022_cv}
\end{figure}

A similar approach was also used to evaluate the fidelity of consonant vowel transitions in \cite{mathad2022consonant}. Imprecise articulation of specific speech sounds can affect neighboring sounds, particularly at consonant-vowel transitions, which are more challenging for production due to the rapid changes in articulatory coordination required. These transitions have been shown to be sensitive to certain conditions \cite{stevens1981evidence, hedrick1993effect}. In \cite{mathad2022consonant}, articulation precision was measured from consonant-vowel (CV) co-articulated segments. A convolutional neural network (CNN) was trained on mel-spectrogram representations of speech from a general population dataset to classify between 20 English consonants based on the CV segments. Once trained, the model was applied to clinical speech data to assess articulation precision. An objective articulation measure (OAM) was derived from the posterior probabilities output by the trained CNN. To validate the clinical relevance of the OAM, the scores for individual speakers were compared with ground-truth perceptual ratings of articulation precision and speech acceptability across both dysarthria and cleft speech. The results showed strong absolute correlations of 0.8 for dysarthric speech and 0.7 for cleft palate speech.

\begin{figure}[h!]
  \setlength\belowcaptionskip{-0.8\baselineskip}
  \centering
  \includegraphics[width=0.90\linewidth]{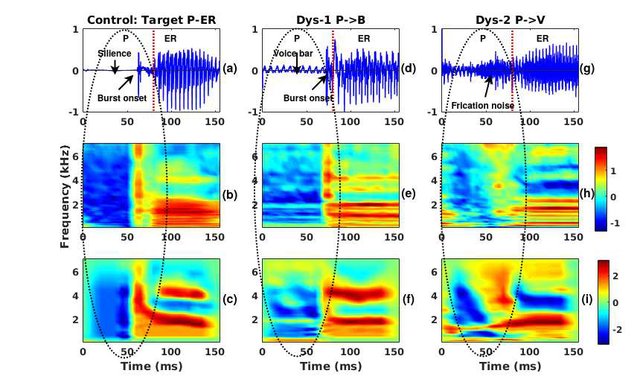}
  \caption{An example of utilizing saliency maps to explain how a CNN acoustic model processes consonant-vowel speech segments produced by the control speaker and the dysarthric speakers \cite{mathad2022consonant}. (a), (b) and (d) illustrates the speech signals. Their time-frequency representations are illustrated in (b), (e), and (h) respectively. The saliency maps produced by the CNN are depicted in (c), (f), and (i).} 
  \label{fig:mathad_2022}
\end{figure}

Figure \ref{fig:mathad_2022} illustrates saliency maps generated by the CNN-based acoustic model when processing consonant-vowel speech segments produced by a control speaker and a speaker with dysarthria. The same consonant-vowel sequence was articulated differently due to the clinical condition, and the visualization highlighted the areas of the segment that reflected problematic articulation.
In Figure \ref{fig:mathad_2022} (d)-(e), the substitution error of /p/\textrightarrow/b/ is characterized by the difference in voicing characteristics, which can be observed in low-frequency region of the spectrogram. The saliency map in Figure \ref{fig:mathad_2022} (f) showed the CNN was focused on the low frequency region when processing the substitution error. 
In Figure \ref{fig:mathad_2022} (g)-(h), the substitution error of /p/\textrightarrow/v/ is distinguished by the continuous turbulent airflow rather than a sudden burst in consonant production, along with the differences in consonant-vowel transition. Both aspects were highlighted by the saliency map in Figure \ref{fig:mathad_2022} (i).
These visualizations provide evidence that the OAM was successfully capturing clinically relevant speech features. Additionally, the OAM was shown to outperform the goodness of pronunciation (GOP) measure across different speech datasets, achieving smaller coefficients of variation. This suggests that the OAM is more reliable than the GOP for measuring related clinical constructs.
%segment is /p/-/e/. 
%In speaker ‘Dys-1’, the /p/ was substituted by /b/. The two consonants differ in terms of the voicing characteristics. The sailency map showed that the voiced component of /b/, which is distributed in lower-freqnecy region, was captured by the CNN. In speaker ‘Dys-2’, the /p/ was substituted by /v/. The burst in /p/ was not inducated by the saliency map due to the substitution, while the characteristic changes in formant-transition caused by /v/ was captured by the CNN, as shown in Figure \cite{mathad2022consonant}(f). 
%was shown to significantly correlate to ground-truth measures of articulation precision in dysarthria speech and acceptability in cleft palate speech, thus confirming the validity of the OAM proposed. 

Similar approaches for deriving speech measures based on posterior features from acoustic models have been applied in studies assessing Parkinson’s disease \cite{cernak2017characterisation}, voice disorders \cite{liu2019acoustical}, and children’s speech disorders \cite{shahin2019anomaly}. 
%More recently, pre-trained models like WAV2VEC... 
These examples demonstrate that the use of acoustic models to generate speech measures has broad applicability and potential for assessing a wide range of speech-related conditions.

A general pipeline for deriving speech measures based on acoustic modeling from the general population is shown in Figure \ref{fig:Data_driven}. In this approach, a pre-trained model — trained on speech from the general population—is used to evaluate clinical speech samples. The outputs of these models are validated against existing clinical constructs. For some representations, this process is straightforward. For example, the outputs 
%posteriors 
of the OAM, OHM, and GOP models described above are inherently interpretable. These outputs correspond to well-understood speech classes in the clinical community, making it possible to reason about the relationships between model predictions and clinical conditions. For instance, since dysarthria often results in hypernasality, the OHM model would be expected to yield higher posterior probabilities for nasal vowel and consonant classes in dysarthric speakers that present with this symptom.

%Herman: Move the intro of wav2vec2 to here so it looks more connected.
More recently, self-supervised pre-trained models such as wav2vec2.0, HuBERT and WavLM, etc. have been increasingly applied to clinical speech. These models have been adopted to predicting progression of amyotrophic lateral sclerosis \cite{wang24e_interspeech}, severity of depression \cite{dumpala24b_interspeech}, speech production skills in children \cite{lee24e_interspeech}, etc. In these approaches, high-dimensional speech representations are extracted to encapsulate the clinical condition carried in speech signal, with the expectation that they will be effective in the downstream clinical prediction/classification tasks.  
However, interpretability becomes more challenge with these models.  
% However, interpretability becomes more challenging with these self-supervised learning models. 
%However, interpretability becomes more challenging with models like Wav2vec2.0, which is based on self-supervised learning. 
Unlike the explicitly defined outputs of the OAM, OHM, or GOP models, self-supervised learning models generate latent speech representations that are not directly linked to clinically relevant classes or features. These representations capture a wide array of phonetic and acoustic information in an unsupervised manner, making it difficult to understand exactly how these features relate to specific clinical conditions. Without clear, interpretable outputs, linking these latent features to clinical constructs requires additional steps, such as mapping them to known speech measures or performing post-hoc analyses. This adds complexity to both the interpretability and clinical application of the model, as clinicians cannot easily reason about what aspects of the speech signal the model is using to make its predictions.
%Parts about Trillsson and similar models...
In addition to designing models focused on articulatory features which primarily address characteristics in short segments of speech (such as OAM, OHM, and GOP as previously discussed), recently there are models designed to analyze the suprasegmental characteristics of speech. For example, conformer-based paralinguistic speech models have been trained to process speech signal using a longer window (i.e. a width of 2 seconds) \cite{shor2022universal, shor22_interspeech, lee2024distilled}. Evaluations on various non-semantic speech tasks \cite{shor20_interspeech} showed that these paralinguistic speech models outperformed those using segmental speech analysis windows (e.g. 20ms), such as wav2vec2.0. While applying these paralinguistic speech models directly to clinical speech still faces the interpretability challenges as with other self-supervised learning models, understanding how the models can be linked to clinical constructs can help derive clinically interpretable measures related to paralinguistics of speech.

% \cite{vikram2018estimation, fritsch2021utterance, arai2020predicting, wang2019child}. 

\subsection{Using speech and language to predict existing speech-based perceptual scales}

\begin{figure}[h!]
  \setlength\belowcaptionskip{-0.8\baselineskip}
  \centering
  \includegraphics[width=\linewidth]{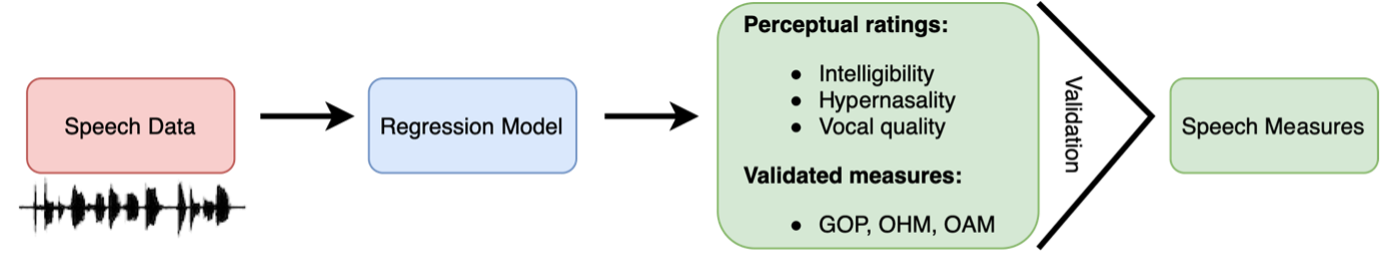}
  \caption{Using AI to predict perceptual ratings as speech measures based on input speech data.} 
  \label{fig:predict_validated_measures}
\end{figure}

Assessment in many clinical speech applications has traditionally relied on perceptual judgments made by experienced clinicians or self-reports by patients or caregivers, focusing on constructs such as articulation precision in dysarthria \cite{kent1996hearing}, social communication skills in schizophrenia or bipolar disorder \cite{goldstein2006social, lee2013social}, and vocal quality in dysphonia \cite{wuyts2000dysphonia}. Since these scales are closely tied to the speech signal, it is reasonable to expect that AI models can learn key perceptual cues during training to perform these predictions. Properly trained and validated AI models hold significant potential to automate these assessments by predicting existing speech-based perceptual scales directly from speech and language data. This approach seeks to develop objective proxies for these well-validated clinical assessment.

We note that this approach is in contrast to other efforts that aim to predict diagnostic labels, such as Parkinson's disease or schizophrenia diagnosis, directly from speech data. Approaches focused on prediction of diagnostic labels face questionable feasibility, as such complex conditions likely require more than just speech data for accurate diagnosis \cite{berisha2024responsible}. In contrast, predicting predicting clinically-important perceptual outcomes that are inherently tied to the speech signal itself makes the problem more tractable and increases the likelihood of success.

\noindent{\bf Predicting Perceptual Measures from Speech Acoustics:} For several conditions, perceptual assessment of patient-produced speech is clinically valuable. For instance, in the assessment of dysarthria, perceptual ratings such as articulation precision and speech intelligibility, provided by trained clinicians, are important for differential diagnosis and tracking patient progress \cite{bunton2007listener), borrie_2012_perceptual}. Given their importance, these perceptual measures offer valuable targets for AI models. The goal of these models is to replicate how professional listeners assess clinically relevant acoustic properties of speech directly from acoustic data.

As illustrated in Figure \ref{fig:predict_validated_measures}, once clinical speech data and corresponding perceptual ratings are collected, a supervised learning model can be trained to predict these ratings from the speech’s acoustic features. With sufficient validation, these predicted ratings can serve as reliable and objective speech measures for clinical assessment of conditions like dysarthria.

\begin{figure}[h!]
  \setlength\belowcaptionskip{-0.85\baselineskip}
  \centering
  \includegraphics[width=0.80\linewidth]{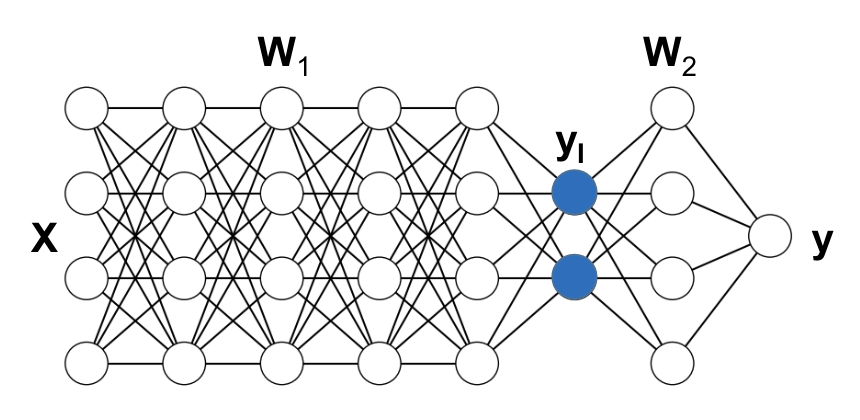}
  \caption{An example of inserting an interpretable DNN layer (highligthed in blue) that contains predicted perceptual measures of speech \cite{tu2017interpretable} $W_{1}$ is responsible for predicting speech measures given input $X$, whereas $W_{2}$ is responsible for performing clinical diagnosis based on the predicted measures $y_{1}$.} 
  \label{fig:tu_is_2017}
\end{figure}

Tu et al. \cite{tu2017interpretable} demonstrated this approach by utilizing a deep neural network (DNN) to predict perceptual ratings for the objective assessment of dysarthric speech. As shown in Figure \ref{fig:tu_is_2017}, an intermediate layer in the DNN served as an information bottleneck, trained using regression loss to predict multiple perceptual dimensions such as nasality, vocal quality, articulatory precision, and prosody. This interpretable layer acted as a regularizer, constraining the solution space of the neural network and preventing overfitting, especially when training data is limited.

Using the predicted perceptual ratings from the bottleneck layer, a shallow neural network was further trained to predict the severity of dysarthria as the diagnostic output. The predicted ratings showed strong correlations with manual perceptual ratings, achieving Pearson correlations between 0.7 and 0.8. The severity scores predicted using these features also correlated well with ground truth severity levels, with a Pearson correlation of 0.8, supporting their clinical validity.

When using human-provided perceptual labels as prediction targets, they must be carefully obtained to improve model robustness during both training and validation. For example, inexperienced listeners have been shown to assess voice quality in pathological speech differently from experienced listeners \cite{helou2010role}. Collecting low-quality labels can lead to misleading validation results and poor performance when the algorithm is deployed in practice. To mitigate this, clinical label collection should adhere to guidance documents from professional organizations \cite{patel2018recommended, vsvec2018tutorial, rusz2021guidelines}.

However, even carefully collected perceptual labels are inherently subjective and will never be perfect. This is why it is critical to evaluate both within-rater and between-rater variability. Within-rater variability refers to the consistency of a single rater’s judgments over time, while between-rater variability reflects differences in judgments across different raters. By characterizing this variability, it becomes possible to account for this variability during model training and validation. Incorporating knowledge about variability can improve the model’s ability to generalize, making it more robust to the inherent noise in human judgments. In practice, this can be done by weighting labels based on their reliability or by introducing uncertainty measures in the training process to ensure that the model learns to handle variable input in a systematic way \cite{karimi2020deep, li2020regularization, ma2020normalized}.

% and the predicted severity score, highly correlated to subjective evaluation of speech-language pathologists, where Pearson correlations of 0.7 at least were achieved.

Following the development of a model that predicts or correlates strongly with existing perceptual measures, we obtain a clinically interpretable measure. While this has clinical value in itself, these objective measures also have the potential to enhance the transparency and interpretability of AI models. For instance, in the assessment of dysarthria, speech measures such as GOP, OAM, OHM, and cepstral peak prominence (CPP) have been shown to correlate strongly with perceptual labels crucial for clinical evaluation \cite{witt2000phone, mathad2021deep, mathad2022consonant, sauder2017predicting}. These measures can be embedded within the middle layers of neural networks to enhance the model’s interpretability.

Xu et al. operationalized this approach by incorporating the information bottleneck principle into dysarthric speech classification \cite{xu2023dysarthria}. In their work, GOP, OHM, OAM, and CPP were used as intermediate prediction targets during DNN training. This serves two key purposes. First, these features guide the model in the right direction during training, as they are known to correlate with perceptual labels important for dysarthria classification. Second, they allow users to interrogate the model after classification to determine which intermediate features were most responsible for the classification decision. The study demonstrated that the use of these intermediate layers improved both the model’s accuracy and its interpretability. The authors applied Shapley values to explain the contribution of each clinically interpretable measure to the assessment \cite{Shapley+1997+69+79, lundberg2017unified}. OAM, measured on consonant-vowel transition segments, emerged as the most significant contributor, followed by GOP, OHM, and CPP.

%either manually or automatically obtain validated measures, and develop a AI algorithm that reliably predict the selected measures.
%The higher CV transition precision derived from speech (denoted in red) indicates the model inclines to classify the speech as “Healthy. The contribution of hypernasality and vocal quality to the model’s predictions is minimal.

\begin{figure}[h!]
  \setlength\belowcaptionskip{-0.8\baselineskip}
  \centering
  \includegraphics[width=\linewidth]{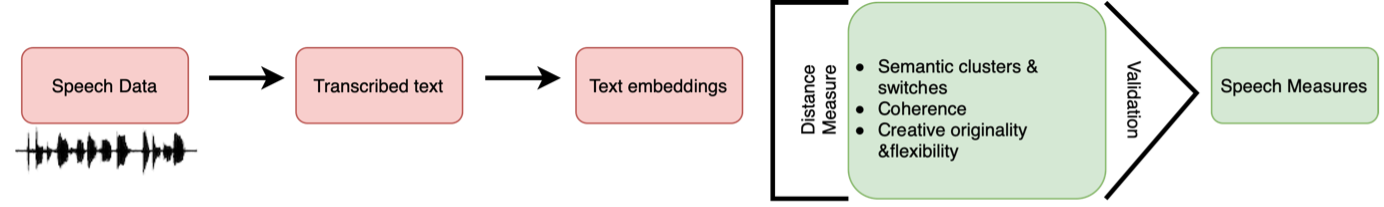}
  \caption{Using natural language processing (NLP) techniques to derive speech measures based on cosine similarities between word/text embeddings.} 
  \label{fig:NLP_measures}
\end{figure}

\noindent {\bf Predicting Clinically-Important Scales Using Natural Language Processing:} While clinical speech data can be processed as acoustic signal, transcribing the speech into text data allows developers to derive speech measures with techniques from natural language processing (NLP). 
While NLP-based speech measures are traditionally derived from word or phrase statistics, such as frequency and types of words used, semantic relevance, syntactic complexity, etc. \cite{voleti2019review, yeung2021correlating}, recent state-of-the-art text analysis has shifted from the statistical approach to data-driven approach. DNN embeddings of text extracted from models such as word2vec, GloVe, and Transformer models are shown to capture semantic information from text \cite{mikolov2013distributed, pennington2014glove, kenton2019bert}. 
The relationship between words or phrases can be reflected through the their similarity (e.g. cosine similarity) computed in the embedding space. 
Using the similarity measure, valid and interpretable speech measures can be obtained for the analyses. The overall pipeline of deriving speech measures sing NLP is illustrated in Figure \ref{fig:NLP_measures}.

For example, semantic clusters and switches, which describe  verbal fluency of a speaker, were measured using the cosine similarities between word2vec embeddings \cite{lundin2022semantic}. Semantic clusters reflected how the speaker elicits semantically related groupings of words over time, while the semantic switches measures the frequency of a subject producing words that are semantically unrelated. In the assessment of early-stage psychosis, cosine similarities between adjacent pairs of word embeddings were used to derive the cluster-related semantic similarity and switch-related semantic similarity respectively. To validate the clinical validity, the measured similarities were shown to significantly correlate to Communication Disturbances Index \cite{docherty1996communication}, a clinically-validated summary score used to identify disorganization in natural speech. 
In \cite{xu2020centroid}, coherence, which refers to the semantic similarity between interconnected flow of ideas, was measured from cosine similarities between word embeddings extracted from the FastText model to assess formal thought disorder \cite{joulin2016fasttext}. Experimental results showed that the centroid-derived coherence measures, where each word embedding was compared with the summed embeddings of previous words, achieved satisfactory correlation with the clinically validated Thought and Language Disorder (TALD) ratings performed by human annotators. 
In \cite{tang2023clinical}, to assess speech from patients with schizophrenia, the coherence was measured by cosine similarity between embeddings of adjacent utterances pairs, using the word2vec model and the GloVe model. Statistical analysis showed that the coherence measures were significantly correlated to various social cognition ratings such as emotion processing, mentalizing, and attribution bias. 
The above examples have demonstrated the use of traditional and deep learning based NLP techniques to derive speech measures from the transcriptions. Specifically, the cosine similarities obtained by pairwise comparison between text embeddings could be relevant to various clinical constructs. With rigorous validation, these similarity scores between text have the potential to be reliable and valid speech measures.
%emotion processing (Penn Emotion Recognition Task), mentalizing (Hinting Task), and attribution bias (Ambiguous Intentions and Hostility Questionnaire).  

\subsection{Algorithm design to improve the robustness of learned speech measures}
% In analytical validation, splitting training and test data from the same speech dataset. Such practice means the speech data is likely to be acquired by the same setup, e.g., environmental background, same hardware and firmware. 
During model development, training, validation, and test data are often split from the same speech dataset, meaning that all three subsets are likely to be acquired under similar conditions, such as the same hardware, firmware, and background noise. This setup limits the evaluation of model robustness and generalization, as the validation procedure does not test whether the model can perform similarly on data collected under different conditions, such as a real-world clinical setting. The fundamental assumption in this setup is that the data distribution during training will match what the model encounters during deployment.

To address this limitation, self-supervised learning (SSL) and transfer learning offer promising approaches. Self-supervised learning allows models to learn general speech representations by leveraging large amounts of unlabeled data \cite{mohamed2022self}. The model can capture underlying patterns in the speech signal, such as phonetic structures or speaker characteristics, that are not specific to any one dataset. This results in more robust representations that can be applied to a variety of tasks or environments, improving generalization to unseen data.
Transfer learning \cite{vasquez2021transfer, wang2015transfer}, on the other hand, enables models that are pre-trained in one domain (e.g., clean, lab-recorded speech) to be fine-tuned on new, smaller datasets from different domains (e.g., noisy, clinical speech data). By retaining the knowledge learned from the original domain, the model can adapt more quickly and efficiently to the new domain with fewer labeled samples, while still maintaining strong performance. This reduces the model’s dependency on large, labeled datasets from specific conditions and improves its ability to generalize across diverse settings, such as real-world clinical environments.

However, even these methods cannot solve all generalization issues. If there are differences between clinical populations in the training data that do not contribute meaningfully to the task (e.g., gender imbalance, demographic differences, etc.), the classifier may focus on irrelevant features like speech measures related to gender or other less important factors, rather than on clinically significant differences \cite{bailey2021gender, yang2024deconstructing}. For example, in a classifier designed to distinguish between a clinical condition and healthy controls, a gender imbalance could lead the model to rely on gender-based features, undermining its ability to generalize effectively in clinical settings. In \cite{rusz2021reproducibility}, it was demonstrated that selecting a subset of high-dimensional features with randomized values correlated to clinical labels could still result in high reported diagnostic accuracy.  This implies that, even with noise- or domain- robust models, the lack of awareness regarding high dimensionality and clinical relevance can still lead to misleading, overly optimistic, and non-generalizable diagnostic outcomes.

Although collecting more clinical speech data could enhance model robustness and generalizability,  data collection in clinical speech AI is resource-intensive and slow. There are strategies to improve models beyond simply increasing the dataset size. In \cite{zhang2023robust}, data augmentation was applied to train deep learning models for automatic assessment of dysphonic speech.
In this approach, noisy factors unrelated to vocal quality were simulated using signal processing techniques. For instance, background noise was added to training speech data to simulate different environments, and impulse responses were used to simulate room reverberation and microphone effects. Additionally, the fundamental frequency of the training speech samples was randomly shifted up or down to simulate disordered speech from unseen speakers. These data augmentation methods, as also adopted in speech recognition, speaker verification and other clinical speech tasks \cite{konig2015automatic, ko2017study, vachhani18_interspeech, shahnawazuddin2020domain, prananta22_interspeech}, significantly improved model accuracy in dysphonic speech detection, particularly under cross-corpus evaluation.

Self-supervised learning can sometimes result in correlated embeddings that lack consistency across datasets. To address this issue, regularization techniques are used. Regularization reduces the model’s over-reliance on correlated features, improving its ability to generate consistent and repeatable embeddings, especially in tasks with high variability (e.g., speaker verification or disordered speech classification). In \cite{zhang2024learning}, a regularization term based on the intra-class correlation coefficient (ICC) was introduced during DNN training. This technique reduces variability within each class, leading to more consistent and repeatable embeddings. It also improved repeatability of speech embeddings and performance in downstream tasks such as speaker verification, zero-shot voice style conversion, and dysphonic speech detection.

Additionally, training algorithms can address the issue of co-linearity in neural network embeddings, which is a common cause of overfitting. As in general AI research, The idea of feature decorrelation has been proposed in \cite{zhou2021isobn, peyser22_interspeech, gao2018representation}, where feature decorrelation was achieved through modifying strategies in batch normalization \cite{zhou2021isobn}, objective function design \cite{peyser22_interspeech}, and introducing new regularization term \cite{gao2018representation}, etc. In the context of clinical speech tasks, specifically the dementia speech assessment using NLP, a regularization scheme that increases variance and limits covariance in each embedding dimension was used to fine-tune pre-trained Bidirectional Encoder Representations from Transformers (BERT) \cite{xu2023decorrelating}. This approach resulted in more robust model performance by mitigating the risks associated with co-linearity. Experimental results showed that the proposed regularization reduced the false alarm rate and improved the precision of severity score prediction when compared to human ratings of dementia speech.

In addition to regularization, domain adaptation techniques can improve model generalization across datasets. Methods like adversarial learning or fine-tuning on target domain data enable models to adapt to new environments by learning features that are invariant to domain shifts \cite{perero2019modeling, park2023adversarial, hsu2018robustness, amiri2024test}. This is especially valuable when clinical data comes from a variety of real-world settings with differing recording setups or speaker populations.

Incorporating techniques such as data augmentation, regularization, self-supervised learning, transfer learning, and domain adaptation into the model development pipeline can significantly enhance the robustness, repeatability, and generalization of speech models, particularly in clinical settings. These methods together form a comprehensive approach to improving the reliability and performance of AI models in real-world applications.

\section{Design and Validation of Clinical Speech Model for Clinical Label Prediction}\label{Design_ClinicalML_Model}
Clinical AI models are designed to utilize validated speech parameters to predict clinical labels, aiding clinicians in decision-making (see Figure \ref{fig:speech_measure_pipeline}). 
These clinical labels may indicate the presence of a clinical condition, performance on clinical tests, disease prognosis, disease severity, etc.  
%social competence in psychosis, severity score of dysarthria speech, empathy measure in autism
During the development of clinical models, developers must consider data volume, model interpretability, inference time, and model generalization \cite{chen2019develop}. 
While there is significant flexibility in implementing clinical AI models — such as the dimensionality of model inputs (number of speech measures used) and model complexity (ranging from tens of parameters in traditional AI models to millions in complex deep neural networks) — the availability of clinical speech data for model training often limits this flexibility.
%Several factors need to be considered for clinical implementation, including the volume of data, interpretability of models, inference time as well as the balance between overfitting and underfitting [Chen et al.]. 
% Clinical AI models are often developed with limited amounts of clinical speech data. 
Many existing works have employed high dimensional inputs to train clinical AI models. 
For instance, a set of over 6,000 acoustic parameters has been widely used in various clinical speech AI tasks \cite{schuller2014interspeech}. 
% With the recent rise of speech and language foundation models, 
High-dimensional neural network embeddings from speech and language foundation models are now also used to predict clinical labels through end-to-end approaches. Despite promising results reported, using excessively high-dimensional model inputs can often lead to situations where the small available sample size does not support the training of complex, high-dimensional models. 
\begin{figure}[h!]
  \setlength\belowcaptionskip{\baselineskip}
  \centering
  \includegraphics[width=0.95\linewidth]{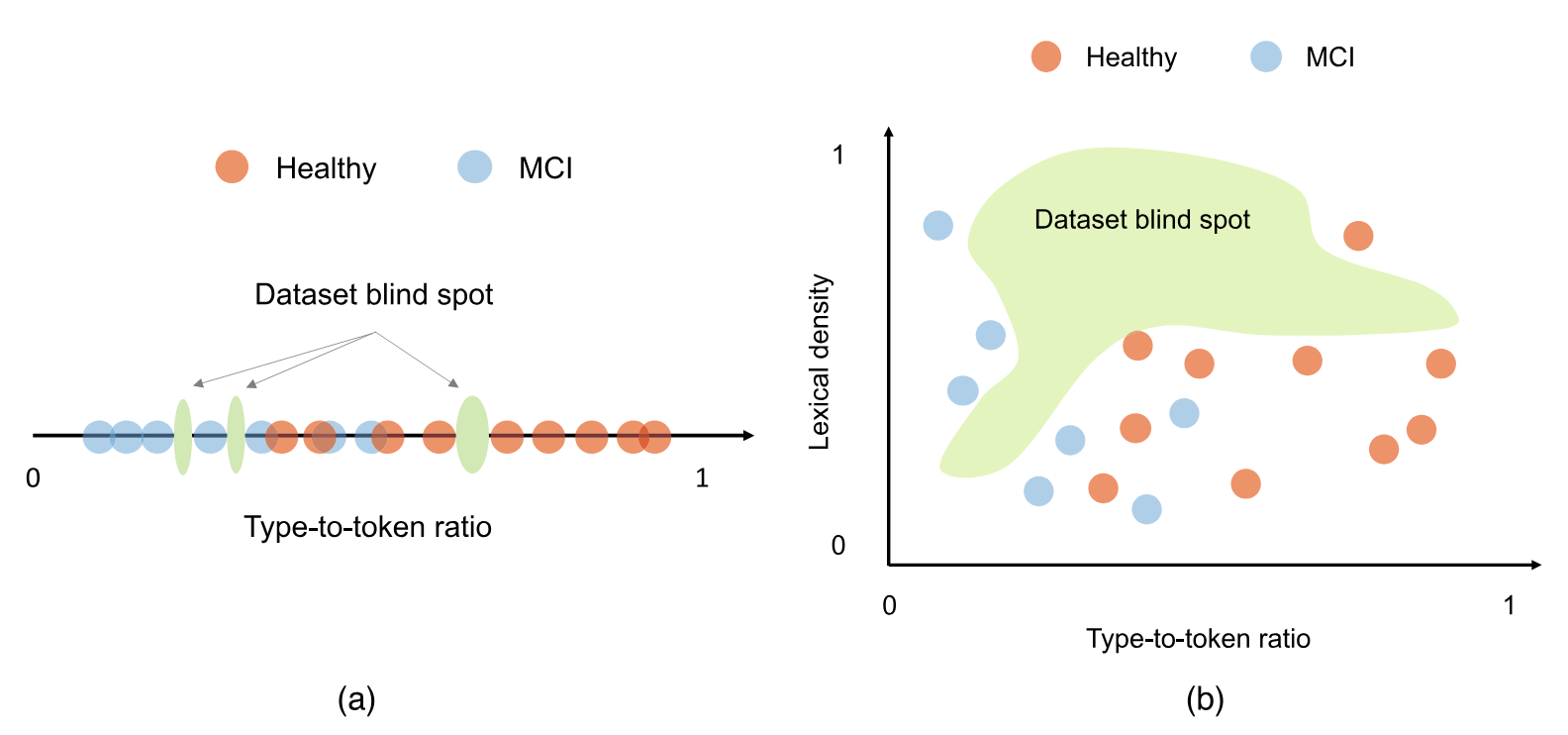}
  \caption{Relationship between dataset blind spot and and increase in feature dimension: (a) Only type-to-token ratio is used to distinguish between healthy controls and patients with mild cognitive impairment (MCI). (b) Both type-to-token ratio and lexical density are used to distinguish between the two clinical groups.} 
  \label{fig:curve-high-dim}
\end{figure}
This is intuitively illustrated in Figure \ref{fig:curve-high-dim}, which demonstrates the examples of using 1- and 2-dimensional speech features in classification of MCI speakers from HC speakers \cite{berisha2021digital}. 
Increasing feature dimensionality (from 1 in panel (a) to 2 in panel (b)) would expand contiguous regions (also known as the {\em blind spots}) that cannot be covered by the limited training data in the feature space. 
These blind spots can diminish model generalizability and lead to overestimation of the performance. The illustration explains and supports the findings in \cite{berisha2021digital, berisha22_interspeech} where reported accuracy of classifying between healthy controls and patients with Alzheimer’s disease, and  between healthy controls and patients with cognitive impairment, decrease with the increased amount of training speech data (see Figure \ref{fig:negative trend}). The result is counter-intuitive from the perspective of classical AI theory. 
Using excessively high-dimensional inputs with very small sample sizes can ultimately result in failures when models are deployed in clinical practice as the model is forced to make predictions on blind-spot data it has never previously been trained on. Unfortunately, there is no silver bullet for mitigating these risks. Clinical speech AI model developers should take great care in thoroughly validating their model in conditions that match those the model will encounter during deployment.

\noindent {\bf Verification, Analytical Validation, and Clinical Validation:} An important approach to mitigating the blind spot issue is through thorough validation at multiple stages, ensuring that the  data collection process, speech measure variability, and model outputs are rigorously evaluated. This can be done using the V3 framework \cite{goldsack2020verification, robin2020evaluation}, which consists of:
\begin{itemize}
\item Verification: This step ensures that the conditions required to extract reliable speech measures are properly validated during model development. By verifying the data collection process, we can define the optimal conditions for capturing high-quality speech data that are consistent with the requirements for accurate model performance. This provides critical guidance for data collection post-deployment, ensuring that the same conditions are met when the model is used in real-world settings. Additionally, this process enables the detection of any mismatch between the conditions encountered during post-deployment data collection and the verified requirements. In such cases, the system can notify users of potential inconsistencies, prompting corrective actions to avoid unreliable predictions and ensure the integrity of the model’s output.
\item Analytical Validation: In this phase, each individual speech measure used in the model is characterized and analyzed. The goal is to define the range of these measures and understand how they vary across different conditions. This allows developers to identify potential blind spots in the data—regions of the input space where the model may not have sufficient training data. By understanding these variations, developers can better ensure that the model remains generalizable and does not overfit to specific conditions.
\item Clinical Validation: This step ensures that the model’s outputs are clinically meaningful and relevant. By evaluating the model against real-world clinical outcomes, developers can confirm that the predictions made by the model—such as disease progression or diagnostic classifications—are both accurate and beneficial to clinical decision-making. This validation phase is crucial in determining whether the model's predictions align with the intended clinical goals and ultimately improve patient care.
\end{itemize}

% Complex AI models is widely believed to be necessary for good performance. This does not hold true when the data is scarce. Insufficient amount of speech data could lead to overfitting of the trained models. For example, in the task of detecting COVID-19 based on speech and cough sound \cite{schuller21_interspeech}, less than 300 speech / cough recordings were provided for model training.
% DNN models were surpassed by support vector machine (SVM) (SummaryComParE, Casanova et al., Klumpp et al., Illium et al.). 
\noindent {\bf Explainable Models:} Clinical AI models, particularly deep learning models used for predicting clinical labels, often exhibit a ``black-box" nature, making their decision-making processes difficult to interpret. This can pose significant challenges in clinical applications where interpretability is important for fostering trust with clinicians \cite{quinn2022three}. To address this issue, various \textit{attribute-based techniques}, such as layer-wise relevance propagation (LRP) \cite{montavon2019layer}, class activation mapping (CAM) \cite{zhou2016learning}, and Shapley values \cite{Shapley+1997+69+79}, aim to highlight the importance of individual input parameters in the model’s predictions. These methods provide insight into which features drive the model's decisions, helping clinicians understand the reasoning behind its outputs.

\textit{Explainer-based techniques} such as Local Interpretable Model-agnostic Explanations (LIME) \cite{ribeiro2016should} take a different approach by constructing simpler, interpretable models that approximate and visualize the decision boundaries of more complex models. While these techniques offer some level of insight, they are approximations and do not fully capture the model’s true mechanisms \cite{rudin2019stop}. For example, LIME has been shown to produce different explanations if the data sampling process is slightly altered, introducing uncertainty in explaining black-box models \cite{alvarez2018towards}. This variability can be particularly problematic in high-stakes applications like clinical speech AI, where even minor errors in model explanations could mislead clinical decision-making.

In contrast, traditional AI models such as linear regression, decision trees, and other parametric models are inherently interpretable when used with validated speech measures \cite{rudin2019stop}. Their decision processes are directly encoded in their parameters and structure, making them easier for humans to understand. Without the need for secondary models to approximate behavior, these inherently interpretable AI models offer better trustworthiness and are often preferred in clinical speech AI where interpretability and reliability are paramount. When combined with well-understood (by clinicians) speech measures, these models can surface a model output and information about which speech measures were most important for the final decision.

In general, both validation and explainability are more straightforward for lower-dimensional models that use well-defined, construct-driven measures which have been clinically validated relative to specific conditions. For example, in predicting clinical labels for Alzheimer’s disease, focusing on a specific set of speech-based features that are clinically understandable, such as the increased number of pauses or reduced vocabulary size, allows the model to rely on clinically relevant inputs grounded in prior research. 

Using lower-dimensional inputs allows developers to build models that are simpler, easier to interpret, and less prone to overfitting, leading to more generalizable solutions. Furthermore, the lower complexity of these models facilitates more thorough validation processes, as each input feature can be individually assessed for its impact on the model’s predictions. 

\begin{figure}[h!]
  \setlength\belowcaptionskip{-0.85\baselineskip}
  \centering
  \includegraphics[width=0.80\linewidth]{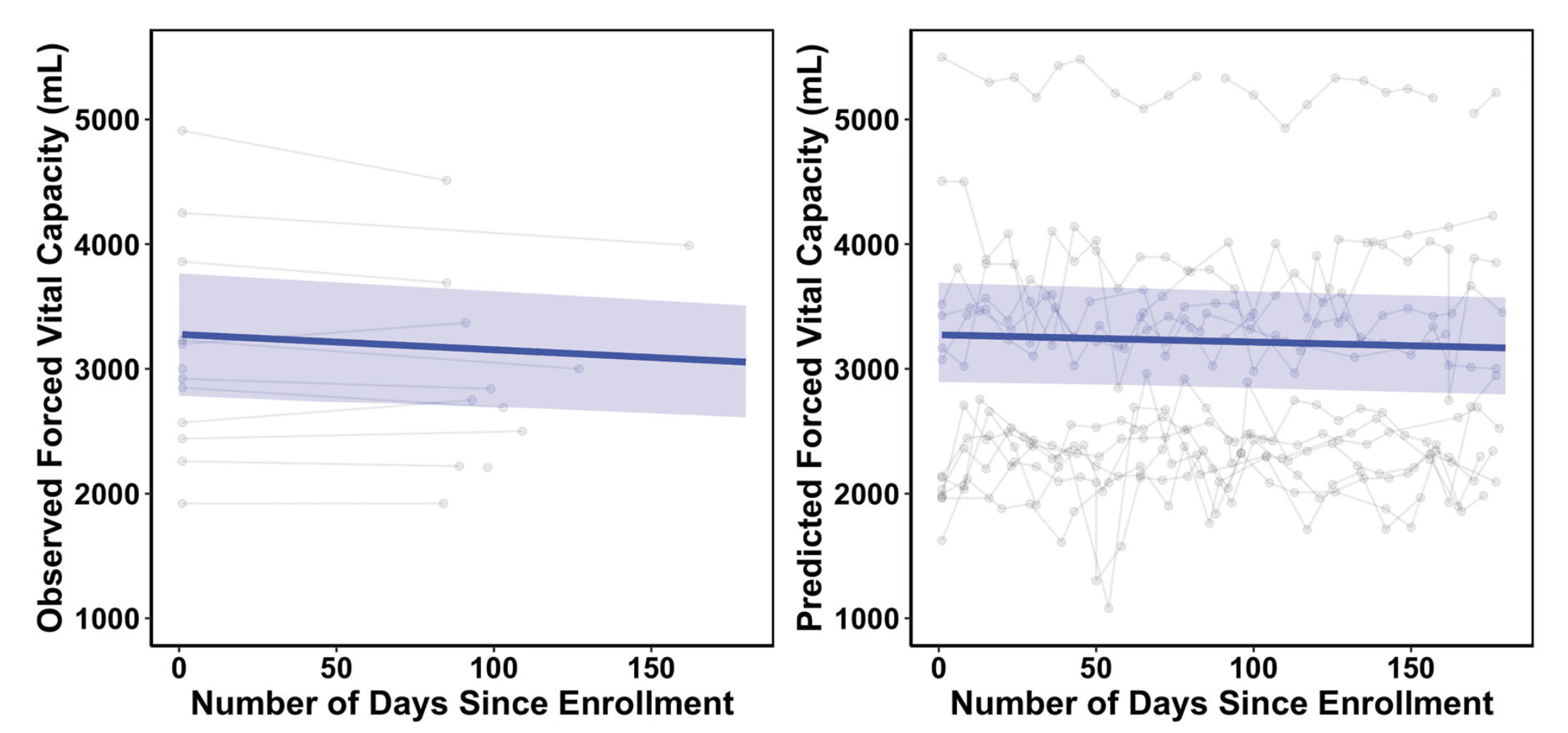}
  \caption{Observed (left) and predicted (right) FVC longitudinal trajectories.} 
  \label{fig:FVC}
\end{figure}
There are several examples of such models in the literature. For example, in \cite{stegmann2021estimation}, the authors develop a model for predicting forced vital capacity (FVC) from speech data collected from maximum phonation task, where the objective is to monitor the FVC of a speaker longitudinally.  Measures of maximum phonation time, pitch, loudness, and vocal quality extracted from the sustained phonation, as well as age, height, gender, and weight were initially selected as candidates for feature inputs. After statistical analyses, maximum phonation time, age, and height was used to construct the 3-dimensional model input. A linear model is then trained to predict the FVC. Following analytical validation, the authors perform clinical validation to determine whether the model reliably and accurately predicts the intended clinical labels (FVC in this case). From a clinical perspective, the validation also aims to validate if the clinical outputs answer the clinical question of interest, and satisfy the predefined context of use. In \cite{stegmann2021estimation}, the predicted FVC was shown to correlate to the ground truth measures with a Pearson correlation of 0.8, while both observed and predicted FVC achieved high repeatability of over 0.9 ICC. Given the objective in \cite{stegmann2021estimation} was longitudinal tracking instead of predicting a single FVC measure, 
%a list of clinical questions is prepared for the validation \cite{goldsack2020verification}. Clinical experts should review the criteria of subject recruitment, measurements made on the subjects, and the diagnostic labels collected during  data collection. Mistakes in any of these steps would lead to biased and misleading model prediction as well as misinterpretation of model performance. 
the longitudinal trajectory of predicted FVC was compared with observed FVC data to validate the model's efficacy in detecting longitudinal decline in FVC, as shown in Figure \ref{fig:FVC}. Both predicted and observed FVC had similar intercepts and declining trajectories. 
The clinical validation provides strong evidence for the efficacy, reliability, and interpretability of the FVC prediction model. 
\begin{figure}[h!]
  \setlength\belowcaptionskip{-0.85\baselineskip}
  \centering
  \includegraphics[width=0.40\linewidth]{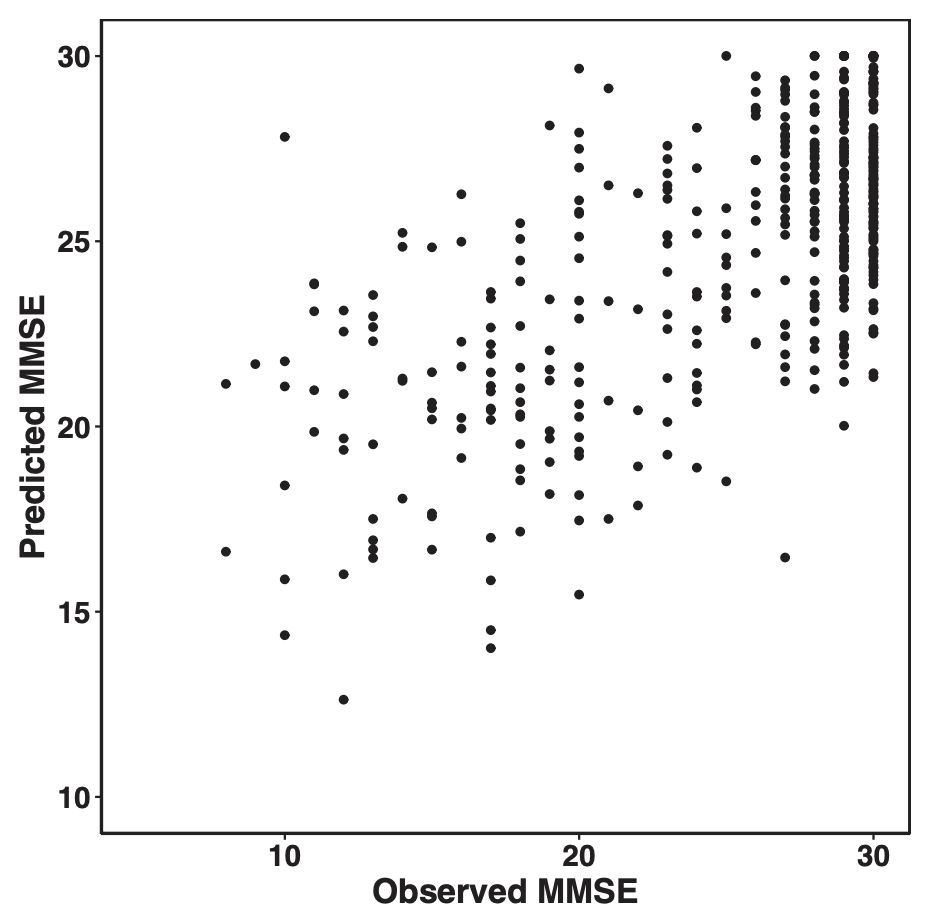}
  \caption{Predicted vs. observed MMSE values.} 
  \label{fig:MMSE}
\end{figure}
\begin{figure}[h!]
  \setlength\belowcaptionskip{-0.85\baselineskip}
  \centering
  \includegraphics[width=0.80\linewidth]{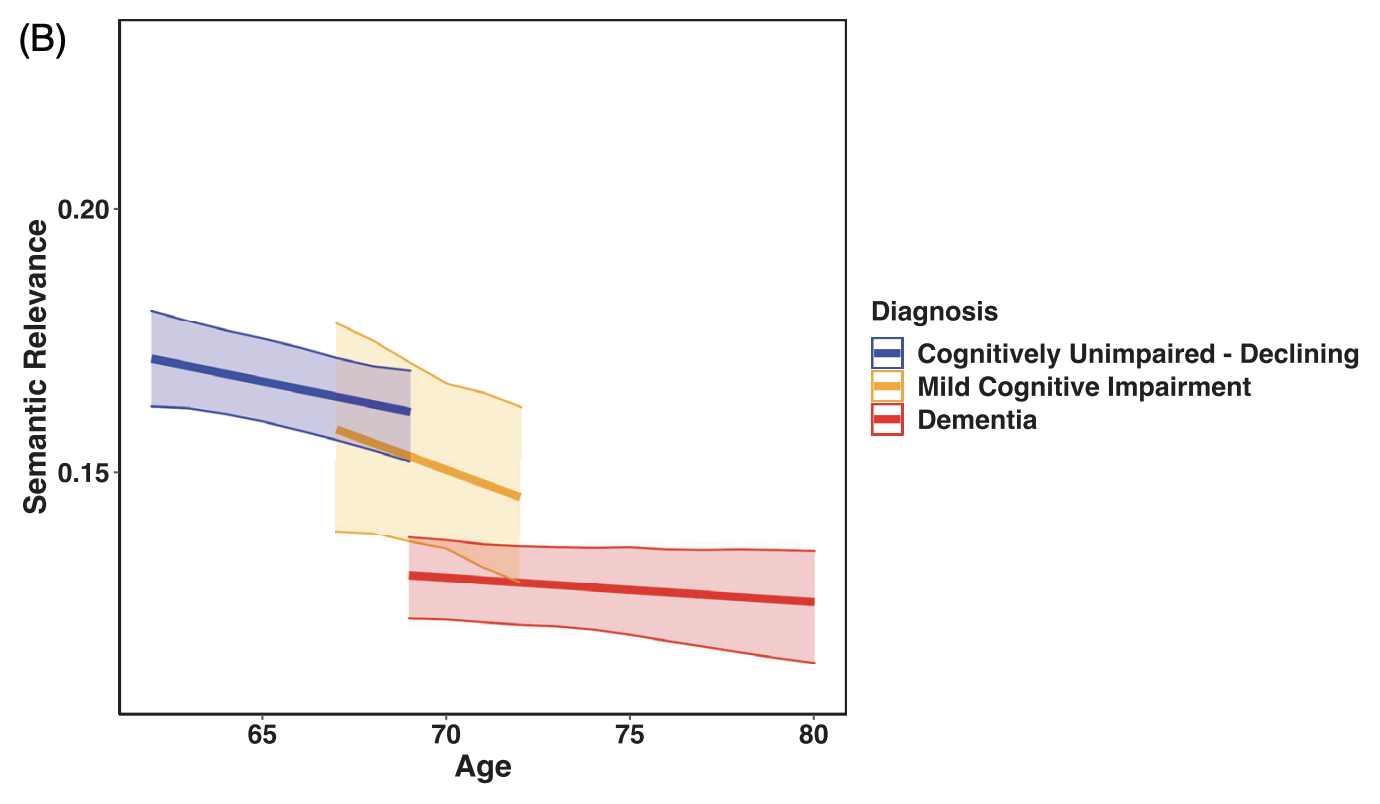}
  \caption{Longitudinal plot visualizing how clinical output of semantic relevance varies as a function of age \cite{stegmann2022automated}.} 
  \label{fig:Declining_MMSE}
\end{figure}

Similarly, in \cite{stegmann2022automated}, the semantic relevance (SemR) of transcribed speech was used to longitudinally predict cognitive decline in speakers. The manual approach to deriving SemR was shown to correlate with automatically-derived SemR, with a Pearson correlation of 0.84. The SemR was shown to achieve a test-retest reliability of 0.73 ICC, of which the reliability was considered moderate to good. The proposed SemR was combined with other language features to predict the clinical score Mini-Mental State Examination (MMSE). As shown in \ref{fig:MMSE}, the use of SemR was able to improve the prediction performance from a correlation of 0.38 to 0.65, thus confirming its efficacy in prediction. To evaluate if the predicted score provide meaningful clinical outputs, its expected longitudinal trajectories for cognitively impaired speakers was illustrated in Figure \ref{fig:Declining_MMSE}, which indicated the SemR longitudinally declined with age and level of cognitive impairment. The predicted output provided a qualitative understanding of semantic relevance can reflect decline of cognitive condition. 

In \cite{holmlund2020applying}, common word types and word mover’s distance (WMD) were derived from speech transcription to assess verbal memory in mental illness. Using the two-dimensional feature as input, a least square linear regression model was used to predict the clinician-rated accuracy scores in story recall task. The predicted score using human transcription and ASR transcription highly agreed with each other, where a Pearson correlation of over 0.95 was achieved. While predicted score using features derived from ASR transcription achieved a correlation of 0.82 with human annotation, the coefficients in the trained regression model were used to interpret the influence of the two input features. The coefficients for common word types and WMD were 0.15 and -0.54 respectively, meaning increasing value of common word types was associated with higher accuracy in story recall tasks, whereas increasing WMD led to lower accuracy as a large WMD indicates significant discrepancy between the recall and the original text.

When the problem dimensionality is well-managed and the model’s decisions are interpretable, model complexity can be adjusted to improve performance while maintaining generalizability. Semenova et al. \cite{semenova2022existence} offer guidelines for balancing complexity and performance. If multiple models perform similarly on validation data—indicating they generalize well—the problem can likely be solved by simpler models. Researchers should continue exploring simpler options within the same model class. However, if models perform differently, a more complex approach may be warranted, as long as the available data is large enough to prevent overfitting and support the model’s complexity.

\noindent {\bf The path from simple to more complex models:}  There is a core tension in model design for clinical speech AI prediction: balancing simplicity and interpretability with the potential to discover new, unknown relationships. By focusing solely on hypothesis-driven, low-dimensional features, we may limit the model’s ability to capture complex patterns or interactions that have not yet been identified. On the other hand, relying on higher-dimensional models with data-driven features introduces risks, such as overfitting, especially when dealing with smaller clinical datasets.

We posit that clinical model development can evolve naturally over time, beginning with simpler, hypothesis-driven models that build trust and demonstrate success. Given the limited clinical sample sizes currently available, these initial models should focus on well-validated, low-dimensional features that are easier to interpret and validate. By achieving reliable clinical outcomes, these models can prove their value in real-world clinical settings, motivating further investment in data collection and infrastructure. As clinical stakeholders recognize their utility, there will be increased motivation to collect larger and more diverse datasets.

With these expanded datasets, model developers can begin incorporating more sophisticated, data-driven approaches. These techniques enable the discovery of new, complex relationships between speech features and clinical conditions that may not have been previously hypothesized. Hybrid approaches can further accelerate these efforts. For instance, developers can begin with well-established, hypothesis-driven features, as demonstrated in \cite{tu2017interpretable} and \cite{xu2023dysarthria}, and gradually expand the model to include additional, potentially high-dimensional data. This strategy allows for the discovery of new patterns while maintaining a foundation of clinically validated intermediate representations.

\noindent {\bf On utilizing large language model (LLM) as clinical diagnostic model:} 
The growing accessibility of large language models (LLMs), such as ChatGPT, GPT-4 \cite{achiam2023gpt}, and Llama-3 \cite{touvron2023llama}, has showcased their potential in the clinical domain. These models outperform traditional AI approaches, particularly in zero-shot or few-shot scenarios with limited clinical data \cite{nori2023capabilities}. Their large size and the information embeeded within these models have resulted in high scores on medical licensing examinations and medical competitions \cite{nori2023capabilities, kasai2023evaluating, rosol2023evaluation, jin2024hidden}. Natural language plays an important role across the clinical domain, from note-taking and documenting patient histories to generating diagnostic reports and supporting patient-provider communications. It is also used for summarizing complex medical records and for aiding decision-making during consultations. It is, therefore, not surprising that GPT-based LLMs have found application across several clinical use-cases. For instance, GPT-4 can recommend stepwise approaches to assist in surgical preparation \cite{bektacs2024chatgpt}, and it can analyze computed tomography (CT) reports to extract oncologic phenotypes and retrieve critical clinical information \cite{fink2023potential}. Recently, however, researchers have also used these models in clinical speech AI applications. 

For cognitive assessment, Wang et al. evaluated GPT-based LLMs for primary screening of mild cognitive impairment (MCI) in a cohort of 174 participants from the DementiaBank database \cite{wang2023text}. By fine-tuning GPT-4 with carefully crafted text prompts, the study achieved promising results, with an AUC of 0.803 in distinguishing individuals with normal cognitive function from those with MCI in the held-out test set. Additionally, GPT-based LLMs have been used to assess cognitive status by scoring responses to a subset of neuropsychological test questions \cite{fei2024evaluating, botelho24_interspeech} and to respond to questions related to dementia caregiving \cite{hristidis2023chatgpt}.

Despite their promising results, LLMs still face several challenges. They are sensitive to input quality and prompts \cite{nori2023capabilities, wang2023can}; even slight modifications to the input can significantly impact the output. Furthermore, the accuracy of such models diminishes in other languages besides English \cite{kasai2023evaluating, rosol2023evaluation}. 
Another critical issue is the tendency of these models to generate hallucinated content. GPT-based LLMs have occasionally offered plausible but suboptimal advice \cite{nori2023capabilities, perlis2023application} or clinically unsafe recommendations \cite{kasai2023evaluating}. When the LLMs are instructed to explain their decision-making logic (via prompts), even the models that make correct decisions offer explanations that often differ from those of clinicians, potentially leading to flawed or misleading justifications and explanations \cite{wang2023can, mohamad2023chatgpt, jin2024hidden}. 
While accessible LLMs show promise by addressing some of the challenges in clinical speech AI, they require rigorous clinical validation to ensure their generalizability, reliability, interpretability, and trustworthiness in real-world clinical applications 
%e.g. explanation of decision-making and scarcity of clinical data in model development, their limitations discussed above highlight the need for rigorous clinical validation to ensure their generalizability, reliability, interpretability, and trustworthiness in real-world clinical applications.

LLMs are typically trained on vast, diverse datasets that may not fully represent the nuances of clinical speech data. In clinical environments, the variability in data—due to differences in patient demographics, speech patterns, and clinical conditions—can lead to inconsistencies in model performance. Unlike more straightforward models that rely on predefined, clinically validated features, LLMs process vast amounts of unstructured text, making it harder to ensure that the data used for training and testing adequately covers all relevant clinical contexts. Additionally, validation processes for LLMs must address their sensitivity to input variations \cite{kasai2023evaluating, rosol2023evaluation}. Therefore, rigorous clinical validation with representative datasets is essential to ensure the model’s performance is not only accurate but also reliable across different clinical scenarios.

\section{Post-deployment Monitoring of Clinical Speech AIModels}\label{Post_deployment_ClinicalMLModels}
A clinical AI model should only be deployed after meeting clinical validation requirements, with continual post-deployment monitoring to ensure long-term reliability and safety in real-world clinical settings. A recent report by U.S. Food and Drug Administration (FDA) highlights a significant growth in approvals for medical devices embedded with AI functionalities, with a ten fold increase from 2018 to 2023 \cite{us2022artificial}. However, studies have also indicated that evaluations conducted during the pre-deployment stage are inadequate for predicting the long-term performance of these AI-embedded devices. One study, which assessed 32 datasets from 4 different modalities, found that 91\% of deployed AI models exhibited a performance degradation over time \cite{vela2022temporal}. 
In \cite{wong2021quantification}, AI models designed for early sepsis detection had a 43\% increase in alerts during the COVID-19 outbreak. It was attributed to a significant domain shift in data that was not observed during model development. In \cite{wang2024performance}, the performance of auto-segmentation model for cancer radiotherapy exhibited performance decline three years after deployment. In speech applications, \cite{cao2023comparative} demonstrated that off-the-shelf ASR models, such as Google Speech and OpenAI Whisper, which are optimized for recognizing clean human speech,  yielded high error rates when deployed in noisy real-world classrooms environments. 
Numerous factors may impact model behavior post-deployment, including changes in the demographic background of the user population, unseen clinical environments, changing hardware or firmware, evolving clinical data collection protocol, among others. 
These examples illustrate that pre-deployment evaluations, typically based on curated test data, are insufficient to ensure the long-term reliability of AI models, as a finite test set cannot account for every possible domain encountered post-deployment. This highlights the necessity for strategies to monitor model performance over time, while leveraging existing knowledge of factors that could contribute to long-term model degradation.
%These examples have indicated that a pre-deployment evaluation, mostly relying on curated test data is insufficient to guarantee long-term reliability of the AI models, since the finite amount of test data is impossible to cover every possible data domain in post-deployment. It necessitate to need for strategies to monitor model performance in the long performance, with existng knowledge of potential factors that deteriorate the models. 

%Bias
Recent studies in deep learning-based computational pathology have demonstrated that existing models exhibit performance gaps across different demographic groups. For instance, in medical imaging research, the performance gap between demographic groups in diagnosing various cancer types ranges from 3\% to 16\%, as seen in studies of pathology models across patient populations with varying ethnic backgrounds \cite{vaidya2024demographic}. Similar biases have been reported in clinical speech AI of mental health issues, where models underperform for certain demographic groups \cite{straw2020artificial}.
%Ways for continual monitoring (real-time...feedback mechanism)
A feedback policy is necessary for real users reporting problematic model behavior that allows developers to backtrack the relevant component in the model development pipeline \cite{lu2021deploying} (refer to Figure \ref{fig:model_training}). In addition, statistical approaches which perform timely and automatic monitoring model performance can also be used \cite{yu2018request, ginart2022mldemon}. 
%Data-drifting, software/hardware version update

The discussion above highlights general challenges associated with deploying AI tools in healthcare, many of which directly apply to clinical speech AI. A key cause of performance decline post-deployment is data distribution shift, which can result from changes in the data collection process, software updates, input demographics, or other factors. For instance, variations in algorithmic implementation between software versions may alter speech measures, ultimately affecting model outputs. Distribution shifts may also arise from revisions or unexpected changes in speech data collection protocols \cite{lu2021deploying}. Even changes in background noise conditions can significantly impact model reliability. Some of these shifts can be automatically tracked and mitigated. For example, software versioning can flag changes in system configuration, while background noise monitoring before data collection can alert users to environmental conditions that may affect model performance, recommending adjustments like relocating to a quieter space.

Other changes in input data distribution, such as demographic shifts in patient populations, are more challenging to address. For instance, the inclusion of new dialects or speech patterns not represented in the training data can significantly alter the distribution of speech features, making the model less reliable for certain populations. These conditions are examples of the blind spot phenomenon described in Section \ref{Design_ClinicalML_Model}.

All of these differences contribute to domain shifts, increasing the disparity between the data used in training and the conditions encountered in deployment, ultimately affecting the reliability of model outputs. For model developers, it is important to routinely validate clinical models after hardware or software upgrades. New efforts are also emerging to support continuous model development post-deployment. Algorithms that can automatically detect distribution shifts and alert developers of potential performance declines \cite{ovadia2019can, koch2024distribution}, or even automatically re-calibrate models to adjust for the new data \cite{ma2022test, davis2020detection}, will be key to ensuring ongoing model reliability in clinical environments.

\section{Ethical, privacy and security considerations in Clinical speech AI}\label{Ethics_privacy_security}

%Clinical speech AI are intended to process highly-sensitive data of the clinical populations and make high-stake clinical decisions. Issues such as biased decision-making process, security risks of AI models, and leakage of privacy information of the patients, etc. could lead to harmful consequences to individuals. These topics should be addressed during the technical pipeline of model development.  

Clinical speech AI has the potential to revolutionize patient care by processing highly sensitive data and facilitating high-stakes clinical decisions. However, the deployment of such technology brings forth significant ethical, privacy, and security challenges. Issues like biased decision-making processes, security risks of AI  models, and leakage of patient information can lead to harmful consequences for individuals. It is imperative to address these concerns throughout the technical pipeline of model development.

A critical ethical concern in clinical speech AI is the potential for biased decision-making stemming from homogeneous data collection. Insufficient demographic diversity in speech data can introduce bias into trained AI models~\cite{leschanowsky24_spsc}. It is known that speech characteristics are influenced by biological and socio-cultural factors. For example, F0 and formant frequencies vary between genders,  \cite{gelfer2005relative}, age groups \cite{reubold2010vocal}, and lifestyles \cite{berg2017speaking}. Features such as articulation rate, pausing, and phrasing also differ across dialects \cite{bertelsen2018sociodemographic, clopper2011effects}.
Ignoring these differences during data collection can result in models that perform well only on certain socio-demographic groups, leading to biased outcomes in applications like speech recognition and speaker recognition \cite{feng2024towards, hutiri2022bias}.
In clinical contexts, similar biases during data collection, analytical validation, and/or clinical validation can compromise the accuracy of assessments. A recent study revealed that 71\% of training data in published clinical AI models targeting American English patients came from just three states—California, Massachusetts, and New York-omitting data from the remaining states \cite{kaushal2020geographic}. Such skewed representation can lead to inaccurate results for test subjects outside the specific populations included in the training data.

While speech data and AI models for lower-stake applications (e.g. ASR) are often open-sourced \cite{panayotov2015librispeech, wolf2019huggingface}, 
clinical speech AI demands strict confidentiality due to the sensitive nature of clinical data. The high clinical value and impact of these models make them vulnerable to cybersecurity attacks. For example, training speech data can be poisoned, significantly manipulating model behavior even when only a small portion of labels are altered  \cite{li2023defend}. Techniques like active learning and knowledge distillation can be exploited to steal information from the original model, including hyperparameters, learned parameters, and model behaviors \cite{oliynyk2023know}. Once confidential information is compromised, attacks such as adversarial attacks, model inversion, and membership inference become possible~\cite{VERDE20212624}. 
Adversarial attacks involve injecting hand-crafted noise signals into the original input without being detectable by humans, misleading or manipulating the model’s decision-making process \cite{kwon2019selective, zelasko2021adversarial}. In the clinical setting, such attacks could produce distorted speech measures and misleading clinical scores, leading to misallocation of resources, monetary waste, and significant threats to patient health, including delayed or invalid treatment and worsened symptoms. 
Model inversion attempts to reconstruct speech data from model parameters, potentially leaking patient privacy information during the process \cite{pizzi2023introducing}. 
On the other hand, membership inference attacks determine if an AI model was trained on a specific data sample, exploiting the model to reveal the identity and related information of speakers in the training dataset \cite{shah21_interspeech}. These threats necessitate that algorithm developers incorporate robust detection and prevention strategies against such attacks in clinical speech AI.

Another emerging approach in speech technology is the development of voice anonymization methods, which are gaining significant attention~\cite{TOMASHENKO2022101362,rahman24_spsc}. Anonymization refers to the goal of suppressing personally identifiable information in the speech signal, while preserving other attributes such as linguistic content and speaker intent. This ensures that the speech can still be used for analysis or system interaction without compromising privacy. Recent approaches to voice anonymization include voice conversion~\cite{ghosh24_interspeech}, where one speaker's voice is transformed to sound like another speaker, and disentangled representation learning~\cite{10.1145/3411495.3421355}, which aims to separate the identifiable characteristics of the voice from its other features, allowing for selective suppression of speaker identity.

Speech carries clinically rich information, and our daily interactions with mobile devices leave extensive footprints of our health status. Recent clinical studies have utilized publicly available social media data for developing clinical applications. For example, users’ text data on Twitter has been used to detect depression \cite{orabi2018deep}, and speech data from YouTube has been employed to develop AI models for evaluating oral cancer speech, dysarthria, and dementia \cite{petti2023much, halpern2023automatic}. While publicly available data offers cost-effective opportunities for algorithm developers and researchers, it poses considerable risks such as the quality of speech data and the reliability of labels. 
Claims about users’ health statuses might be made without their consent. Analyzing high-profile public figures without consent can lead to the widespread dissemination of false narratives regarding their health conditions, potentially causing social and political crises beyond the control of researchers and clinicians.

The ethical, privacy, and security considerations in clinical speech AI are multifaceted and demand comprehensive attention. It is crucial for academia, industry, and regulatory bodies to collaborate on stringent data protection regulations and develop ethical guidelines to safeguard individuals’ health information. By proactively addressing these challenges, we can harness the benefits of clinical speech AI while minimizing potential harms.

\section{Conclusion}\label{Conclusion}
The tutorial paper presents a comprehensive pipeline for developing clinical speech AI models, from speech data collection and feature design to diagnostic model development and validation at each stage. 
We begin by reviewing the human speech production mechanism, and discussing how various clinical conditions are linked to the speech production process. We compare different speech elicitation tasks based on their complexity and the computational load place on speakers. Using clinical examples, we demonstrate how the careful task selection can effectively magnify the symptoms in the target clinical population. We then discuss the considerations for speech data collection, emphasizing the importance of hardware validation and proper microphone setups to produce high-quality data that accurately captures the clinical conditions.  

Moving forward, we analyze conventional speech representations used in clinical speech AI and introduce the concept of ``speech measures" as a more interpretable and clinically meaningful alternative for feature design. With an introduction to measurement theory and a review of selected studies, we summarize several approaches that integrate speech-language processing techniques and clinical speech science to develop these measures. We also highlight the importance of analytical validation for ensuring reliability and validity.
In terms of developing clinical speech models for clinical label prediction, we discuss the risks of overfitting in clinical speech models and outline the requirements for interpretability and generalizability, supported by examples from published studies. We stress the need for clinical validation to ensure the reliability and meaningfulness of model predictions. We also discuss the need for post-deployment monitoring to address model degradation, outlining current strategies to mitigate these risks. Finally, we address ongoing debates surrounding ethical, privacy, and security concerns in the development of clinical speech AI. We highlight vulnerabilities in the current development pipeline and application scenarios. In light of these challenges, we advocate for collaboration between academia, industry, and regulatory bodies to maximizing the benefits of clinical speech AI with minimizing potential risks to the public.

\bibliographystyle{IEEEtran}
\bibliography{mybib}

\begin{thebibliography}{100}
\providecommand{\url}[1]{#1}
\csname url@samestyle\endcsname
\providecommand{\newblock}{\relax}
\providecommand{\bibinfo}[2]{#2}
\providecommand{\BIBentrySTDinterwordspacing}{\spaceskip=0pt\relax}
\providecommand{\BIBentryALTinterwordstretchfactor}{4}
\providecommand{\BIBentryALTinterwordspacing}{\spaceskip=\fontdimen2\font plus
\BIBentryALTinterwordstretchfactor\fontdimen3\font minus \fontdimen4\font\relax}
\providecommand{\BIBforeignlanguage}[2]{{%
\expandafter\ifx\csname l@#1\endcsname\relax
\typeout{** WARNING: IEEEtran.bst: No hyphenation pattern has been}%
\typeout{** loaded for the language `#1'. Using the pattern for}%
\typeout{** the default language instead.}%
\else
\language=\csname l@#1\endcsname
\fi
#2}}
\providecommand{\BIBdecl}{\relax}
\BIBdecl

\bibitem{vikram2022digital}
V.~Ramanarayanan, A.~C. Lammert, H.~P. Rowe, T.~F. Quatieri, and J.~R. Green, ``Speech as a biomarker: Opportunities, interpretability, and challenges,'' \emph{Perspectives of the ASHA Special Interest Groups}, vol.~7, no.~1, pp. 276--283, 2022.

\bibitem{faurholt2016voice}
M.~Faurholt-Jepsen, J.~Busk, M.~Frost, M.~Vinberg, E.~M. Christensen, O.~Winther, J.~E. Bardram, and L.~V. Kessing, ``Voice analysis as an objective state marker in bipolar disorder,'' \emph{Translational psychiatry}, vol.~6, no.~7, pp. e856--e856, 2016.

\bibitem{rapcan2010acoustic}
V.~Rapcan, S.~D’Arcy, S.~Yeap, N.~Afzal, J.~Thakore, and R.~B. Reilly, ``Acoustic and temporal analysis of speech: A potential biomarker for schizophrenia,'' \emph{Medical engineering \& physics}, vol.~32, no.~9, pp. 1074--1079, 2010.

\bibitem{luz2021detecting}
S.~Luz, F.~Haider, S.~de~la Fuente, D.~Fromm, and B.~MacWhinney, ``Detecting cognitive decline using speech only: The adresso challenge,'' in \emph{INTERSPEECH 2021}.\hskip 1em plus 0.5em minus 0.4em\relax ISCA, 2021.

\bibitem{vasquez2018multimodal}
J.~C. V{\'a}squez-Correa, T.~Arias-Vergara, J.~R. Orozco-Arroyave, B.~Eskofier, J.~Klucken, and E.~N{\"o}th, ``Multimodal assessment of parkinson's disease: a deep learning approach,'' \emph{IEEE journal of biomedical and health informatics}, vol.~23, no.~4, pp. 1618--1630, 2018.

\bibitem{quintas2024automatic}
S.~Quintas, M.~Balaguer, J.~Mauclair, V.~Woisard, and J.~Pinquier, ``Automatic modelling of perceptual judges in the context of head and neck cancer speech intelligibility,'' \emph{International Journal of Language \& Communication Disorders}.

\bibitem{stegmann2023speech}
G.~Stegmann, S.~Charles, J.~Liss, J.~Shefner, S.~Rutkove, and V.~Berisha, ``A speech-based prognostic model for dysarthria progression in als,'' \emph{Amyotrophic Lateral Sclerosis and Frontotemporal Degeneration}, vol.~24, no. 7-8, pp. 599--604, 2023.

\bibitem{martens2015automated}
H.~Martens, T.~Dekens, G.~Van~Nuffelen, L.~Latacz, W.~Verhelst, and M.~De~Bodt, ``Automated speech rate measurement in dysarthria,'' \emph{Journal of Speech, Language, and Hearing Research}, vol.~58, no.~3, pp. 698--712, 2015.

\bibitem{cummins2015review}
N.~Cummins, S.~Scherer, J.~Krajewski, S.~Schnieder, J.~Epps, and T.~F. Quatieri, ``A review of depression and suicide risk assessment using speech analysis,'' \emph{Speech communication}, vol.~71, pp. 10--49, 2015.

\bibitem{benway2024artificial}
N.~R. Benway and J.~L. Preston, ``Artificial intelligence--assisted speech therapy for /\textturnr/: A single-case experimental study,'' \emph{American Journal of Speech-Language Pathology}, vol.~33, no.~5, pp. 2461--2486, 2024.

\bibitem{ng2024ssd}
S.-I. Ng, C.~W.-Y. Ng, J.~Wang, and T.~Lee, ``Automatic detection of speech sound disorder in cantonese-speaking pre-school children,'' \emph{IEEE/ACM Transactions on Audio, Speech, and Language Processing}, vol.~32, pp. 4355--4368, 2024.

\bibitem{voleti2023language}
R.~Voleti, S.~M. Woolridge, J.~M. Liss, M.~Milanovic, G.~Stegmann, S.~Hahn, P.~D. Harvey, T.~L. Patterson, C.~R. Bowie, and V.~Berisha, ``Language analytics for assessment of mental health status and functional competency,'' \emph{Schizophrenia bulletin}, vol.~49, no. Supplement\_2, pp. S183--S195, 2023.

\bibitem{bedi2015automated}
G.~Bedi, F.~Carrillo, G.~A. Cecchi, D.~F. Slezak, M.~Sigman, N.~B. Mota, S.~Ribeiro, D.~C. Javitt, M.~Copelli, and C.~M. Corcoran, ``Automated analysis of free speech predicts psychosis onset in high-risk youths,'' \emph{npj Schizophrenia}, vol.~1, no.~1, pp. 1--7, 2015.

\bibitem{de2020artificial}
S.~De~la Fuente~Garcia, C.~W. Ritchie, and S.~Luz, ``Artificial intelligence, speech, and language processing approaches to monitoring alzheimer’s disease: a systematic review,'' \emph{Journal of Alzheimer's Disease}, vol.~78, no.~4, pp. 1547--1574, 2020.

\bibitem{moro2021advances}
L.~Moro-Velazquez, J.~A. Gomez-Garcia, J.~D. Arias-Londo{\~n}o, N.~Dehak, and J.~I. Godino-Llorente, ``Advances in parkinson's disease detection and assessment using voice and speech: A review of the articulatory and phonatory aspects,'' \emph{Biomedical Signal Processing and Control}, vol.~66, p. 102418, 2021.

\bibitem{deng2013machine}
L.~Deng and X.~Li, ``Machine learning paradigms for speech recognition: An overview,'' \emph{IEEE Transactions on Audio, Speech, and Language Processing}, vol.~21, no.~5, pp. 1060--1089, 2013.

\bibitem{shi2023speech}
M.~Shi, G.~Cheung, and S.~R. Shahamiri, ``Speech and language processing with deep learning for dementia diagnosis: A systematic review,'' \emph{Psychiatry Research}, p. 115538, 2023.

\bibitem{ardila-etal-2020-common}
R.~Ardila, M.~Branson, K.~Davis, M.~Kohler, J.~Meyer, M.~Henretty, R.~Morais, L.~Saunders, F.~Tyers, and G.~Weber, ``Common voice: A massively-multilingual speech corpus,'' in \emph{Proceedings of the Twelfth Language Resources and Evaluation Conference}, N.~Calzolari, F.~B{\'e}chet, P.~Blache, K.~Choukri, C.~Cieri, T.~Declerck, S.~Goggi, H.~Isahara, B.~Maegaard, J.~Mariani, H.~Mazo, A.~Moreno, J.~Odijk, and S.~Piperidis, Eds.\hskip 1em plus 0.5em minus 0.4em\relax Marseille, France: European Language Resources Association, May 2020, pp. 4218--4222.

\bibitem{panayotov2015librispeech}
V.~Panayotov, G.~Chen, D.~Povey, and S.~Khudanpur, ``Librispeech: an asr corpus based on public domain audio books,'' in \emph{2015 IEEE international conference on acoustics, speech and signal processing (ICASSP)}.\hskip 1em plus 0.5em minus 0.4em\relax IEEE, 2015, pp. 5206--5210.

\bibitem{arbabshirani2017single}
M.~R. Arbabshirani, S.~Plis, J.~Sui, and V.~D. Calhoun, ``Single subject prediction of brain disorders in neuroimaging: Promises and pitfalls,'' \emph{Neuroimage}, vol. 145, pp. 137--165, 2017.

\bibitem{vabalas2019machine}
A.~Vabalas, E.~Gowen, E.~Poliakoff, and A.~J. Casson, ``Machine learning algorithm validation with a limited sample size,'' \emph{PloS one}, vol.~14, no.~11, p. e0224365, 2019.

\bibitem{berisha22_interspeech}
V.~Berisha, C.~Krantsevich, G.~Stegmann, S.~Hahn, and J.~Liss, ``Are reported accuracies in the clinical speech machine learning literature overoptimistic?'' in \emph{Interspeech 2022}, 2022, pp. 2453--2457.

\bibitem{viering2022shape}
T.~Viering and M.~Loog, ``The shape of learning curves: a review,'' \emph{IEEE Transactions on Pattern Analysis and Machine Intelligence}, vol.~45, no.~6, pp. 7799--7819, 2022.

\bibitem{kapoor2023leakage}
S.~Kapoor and A.~Narayanan, ``Leakage and the reproducibility crisis in machine-learning-based science,'' \emph{Patterns}, vol.~4, no.~9, 2023.

\bibitem{coretta2023multidimensional}
S.~Coretta, J.~V. Casillas, S.~Roessig, M.~Franke, B.~Ahn, A.~H. Al-Hoorie, J.~Al-Tamimi, N.~E. Alotaibi, M.~K. AlShakhori, R.~M. Altmiller \emph{et~al.}, ``Multidimensional signals and analytic flexibility: Estimating degrees of freedom in human-speech analyses,'' \emph{Advances in Methods and Practices in Psychological Science}, vol.~6, no.~3, p. 25152459231162567, 2023.

\bibitem{food2019proposed}
Food, D.~Administration \emph{et~al.}, ``Proposed regulatory framework for modifications to artificial intelligence/machine learning (ai/ml)-based software as a medical device (samd),'' 2019.

\bibitem{goldsack2020verification}
J.~C. Goldsack, A.~Coravos, J.~P. Bakker, B.~Bent, A.~V. Dowling, C.~Fitzer-Attas, A.~Godfrey, J.~G. Godino, N.~Gujar, E.~Izmailova \emph{et~al.}, ``Verification, analytical validation, and clinical validation (v3): the foundation of determining fit-for-purpose for biometric monitoring technologies (biomets),'' \emph{npj digital Medicine}, vol.~3, no.~1, p.~55, 2020.

\bibitem{levelt1993speaking}
W.~J. Levelt, \emph{Speaking: From intention to articulation}.\hskip 1em plus 0.5em minus 0.4em\relax MIT press, 1993.

\bibitem{moll2003morals}
J.~Moll, R.~de~Oliveira-Souza, and P.~J. Eslinger, ``Morals and the human brain: a working model,'' \emph{Neuroreport}, vol.~14, no.~3, pp. 299--305, 2003.

\bibitem{sapir2007voice}
S.~Sapir, L.~O. Ramig, and C.~Fox, ``Voice, speech, and swallowing disorders,'' in \emph{Handbook of Parkinson's disease}.\hskip 1em plus 0.5em minus 0.4em\relax CRC Press, 2007, pp. 469--492.

\bibitem{harvey1983speech}
P.~D. Harvey, ``Speech competence in manic and schizophrenic psychoses: the association between clinically rated thought disorder and cohesion and reference performance.'' \emph{Journal of Abnormal Psychology}, vol.~92, no.~3, p. 368, 1983.

\bibitem{jacobs1995neuropsychological}
D.~M. Jacobs, M.~Sano, G.~Dooneief, K.~Marder, K.~L. Bell, and Y.~Stern, ``Neuropsychological detection and characterization of preclinical alzheimer's disease,'' \emph{Neurology}, vol.~45, no.~5, pp. 957--962, 1995.

\bibitem{duffy2012motor}
J.~R. Duffy \emph{et~al.}, \emph{Motor speech disorders: Substrates, differential diagnosis, and management}.\hskip 1em plus 0.5em minus 0.4em\relax Elsevier Health Sciences, 2012.

\bibitem{pereira2023differences}
V.~J. Pereira and D.~Sell, ``How differences in anatomy and physiology and other aetiology affect the way we label and describe speech in individuals with cleft lip and palate,'' \emph{International Journal of Language \& Communication Disorders}, 2023.

\bibitem{orlikoff1996dysphonia}
R.~F. Orlikoff and D.~H. Kraus, ``Dysphonia following nonsurgical management of advanced laryngeal carcinoma,'' \emph{American Journal of Speech-Language Pathology}, vol.~5, no.~3, pp. 47--52, 1996.

\bibitem{binazzi2011dyspnea}
B.~Binazzi, B.~Lanini, I.~Romagnoli, S.~Garuglieri, L.~Stendardi, R.~Bianchi, F.~Gigliotti, and G.~Scano, ``Dyspnea during speech in chronic obstructive pulmonary disease patients: effects of pulmonary rehabilitation,'' \emph{Respiration}, vol.~81, no.~5, pp. 379--385, 2011.

\bibitem{folsom2006diagnostic}
D.~P. Folsom, L.~Lindamer, L.~P. Montross, W.~Hawthorne, S.~Golshan, R.~Hough, J.~Shale, and D.~V. Jeste, ``Diagnostic variability for schizophrenia and major depression in a large public mental health care system dataset,'' \emph{Psychiatry Research}, vol. 144, no. 2-3, pp. 167--175, 2006.

\bibitem{weinstein2022diagnostic}
A.~M. Weinstein, S.~Gujral, M.~A. Butters, C.~R. Bowie, C.~E. Fischer, A.~J. Flint, N.~Herrmann, J.~L. Kennedy, L.~Mah, S.~Ovaysikia \emph{et~al.}, ``Diagnostic precision in the detection of mild cognitive impairment: a comparison of two approaches,'' \emph{The American Journal of Geriatric Psychiatry}, vol.~30, no.~1, pp. 54--64, 2022.

\bibitem{beach2018importance}
T.~G. Beach and C.~H. Adler, ``Importance of low diagnostic accuracy for early parkinson's disease,'' \emph{Movement Disorders}, vol.~33, no.~10, pp. 1551--1554, 2018.

\bibitem{xiong2017toward}
W.~Xiong, J.~Droppo, X.~Huang, F.~Seide, M.~L. Seltzer, A.~Stolcke, D.~Yu, and G.~Zweig, ``Toward human parity in conversational speech recognition,'' \emph{IEEE/ACM Transactions on Audio, Speech, and Language Processing}, vol.~25, no.~12, pp. 2410--2423, 2017.

\bibitem{schuller16_interspeech}
B.~Schuller, S.~Steidl, A.~Batliner, J.~Hirschberg, J.~K. Burgoon, A.~Baird, A.~Elkins, Y.~Zhang, E.~Coutinho, and K.~Evanini, ``The interspeech 2016 computational paralinguistics challenge: Deception, sincerity \& native language,'' in \emph{Interspeech 2016}, 2016, pp. 2001--2005.

\bibitem{mohamed2022self}
A.~Mohamed, H.-y. Lee, L.~Borgholt, J.~D. Havtorn, J.~Edin, C.~Igel, K.~Kirchhoff, S.-W. Li, K.~Livescu, L.~Maal{\o}e \emph{et~al.}, ``Self-supervised speech representation learning: A review,'' \emph{IEEE Journal of Selected Topics in Signal Processing}, vol.~16, no.~6, pp. 1179--1210, 2022.

\bibitem{murton2020cepstral}
O.~Murton, R.~Hillman, and D.~Mehta, ``Cepstral peak prominence values for clinical voice evaluation,'' \emph{American Journal of Speech-Language Pathology}, vol.~29, no.~3, pp. 1596--1607, 2020.

\bibitem{bunton2007listener)}
\BIBentryALTinterwordspacing
K.~Bunton, R.~D. Kent, J.~R. Duffy, J.~C. Rosenbek, and J.~F. Kent, ``Listener agreement for auditory-perceptual ratings of dysarthria,'' \emph{Journal of Speech, Language, and Hearing Research}, vol.~50, no.~6, pp. 1481--1495, 2007. [Online]. Available: \url{https://pubs.asha.org/doi/abs/10.1044/1092-4388%282007/102%29}
\BIBentrySTDinterwordspacing

\bibitem{us2022clinical}
U.~Food, D.~Administration \emph{et~al.}, ``Clinical decision support software: guidance for industry and food and drug administration staff,'' \emph{FDA digirepo. nlm. nih. gov/master/borndig/9918504188706676/9918504188706676. pdf}, 2022.

\bibitem{reich1989factors}
A.~R. Reich, J.~A. Mason, R.~R. Frederickson, and R.~S. Schlauch, ``Factors influencing fundamental frequency range estimates in children,'' \emph{Journal of Speech and Hearing Disorders}, vol.~54, no.~3, pp. 429--438, 1989.

\bibitem{wit1993maximum}
J.~Wit, B.~Maassen, F.~Gabreels, and G.~Thoonen, ``Maximum performance tests in children with developmental spastic dysarthria,'' \emph{Journal of Speech, Language, and Hearing Research}, vol.~36, no.~3, pp. 452--459, 1993.

\bibitem{ordin2017cross}
M.~Ordin and I.~Mennen, ``Cross-linguistic differences in bilinguals' fundamental frequency ranges,'' \emph{Journal of Speech, Language, and Hearing Research}, vol.~60, no.~6, pp. 1493--1506, 2017.

\bibitem{shear2001reliability}
M.~K. Shear, J.~Vander~Bilt, P.~Rucci, J.~Endicott, B.~Lydiard, M.~W. Otto, M.~H. Pollack, L.~Chandler, J.~Williams, A.~Ali \emph{et~al.}, ``Reliability and validity of a structured interview guide for the hamilton anxiety rating scale (sigh-a),'' \emph{Depression and anxiety}, vol.~13, no.~4, pp. 166--178, 2001.

\bibitem{stassen1991speech}
H.~Stassen, G.~Bomben, and E.~G{\"u}nther, ``Speech characteristics in depression,'' \emph{Psychopathology}, vol.~24, no.~2, pp. 88--105, 1991.

\bibitem{cummings2019describing}
L.~Cummings, ``Describing the cookie theft picture: Sources of breakdown in alzheimer’s dementia,'' \emph{Pragmatics and Society}, vol.~10, no.~2, pp. 153--176, 2019.

\bibitem{goldman1969goldman}
R.~Goldman and M.~Fristoe, ``Goldman-fristoe test of articulation,'' 1969.

\bibitem{cheung2006hong}
P.~Cheung, A.~Ng, and C.~To, ``Hong kong cantonese articulation test,'' \emph{Language Information Sciences Research Centre, City University of Hong Kong}, 2006.

\bibitem{benway2020differences}
N.~R. Benway and J.~L. Preston, ``Differences between school-age children with apraxia of speech and other speech sound disorders on multisyllable repetition,'' \emph{Perspectives of the ASHA Special Interest Groups}, vol.~5, no.~4, pp. 794--808, 2020.

\bibitem{sell2005issues}
D.~Sell, ``Issues in perceptual speech analysis in cleft palate and related disorders: a review,'' \emph{International Journal of Language \& Communication Disorders}, vol.~40, no.~2, pp. 103--121, 2005.

\bibitem{woisard2021c2si}
V.~Woisard, C.~Ast{\'e}sano, M.~Balaguer, J.~Farinas, C.~Fredouille, P.~Gaillard, A.~Ghio, L.~Giusti, I.~Laaridh, M.~Lalain \emph{et~al.}, ``C2si corpus: a database of speech disorder productions to assess intelligibility and quality of life in head and neck cancers,'' \emph{Language Resources and Evaluation}, vol.~55, no.~1, pp. 173--190, 2021.

\bibitem{omori2011diagnosis}
K.~Omori, ``Diagnosis of voice disorders,'' \emph{JMAJ}, vol.~54, no.~4, pp. 248--253, 2011.

\bibitem{karlsen2020acoustic}
T.~Karlsen, L.~Sandvik, J.-H. Heimdal, and H.~J. Aarstad, ``Acoustic voice analysis and maximum phonation time in relation to voice handicap index score and larynx disease,'' \emph{Journal of Voice}, vol.~34, no.~1, pp. 161--e27, 2020.

\bibitem{mulligan1994intelligibility}
M.~Mulligan, J.~Carpenter, J.~Riddel, M.~K. Delaney, G.~Badger, P.~Krusinski, and R.~Tandan, ``Intelligibility and the acoustic characteristics of speech in amyotrophic lateral sclerosis (als),'' \emph{Journal of Speech, Language, and Hearing Research}, vol.~37, no.~3, pp. 496--503, 1994.

\bibitem{ball2004communication}
L.~J. Ball, D.~R. Beukelman, and G.~L. Pattee, ``Communication effectiveness of individuals with amyotrophic lateral sclerosis,'' \emph{Journal of Communication Disorders}, vol.~37, no.~3, pp. 197--215, 2004.

\bibitem{usita1998narrative}
P.~M. Usita, I.~E. Hyman~Jr, and K.~C. Herman, ``Narrative intentions: Listening to life stories in alzheimer's disease,'' \emph{Journal of Aging Studies}, vol.~12, no.~2, pp. 185--197, 1998.

\bibitem{leyton2014verbal}
C.~E. Leyton, S.~Savage, M.~Irish, S.~Schubert, O.~Piguet, K.~J. Ballard, and J.~R. Hodges, ``Verbal repetition in primary progressive aphasia and alzheimer's disease,'' \emph{Journal of Alzheimer's Disease}, vol.~41, no.~2, pp. 575--585, 2014.

\bibitem{cohen2014speech}
A.~S. Cohen, J.~E. McGovern, T.~J. Dinzeo, and M.~A. Covington, ``Speech deficits in serious mental illness: a cognitive resource issue?'' \emph{Schizophrenia research}, vol. 160, no. 1-3, pp. 173--179, 2014.

\bibitem{schnur2024differences}
T.~T. Schnur and S.~Wang, ``Differences in connected speech outcomes across elicitation methods,'' \emph{Aphasiology}, vol.~38, no.~5, pp. 816--837, 2024.

\bibitem{mayer2003functional}
J.~Mayer and L.~Murray, ``Functional measures of naming in aphasia: Word retrieval in confrontation naming versus connected speech,'' \emph{Aphasiology}, vol.~17, no.~5, pp. 481--497, 2003.

\bibitem{tykalova2021effect}
T.~Tykalova, D.~Skrabal, T.~Boril, R.~Cmejla, J.~Volin, and J.~Rusz, ``Effect of ageing on acoustic characteristics of voice pitch and formants in czech vowels,'' \emph{Journal of Voice}, vol.~35, no.~6, pp. 931--e21, 2021.

\bibitem{lee2012variability}
G.-S. Lee, ``Variability in voice fundamental frequency of sustained vowels in speakers with sensorineural hearing loss,'' \emph{Journal of Voice}, vol.~26, no.~1, pp. 24--29, 2012.

\bibitem{abbiati2023speech}
C.~I. Abbiati, K.~R. Bauerly, and S.~L. Velleman, ``Speech elicitation methods for measuring articulatory control,'' \emph{Journal of Speech, Language, and Hearing Research}, pp. 1--8, 2023.

\bibitem{tran2022investigating}
K.~Tran, L.~Xu, G.~Stegmann, J.~Liss, V.~Berisha, and R.~Utianski, ``Investigating the impact of speech compression on the acoustics of dysarthric speech.'' in \emph{Proc. Interspeech}, 2022, pp. 2263--2267.

\bibitem{ge21b_interspeech}
C.~Ge, Y.~Xiong, and P.~Mok, ``{How Reliable Are Phonetic Data Collected Remotely? Comparison of Recording Devices and Environments on Acoustic Measurements},'' in \emph{Proc. Interspeech 2021}, 2021, pp. 3984--3988.

\bibitem{JHMDAGA:2019}
J.~Höbel-Müller, I.~Siegert, R.~Heinemann, A.~F. Requardt, M.~Tornow, and A.~Wendemuth, ``Analysis of the influence of different room acoustics on acoustic emotion features and emotion recognition performance,'' in \emph{Tagungsband - DAGA 2019}, Rostock, Germany, 2019, pp. 886--889.

\bibitem{fahed2022comparison}
V.~S. Fahed, E.~P. Doheny, M.~Busse, J.~Hoblyn, and M.~M. Lowery, ``Comparison of acoustic voice features derived from mobile devices and studio microphone recordings,'' \emph{Journal of Voice}, 2022.

\bibitem{szabo2001voice}
A.~Szabo, B.~Hammarberg, A.~Hakansson, and M.~Sodersten, ``A voice accumulator device: Evaluation based on studio and field recordings,'' \emph{Logopedics Phoniatrics Vocology}, vol.~26, no.~3, pp. 102--117, 2001.

\bibitem{printz2018test}
T.~Printz, J.~R. Sorensen, C.~Godballe, and {\AA}.~M. Gr{\o}ntved, ``Test-retest reliability of the dual-microphone voice range profile,'' \emph{Journal of Voice}, vol.~32, no.~1, pp. 32--37, 2018.

\bibitem{rusz2021guidelines}
J.~Rusz, T.~Tykalova, L.~O. Ramig, and E.~Tripoliti, ``Guidelines for speech recording and acoustic analyses in dysarthrias of movement disorders,'' \emph{Movement Disorders}, vol.~36, no.~4, pp. 803--814, 2021.

\bibitem{svec-ajslp}
J.~G. Švec and S.~Granqvist, ``Guidelines for selecting microphones for human voice production research,'' \emph{American Journal of Speech-Language Pathology}, vol.~19, no.~4, pp. 356--368, 2010.

\bibitem{pan2000effects}
Y.~Pan and A.~Waibel, ``The effects of room acoustics on mfcc speech parameter.'' in \emph{Proc. Interspeech}, 2000, pp. 129--132.

\bibitem{dineley23_interspeech}
J.~Dineley, E.~Carr, F.~Matcham, J.~Downs, R.~J.~B. Dobson, T.~F. Quatieri, and N.~Cummins, ``Towards robust paralinguistic assessment for real-world mobile health (mhealth) monitoring: an initial study of reverberation effects on speech,'' in \emph{Proc. Interspeech}, 2023, pp. 2373--2377.

\bibitem{kim1994monotony}
H.-H. Kim, \emph{Monotony of speech production in Parkinson's disease: Acoustic characteristics and their perceptual relations}.\hskip 1em plus 0.5em minus 0.4em\relax The University of Wisconsin-Madison, 1994.

\bibitem{skodda2008speech}
S.~Skodda and U.~Schlegel, ``Speech rate and rhythm in parkinson's disease,'' \emph{Movement disorders: official journal of the Movement Disorder Society}, vol.~23, no.~7, pp. 985--992, 2008.

\bibitem{turner1995influence}
G.~S. Turner, K.~Tjaden, and G.~Weismer, ``The influence of speaking rate on vowel space and speech intelligibility for individuals with amyotrophic lateral sclerosis,'' \emph{Journal of Speech, Language, and Hearing Research}, vol.~38, no.~5, pp. 1001--1013, 1995.

\bibitem{weismer2001acoustic}
G.~Weismer, J.-Y. Jeng, J.~S. Laures, R.~D. Kent, and J.~F. Kent, ``Acoustic and intelligibility characteristics of sentence production in neurogenic speech disorders,'' \emph{Folia Phoniatrica et Logopaedica}, vol.~53, no.~1, pp. 1--18, 2001.

\bibitem{berisha2014characterizing}
V.~Berisha, S.~Sandoval, R.~Utianski, J.~Liss, and A.~Spanias, ``Characterizing the distribution of the quadrilateral vowel space area,'' \emph{The Journal of the Acoustical Society of America}, vol. 135, no.~1, pp. 421--427, 2014.

\bibitem{mueller2018connected}
K.~D. Mueller, B.~Hermann, J.~Mecollari, and L.~S. Turkstra, ``Connected speech and language in mild cognitive impairment and alzheimer’s disease: A review of picture description tasks,'' \emph{Journal of clinical and experimental neuropsychology}, vol.~40, no.~9, pp. 917--939, 2018.

\bibitem{vellido2020importance}
A.~Vellido, ``The importance of interpretability and visualization in machine learning for applications in medicine and health care,'' \emph{Neural computing and applications}, vol.~32, no.~24, pp. 18\,069--18\,083, 2020.

\bibitem{liss2024operationalizing}
J.~Liss and V.~Berisha, ``Operationalizing clinical speech analytics: Moving from features to measures for real-world clinical impact,'' \emph{Journal of Speech, Language, and Hearing Research}, pp. 1--7, 2024.

\bibitem{halpern20_interspeech}
B.~M. Halpern, R.~van Son, M.~van~den Brekel, and O.~Scharenborg, ``{Detecting and Analysing Spontaneous Oral Cancer Speech in the Wild},'' in \emph{Proc. Interspeech}, 2020, pp. 4826--4830.

\bibitem{moller1999analysis}
S.~M{\"o}ller and R.~Sch{\"o}nweiler, ``Analysis of infant cries for the early detection of hearing impairment,'' \emph{Speech communication}, vol.~28, no.~3, pp. 175--193, 1999.

\bibitem{quintas20_interspeech}
S.~Quintas, J.~Mauclair, V.~Woisard, and J.~Pinquier, ``{Automatic Prediction of Speech Intelligibility Based on X-Vectors in the Context of Head and Neck Cancer},'' in \emph{Proc. Interspeech 2020}, 2020, pp. 4976--4980.

\bibitem{bayerl22b_interspeech}
S.~P. Bayerl, D.~Wagner, E.~Noeth, and K.~Riedhammer, ``{Detecting Dysfluencies in Stuttering Therapy Using wav2vec 2.0},'' in \emph{Proc. Interspeech}, 2022, pp. 2868--2872.

\bibitem{braun23_interspeech}
F.~Braun, S.~P. Bayerl, P.~A. Pérez-Toro, F.~Hönig, H.~Lehfeld, T.~Hillemacher, E.~Nöth, T.~Bocklet, and K.~Riedhammer, ``{Classifying Dementia in the Presence of Depression: A Cross-Corpus Study},'' in \emph{Proc. INTERSPEECH 2023}, 2023, pp. 2308--2312.

\bibitem{van2008visualizing}
L.~Van~der Maaten and G.~Hinton, ``Visualizing data using t-sne.'' \emph{Journal of machine learning research}, vol.~9, no.~11, 2008.

\bibitem{zheng2001comparison}
F.~Zheng, G.~Zhang, and Z.~Song, ``Comparison of different implementations of mfcc,'' \emph{Journal of Computer science and Technology}, vol.~16, pp. 582--589, 2001.

\bibitem{meyer15_interspeech}
B.~T. Meyer, B.~Kollmeier, and J.~Ooster, ``Autonomous measurement of speech intelligibility utilizing automatic speech recognition,'' in \emph{Proc. Interspeech}, 2015, pp. 2982--2986.

\bibitem{schuster2006evaluation}
M.~Schuster, A.~Maier, T.~Haderlein, E.~Nkenke, U.~Wohlleben, F.~Rosanowski, U.~Eysholdt, and E.~N{\"o}th, ``Evaluation of speech intelligibility for children with cleft lip and palate by means of automatic speech recognition,'' \emph{International Journal of Pediatric Otorhinolaryngology}, vol.~70, no.~10, pp. 1741--1747, 2006.

\bibitem{kim2011vowel}
H.~Kim, M.~Hasegawa-Johnson, and A.~Perlman, ``Vowel contrast and speech intelligibility in dysarthria,'' \emph{Folia Phoniatrica et Logopaedica}, vol.~63, no.~4, pp. 187--194, 2011.

\bibitem{kim2011acoustic}
Y.~Kim, R.~D. Kent, and G.~Weismer, ``An acoustic study of the relationships among neurologic disease, dysarthria type, and severity of dysarthria,'' \emph{Journal of Speech, Language, and Hearing Research}, vol.~54, no.~2, pp. 417--429, 2011.

\bibitem{lansford2014dysarthria}
\BIBentryALTinterwordspacing
K.~L. Lansford and J.~M. Liss, ``Vowel acoustics in dysarthria: Speech disorder diagnosis and classification,'' \emph{Journal of Speech, Language, and Hearing Research}, vol.~57, no.~1, pp. 57--67, 2014. [Online]. Available: \url{https://pubs.asha.org/doi/abs/10.1044/1092-4388%282013/12-0262%29}
\BIBentrySTDinterwordspacing

\bibitem{darley1969differential}
F.~L. Darley, A.~E. Aronson, and J.~R. Brown, ``Differential diagnostic patterns of dysarthria,'' \emph{Journal of speech and hearing research}, vol.~12, no.~2, pp. 246--269, 1969.

\bibitem{allen2001introduction}
M.~J. Allen and W.~M. Yen, \emph{Introduction to measurement theory}.\hskip 1em plus 0.5em minus 0.4em\relax Waveland Press, 2001.

\bibitem{stegmann2020repeatability}
G.~M. Stegmann, S.~Hahn, J.~Liss, J.~Shefner, S.~B. Rutkove, K.~Kawabata, S.~Bhandari, K.~Shelton, C.~J. Duncan, and V.~Berisha, ``Repeatability of commonly used speech and language features for clinical applications,'' \emph{Digital biomarkers}, vol.~4, no.~3, pp. 109--122, 2020.

\bibitem{eyben2010opensmile}
F.~Eyben, M.~W{\"o}llmer, and B.~Schuller, ``Opensmile: the munich versatile and fast open-source audio feature extractor,'' in \emph{Proceedings of the 18th ACM international conference on Multimedia}, 2010, pp. 1459--1462.

\bibitem{boersma2018praat}
P.~Boersma and D.~Weenink, ``Praat: Doing phonetics by computer [computer program]. version 6.0. 37,'' \emph{Retrieved February 2018 from http://www.praat.org/}.

\bibitem{fleiss2011design}
J.~L. Fleiss, \emph{Design and analysis of clinical experiments}.\hskip 1em plus 0.5em minus 0.4em\relax John Wiley \& Sons, 2011.

\bibitem{portney2009foundations}
L.~G. Portney, M.~P. Watkins \emph{et~al.}, \emph{Foundations of clinical research: applications to practice}.\hskip 1em plus 0.5em minus 0.4em\relax Pearson/Prentice Hall Upper Saddle River, NJ, 2009, vol. 892.

\bibitem{iter2018automatic}
D.~Iter, J.~Yoon, and D.~Jurafsky, ``Automatic detection of incoherent speech for diagnosing schizophrenia,'' in \emph{Proceedings of the Fifth Workshop on Computational Linguistics and Clinical Psychology: From Keyboard to Clinic}, 2018, pp. 136--146.

\bibitem{hitczenko2021automated}
K.~Hitczenko, H.~Cowan, V.~Mittal, and M.~Goldrick, ``Automated coherence measures fail to index thought disorder in individuals at risk for psychosis,'' in \emph{Proceedings of the seventh workshop on computational linguistics and clinical psychology: improving access}, 2021, pp. 129--150.

\bibitem{park2019test}
Y.~Park and C.~E. Stepp, ``Test--retest reliability of relative fundamental frequency and conventional acoustic, aerodynamic, and perceptual measures in individuals with healthy voices,'' \emph{Journal of Speech, Language, and Hearing Research}, vol.~62, no.~6, pp. 1707--1718, 2019.

\bibitem{dos2016protocol}
S.~dos Santos~Barreto and K.~Zazo~Ortiz, ``Protocol for the evaluation of speech intelligibility in dysarthrias: evidence of reliability and validity,'' \emph{Folia Phoniatrica et Logopaedica}, vol.~67, no.~4, pp. 212--218, 2016.

\bibitem{rowe2021validation}
H.~P. Rowe, K.~L. Stipancic, A.~C. Lammert, and J.~R. Green, ``Validation of an acoustic-based framework of speech motor control: Assessing criterion and construct validity using kinematic and perceptual measures,'' \emph{Journal of Speech, Language, and Hearing Research}, vol.~64, no.~12, pp. 4736--4753, 2021.

\bibitem{yawer2023reliability}
B.~A. Yawer, J.~Liss, and V.~Berisha, ``Reliability and validity of a widely-available ai tool for assessment of stress based on speech,'' \emph{Scientific reports}, vol.~13, no.~1, p. 20224, 2023.

\bibitem{reis2010perceived}
R.~S. Reis, A.~Hino, and C.~A{\~n}ez, ``Perceived stress scale,'' \emph{J. health Psychol}, vol.~15, no.~1, pp. 107--114, 2010.

\bibitem{darling2020impact}
M.~Darling-White and J.~E. Huber, ``The impact of parkinson's disease on breath pauses and their relationship to speech impairment: A longitudinal study,'' \emph{American Journal of Speech-Language Pathology}, vol.~29, no.~4, pp. 1910--1922, 2020.

\bibitem{darling2022longitudinal}
M.~Darling-White, Z.~Anspach, and J.~E. Huber, ``Longitudinal effects of parkinson's disease on speech breathing during an extemporaneous connected speech task,'' \emph{Journal of Speech, Language, and Hearing Research}, vol.~65, no.~4, pp. 1402--1415, 2022.

\bibitem{solomon1993speech}
N.~P. Solomon and T.~J. Hixon, ``Speech breathing in parkinson’s disease,'' \emph{Journal of Speech, Language, and Hearing Research}, vol.~36, no.~2, pp. 294--310, 1993.

\bibitem{voleti2019review}
R.~Voleti, J.~M. Liss, and V.~Berisha, ``A review of automated speech and language features for assessment of cognitive and thought disorders,'' \emph{IEEE journal of selected topics in signal processing}, vol.~14, no.~2, pp. 282--298, 2019.

\bibitem{roark2011spoken}
B.~Roark, M.~Mitchell, J.-P. Hosom, K.~Hollingshead, and J.~Kaye, ``Spoken language derived measures for detecting mild cognitive impairment,'' \emph{IEEE transactions on audio, speech, and language processing}, vol.~19, no.~7, pp. 2081--2090, 2011.

\bibitem{konig2015automatic}
A.~K{\"o}nig, A.~Satt, A.~Sorin, R.~Hoory, O.~Toledo-Ronen, A.~Derreumaux, V.~Manera, F.~Verhey, P.~Aalten, P.~H. Robert \emph{et~al.}, ``Automatic speech analysis for the assessment of patients with predementia and alzheimer's disease,'' \emph{Alzheimer's \& Dementia: Diagnosis, Assessment \& Disease Monitoring}, vol.~1, no.~1, pp. 112--124, 2015.

\bibitem{horwitz2016relation}
R.~L. Horwitz-Martin, T.~F. Quatieri, A.~C. Lammert, J.~R. Williamson, Y.~Yunusova, E.~Godoy, D.~D. Mehta, and J.~R. Green, ``Relation of automatically extracted formant trajectories with intelligibility loss and speaking rate decline in amyotrophic lateral sclerosis.'' in \emph{Proc. Interspeech}, 2016, pp. 1205--1209.

\bibitem{brunet1978vocabulaire}
{\'E}.~Brunet \emph{et~al.}, \emph{Le vocabulaire de Jean Giraudoux structure et {\'e}volution}.\hskip 1em plus 0.5em minus 0.4em\relax Slatkine, 1978.

\bibitem{yngve1960model}
V.~H. Yngve, ``A model and an hypothesis for language structure,'' \emph{Proceedings of the American philosophical society}, vol. 104, no.~5, pp. 444--466, 1960.

\bibitem{berg2011structure}
T.~Berg, \emph{Structure in language: A dynamic perspective}.\hskip 1em plus 0.5em minus 0.4em\relax Routledge, 2011.

\bibitem{lee2023knowledge}
S.~Lee, E.~J. Yeo, S.~Kim, and M.~Chung, ``Knowledge-driven speech features for detection of korean-speaking children with autism spectrum disorder,'' \emph{Phonetics and Speech Sciences}, vol.~15, no.~2, pp. 53--59, 2023.

\bibitem{rusz2011quantitative}
J.~Rusz, R.~Cmejla, H.~Ruzickova, and E.~Ruzicka, ``Quantitative acoustic measurements for characterization of speech and voice disorders in early untreated parkinson’s disease,'' \emph{The journal of the Acoustical Society of America}, vol. 129, no.~1, pp. 350--367, 2011.

\bibitem{harel2004acoustic}
B.~T. Harel, M.~S. Cannizzaro, H.~Cohen, N.~Reilly, and P.~J. Snyder, ``Acoustic characteristics of parkinsonian speech: a potential biomarker of early disease progression and treatment,'' \emph{Journal of Neurolinguistics}, vol.~17, no.~6, pp. 439--453, 2004.

\bibitem{tsanas2013objective}
A.~Tsanas, M.~A. Little, C.~Fox, and L.~O. Ramig, ``Objective automatic assessment of rehabilitative speech treatment in parkinson's disease,'' \emph{IEEE Transactions on Neural Systems and Rehabilitation Engineering}, vol.~22, no.~1, pp. 181--190, 2013.

\bibitem{romana2020classification}
A.~Romana, J.~Bandon, N.~Carlozzi, A.~Roberts, and E.~M. Provost, ``Classification of manifest huntington disease using vowel distortion measures,'' in \emph{Proc. Interspeech}, vol. 2020, 2020, p. 4966.

\bibitem{perez2018classification}
M.~Perez, W.~Jin, D.~Le, N.~Carlozzi, P.~Dayalu, A.~Roberts, and E.~M. Provost, ``Classification of huntington disease using acoustic and lexical features,'' in \emph{Proc. Interspeech}, vol. 2018, 2018, p. 1898.

\bibitem{charest2020properties}
M.~Charest, M.~J. Skoczylas, and P.~Schneider, ``Properties of lexical diversity in the narratives of children with typical language development and developmental language disorder,'' \emph{American Journal of Speech-Language Pathology}, vol.~29, no.~4, pp. 1866--1882, 2020.

\bibitem{fergadiotis2013measuring}
G.~Fergadiotis, H.~H. Wright, and T.~M. West, ``Measuring lexical diversity in narrative discourse of people with aphasia.'' \emph{American Journal of Speech-Language Pathology}, vol.~22, no.~2, 2013.

\bibitem{NAGRANI2020101027}
A.~Nagrani, J.~S. Chung, W.~Xie, and A.~Zisserman, ``Voxceleb: Large-scale speaker verification in the wild,'' \emph{Computer Speech \& Language}, vol.~60, p. 101027, 2020.

\bibitem{chen21o_interspeech}
G.~Chen, S.~Chai, G.-B. Wang, J.~Du, W.-Q. Zhang, C.~Weng, D.~Su, D.~Povey, J.~Trmal, J.~Zhang, M.~Jin, S.~Khudanpur, S.~Watanabe, S.~Zhao, W.~Zou, X.~Li, X.~Yao, Y.~Wang, Z.~You, and Z.~Yan, ``Gigaspeech: An evolving, multi-domain asr corpus with 10,000 hours of transcribed audio,'' in \emph{Proc. Interspeech}, 2021, pp. 3670--3674.

\bibitem{changawala24_interspeech}
V.~Changawala and F.~Rudzicz, ``Whister: Using whisper’s representations for stuttering detection,'' in \emph{Interspeech 2024}, 2024, pp. 897--901.

\bibitem{schubert24_interspeech}
M.~Schubert, D.~Duran, and I.~Siegert, ``Challenges of german speech recognition: A study on multi-ethnolectal speech among adolescents,'' in \emph{Interspeech 2024}, 2024, pp. 3045--3049.

\bibitem{KM2021}
V.~Silber-Varod, I.~Siegert, O.~Jokisch, Y.~Sinha, and N.~Geri, ``A cross-language study of selected speech recognition systems,'' \emph{The Online Journal of Applied Knowledge Management: OJAKM}, vol.~9, pp. 1 -- 15, 2021.

\bibitem{witt2000phone}
S.~M. Witt and S.~J. Young, ``Phone-level pronunciation scoring and assessment for interactive language learning,'' \emph{Speech communication}, vol.~30, no. 2-3, pp. 95--108, 2000.

\bibitem{hu2015improved}
W.~Hu, Y.~Qian, F.~K. Soong, and Y.~Wang, ``Improved mispronunciation detection with deep neural network trained acoustic models and transfer learning based logistic regression classifiers,'' \emph{Speech Communication}, vol.~67, pp. 154--166, 2015.

\bibitem{fontan2015predicting}
L.~Fontan, T.~Pellegrini, J.~Olcoz, and A.~Abad, ``Predicting disordered speech comprehensibility from goodness of pronunciation scores,'' in \emph{Proc. SLPAT}, 2015, pp. 42--46.

\bibitem{jiao2017interpretable}
Y.~Jiao, V.~Berisha, and J.~Liss, ``Interpretable phonological features for clinical applications,'' in \emph{Proc. ICASSP}, 2017, pp. 5045--5049.

\bibitem{mathad2021deep}
V.~C. Mathad, N.~Scherer, K.~Chapman, J.~M. Liss, and V.~Berisha, ``A deep learning algorithm for objective assessment of hypernasality in children with cleft palate,'' \emph{IEEE Transactions on Biomedical Engineering}, vol.~68, no.~10, pp. 2986--2996, 2021.

\bibitem{mathad2022consonant}
V.~C. Mathad, J.~M. Liss, K.~Chapman, N.~Scherer, and V.~Berisha, ``Consonant-vowel transition models based on deep learning for objective evaluation of articulation,'' \emph{IEEE/ACM transactions on audio, speech, and language processing}, vol.~31, pp. 86--95, 2022.

\bibitem{stevens1981evidence}
K.~N. Stevens, ``Evidence for the role of acoustic boundaries in the perception of speech sounds,'' \emph{The Journal of the Acoustical Society of America}, vol.~69, no.~S1, pp. S116--S116, 1981.

\bibitem{hedrick1993effect}
M.~S. Hedrick and R.~N. Ohde, ``Effect of relative amplitude of frication on perception of place of articulation,'' \emph{The Journal of the Acoustical Society of America}, vol.~94, no.~4, pp. 2005--2026, 1993.

\bibitem{cernak2017characterisation}
M.~Cernak, J.~R. Orozco-Arroyave, F.~Rudzicz, H.~Christensen, J.~C. V{\'a}squez-Correa, and E.~N{\"o}th, ``Characterisation of voice quality of parkinson’s disease using differential phonological posterior features,'' \emph{Computer Speech \& Language}, vol.~46, pp. 196--208, 2017.

\bibitem{liu2019acoustical}
Y.~Liu, T.~Lee, T.~Law, and K.~Y.-S. Lee, ``Acoustical assessment of voice disorder with continuous speech using asr posterior features,'' \emph{IEEE/ACM Transactions on Audio, Speech, and Language Processing}, vol.~27, no.~6, pp. 1047--1059, 2019.

\bibitem{shahin2019anomaly}
M.~Shahin and B.~Ahmed, ``Anomaly detection based pronunciation verification approach using speech attribute features,'' \emph{Speech Communication}, vol. 111, pp. 29--43, 2019.

\bibitem{wang24e_interspeech}
L.~Wang, Y.~Gong, N.~Dawalatabad, M.~Vilela, K.~Placek, B.~Tracey, Y.~Gong, A.~Premasiri, F.~Vieira, and J.~Glass, ``Automatic prediction of amyotrophic lateral sclerosis progression using longitudinal speech transformer,'' in \emph{Interspeech 2024}, 2024, pp. 2000--2004.

\bibitem{dumpala24b_interspeech}
S.~H. Dumpala, K.~Dikaios, A.~Nunes, F.~Rudzicz, R.~Uher, and S.~Oore, ``Self-supervised embeddings for detecting individual symptoms of depression,'' in \emph{Proc. Interspeech}, 2024, pp. 1450--1454.

\bibitem{lee24e_interspeech}
S.~Lee, S.~Kim, and M.~Chung, ``Automatic assessment of speech production skills for children with cochlear implants using wav2vec2.0 acoustic embeddings,'' in \emph{Interspeech 2024}, 2024, pp. 862--866.

\bibitem{shor2022universal}
J.~Shor, A.~Jansen, W.~Han, D.~Park, and Y.~Zhang, ``Universal paralinguistic speech representations using self-supervised conformers,'' in \emph{Proc. ICASSP}.\hskip 1em plus 0.5em minus 0.4em\relax IEEE, 2022, pp. 3169--3173.

\bibitem{shor22_interspeech}
J.~Shor and S.~Venugopalan, ``Trillsson: Distilled universal paralinguistic speech representations,'' in \emph{Proc. Interspeech}, 2022, pp. 356--360.

\bibitem{lee2024distilled}
H.~Lee and A.~Saeed, ``Distilled non-semantic speech embeddings with binary neural networks for low-resource devices,'' \emph{Pattern Recognition Letters}, vol. 177, pp. 15--19, 2024.

\bibitem{shor20_interspeech}
J.~Shor, A.~Jansen, R.~Maor, O.~Lang, O.~Tuval, F.~de~Chaumont~Quitry, M.~Tagliasacchi, I.~Shavitt, D.~Emanuel, and Y.~Haviv, ``Towards learning a universal non-semantic representation of speech,'' in \emph{Proc. Interspeech}, 2020, pp. 140--144.

\bibitem{kent1996hearing}
R.~D. Kent, ``Hearing and believing: Some limits to the auditory-perceptual assessment of speech and voice disorders,'' \emph{American Journal of Speech-Language Pathology}, vol.~5, no.~3, pp. 7--23, 1996.

\bibitem{goldstein2006social}
T.~R. Goldstein, D.~J. Miklowitz, and K.~L. Mullen, ``Social skills knowledge and performance among adolescents with bipolar disorder,'' \emph{Bipolar disorders}, vol.~8, no.~4, pp. 350--361, 2006.

\bibitem{lee2013social}
J.~Lee, L.~Altshuler, D.~C. Glahn, D.~J. Miklowitz, K.~Ochsner, and M.~F. Green, ``Social and nonsocial cognition in bipolar disorder and schizophrenia: relative levels of impairment,'' \emph{American Journal of Psychiatry}, vol. 170, no.~3, pp. 334--341, 2013.

\bibitem{wuyts2000dysphonia}
F.~L. Wuyts, M.~S.~D. Bodt, G.~Molenberghs, M.~Remacle, L.~Heylen, B.~Millet, K.~V. Lierde, J.~Raes, and P.~H. V.~d. Heyning, ``The dysphonia severity index: an objective measure of vocal quality based on a multiparameter approach,'' \emph{Journal of speech, language, and hearing research}, vol.~43, no.~3, pp. 796--809, 2000.

\bibitem{berisha2024responsible}
V.~Berisha and J.~M. Liss, ``Responsible development of clinical speech ai: Bridging the gap between clinical research and technology,'' \emph{NPJ Digital Medicine}, vol.~7, no.~1, p. 208, 2024.

\bibitem{borrie_2012_perceptual}
\BIBentryALTinterwordspacing
S.~A. Borrie, M.~J. McAuliffe, and J.~M. Liss, ``Perceptual learning of dysarthric speech: A review of experimental studies,'' \emph{Journal of Speech, Language, and Hearing Research}, vol.~55, no.~1, pp. 290--305, 2012. [Online]. Available: \url{https://pubs.asha.org/doi/abs/10.1044/1092-4388%282011/10-0349%29}
\BIBentrySTDinterwordspacing

\bibitem{tu2017interpretable}
M.~Tu, V.~Berisha, and J.~Liss, ``Interpretable objective assessment of dysarthric speech based on deep neural networks.'' in \emph{Proc. Interspeech}, 2017, pp. 1849--1853.

\bibitem{helou2010role}
L.~B. Helou, N.~P. Solomon, L.~R. Henry, G.~L. Coppit, R.~S. Howard, and A.~Stojadinovic, ``The role of listener experience on consensus auditory-perceptual evaluation of voice (cape-v) ratings of postthyroidectomy voice,'' \emph{American Journal of Speech-Language Pathology}, vol.~19, no.~3, p. 248, 2010.

\bibitem{patel2018recommended}
R.~R. Patel, S.~N. Awan, J.~Barkmeier-Kraemer, M.~Courey, D.~Deliyski, T.~Eadie, D.~Paul, J.~G. {\v{S}}vec, and R.~Hillman, ``Recommended protocols for instrumental assessment of voice: American speech-language-hearing association expert panel to develop a protocol for instrumental assessment of vocal function,'' \emph{American journal of speech-language pathology}, vol.~27, no.~3, pp. 887--905, 2018.

\bibitem{vsvec2018tutorial}
J.~G. {\v{S}}vec and S.~Granqvist, ``Tutorial and guidelines on measurement of sound pressure level in voice and speech,'' \emph{Journal of Speech, Language, and Hearing Research}, vol.~61, no.~3, pp. 441--461, 2018.

\bibitem{karimi2020deep}
D.~Karimi, H.~Dou, S.~K. Warfield, and A.~Gholipour, ``Deep learning with noisy labels: Exploring techniques and remedies in medical image analysis,'' \emph{Medical image analysis}, vol.~65, p. 101759, 2020.

\bibitem{li2020regularization}
W.~Li, G.~Dasarathy, and V.~Berisha, ``Regularization via structural label smoothing,'' in \emph{International Conference on Artificial Intelligence and Statistics}.\hskip 1em plus 0.5em minus 0.4em\relax PMLR, 2020, pp. 1453--1463.

\bibitem{ma2020normalized}
X.~Ma, H.~Huang, Y.~Wang, S.~Romano, S.~Erfani, and J.~Bailey, ``Normalized loss functions for deep learning with noisy labels,'' in \emph{International conference on machine learning}.\hskip 1em plus 0.5em minus 0.4em\relax PMLR, 2020, pp. 6543--6553.

\bibitem{sauder2017predicting}
C.~Sauder, M.~Bretl, and T.~Eadie, ``Predicting voice disorder status from smoothed measures of cepstral peak prominence using praat and analysis of dysphonia in speech and voice (adsv),'' \emph{Journal of Voice}, vol.~31, no.~5, pp. 557--566, 2017.

\bibitem{xu2023dysarthria}
L.~Xu, J.~Liss, and V.~Berisha, ``Dysarthria detection based on a deep learning model with a clinically-interpretable layer,'' \emph{JASA Express Letters}, vol.~3, no.~1, 2023.

\bibitem{Shapley+1997+69+79}
L.~Shapley, \emph{A Value for n-Person Games.}\hskip 1em plus 0.5em minus 0.4em\relax Princeton University Press, 1997, pp. 69--79.

\bibitem{lundberg2017unified}
S.~M. Lundberg and S.-I. Lee, ``A unified approach to interpreting model predictions,'' \emph{Proc. NeurIPS}, p. 4768–4777, 2017.

\bibitem{yeung2021correlating}
A.~Yeung, A.~Iaboni, E.~Rochon, M.~Lavoie, C.~Santiago, M.~Yancheva, J.~Novikova, M.~Xu, J.~Robin, L.~D. Kaufman \emph{et~al.}, ``Correlating natural language processing and automated speech analysis with clinician assessment to quantify speech-language changes in mild cognitive impairment and alzheimer’s dementia,'' \emph{Alzheimer's research \& therapy}, vol.~13, no.~1, p. 109, 2021.

\bibitem{mikolov2013distributed}
T.~Mikolov, I.~Sutskever, K.~Chen, G.~S. Corrado, and J.~Dean, ``Distributed representations of words and phrases and their compositionality,'' \emph{Advances in neural information processing systems}, vol.~26, 2013.

\bibitem{pennington2014glove}
J.~Pennington, R.~Socher, and C.~D. Manning, ``Glove: Global vectors for word representation,'' in \emph{Proceedings of the 2014 conference on empirical methods in natural language processing (EMNLP)}, 2014, pp. 1532--1543.

\bibitem{kenton2019bert}
J.~D. M.-W.~C. Kenton and L.~K. Toutanova, ``Bert: Pre-training of deep bidirectional transformers for language understanding,'' in \emph{Proceedings of NAACL-HLT}, 2019, pp. 4171--4186.

\bibitem{lundin2022semantic}
N.~B. Lundin, M.~N. Jones, E.~J. Myers, A.~Breier, and K.~S. Minor, ``Semantic and phonetic similarity of verbal fluency responses in early-stage psychosis,'' \emph{Psychiatry research}, vol. 309, p. 114404, 2022.

\bibitem{docherty1996communication}
N.~M. Docherty, M.~DeRosa, and N.~C. Andreasen, ``Communication disturbances in schizophrenia and mania,'' \emph{Archives of General Psychiatry}, vol.~53, no.~4, pp. 358--364, 1996.

\bibitem{xu2020centroid}
W.~Xu, J.~Portanova, A.~Chander, D.~Ben-Zeev, and T.~Cohen, ``The centroid cannot hold: comparing sequential and global estimates of coherence as indicators of formal thought disorder,'' in \emph{AMIA Annual Symposium Proceedings}, vol. 2020.\hskip 1em plus 0.5em minus 0.4em\relax American Medical Informatics Association, 2020, p. 1315.

\bibitem{joulin2016fasttext}
A.~Joulin, E.~Grave, P.~Bojanowski, M.~Douze, H.~J{\'e}gou, and T.~Mikolov, ``Fasttext.zip: Compressing text classification models,'' \emph{arXiv preprint arXiv:1612.03651}, 2016.

\bibitem{tang2023clinical}
S.~X. Tang, Y.~Cong, A.~H. Nikzad, A.~Mehta, S.~Cho, K.~H{\"a}nsel, S.~Berretta, A.~A. Dhar, J.~M. Kane, and A.~K. Malhotra, ``Clinical and computational speech measures are associated with social cognition in schizophrenia spectrum disorders,'' \emph{Schizophrenia Research}, vol. 259, pp. 28--37, 2023.

\bibitem{vasquez2021transfer}
J.~C. V{\'a}squez-Correa, C.~D. Rios-Urrego, T.~Arias-Vergara, M.~Schuster, J.~Rusz, E.~Noeth, and J.~R. Orozco-Arroyave, ``Transfer learning helps to improve the accuracy to classify patients with different speech disorders in different languages,'' \emph{Pattern Recognition Letters}, vol. 150, pp. 272--279, 2021.

\bibitem{wang2015transfer}
D.~Wang and T.~F. Zheng, ``Transfer learning for speech and language processing,'' in \emph{2015 Asia-Pacific Signal and Information Processing Association Annual Summit and Conference (APSIPA)}.\hskip 1em plus 0.5em minus 0.4em\relax IEEE, 2015, pp. 1225--1237.

\bibitem{bailey2021gender}
A.~Bailey and M.~D. Plumbley, ``Gender bias in depression detection using audio features,'' in \emph{2021 29th European Signal Processing Conference (EUSIPCO)}.\hskip 1em plus 0.5em minus 0.4em\relax IEEE, 2021, pp. 596--600.

\bibitem{yang2024deconstructing}
M.~Yang, A.-A. El-Attar, and T.~Chaspari, ``Deconstructing demographic bias in speech-based machine learning models for digital health,'' \emph{Frontiers in Digital Health}, vol.~6, p. 1351637, 2024.

\bibitem{rusz2021reproducibility}
J.~Rusz, J.~{\v{S}}vihl{\'\i}k, P.~Kr{\`y}{\v{z}}e, M.~Novotn{\`y}, and T.~Tykalov{\'a}, ``Reproducibility of voice analysis with machine learning,'' \emph{Movement Disorders}, vol.~36, no.~5, pp. 1282--1283, 2021.

\bibitem{zhang2023robust}
J.~Zhang, J.~Liss, S.~Jayasuriya, and V.~Berisha, ``Robust vocal quality feature embeddings for dysphonic voice detection,'' \emph{IEEE/ACM transactions on audio, speech, and language processing}, vol.~31, pp. 1348--1359, 2023.

\bibitem{ko2017study}
T.~Ko, V.~Peddinti, D.~Povey, M.~L. Seltzer, and S.~Khudanpur, ``A study on data augmentation of reverberant speech for robust speech recognition,'' in \emph{Proc. ICASSP}.\hskip 1em plus 0.5em minus 0.4em\relax IEEE, 2017, pp. 5220--5224.

\bibitem{vachhani18_interspeech}
B.~Vachhani, C.~Bhat, and S.~K. Kopparapu, ``Data augmentation using healthy speech for dysarthric speech recognition,'' in \emph{Proc. Interspeech}, 2018, pp. 471--475.

\bibitem{shahnawazuddin2020domain}
S.~Shahnawazuddin, W.~Ahmad, N.~Adiga, and A.~Kumar, ``In-domain and out-of-domain data augmentation to improve children’s speaker verification system in limited data scenario,'' in \emph{Proc. ICASSP}.\hskip 1em plus 0.5em minus 0.4em\relax IEEE, 2020, pp. 7554--7558.

\bibitem{prananta22_interspeech}
L.~Prananta, B.~Halpern, S.~Feng, and O.~Scharenborg, ``The effectiveness of time stretching for enhancing dysarthric speech for improved dysarthric speech recognition,'' in \emph{Proc. Interspeech}, 2022, pp. 36--40.

\bibitem{zhang2024learning}
J.~Zhang, S.~Jayasuriya, and V.~Berisha, ``Learning repeatable speech embeddings using an intra-class correlation regularizer,'' \emph{Advances in Neural Information Processing Systems}, vol.~36, 2024.

\bibitem{zhou2021isobn}
W.~Zhou, B.~Y. Lin, and X.~Ren, ``Isobn: Fine-tuning bert with isotropic batch normalization,'' in \emph{Proceedings of the AAAI Conference on Artificial Intelligence}, vol.~35, no.~16, 2021, pp. 14\,621--14\,629.

\bibitem{peyser22_interspeech}
C.~Peyser, W.~R. Huang, A.~Rosenberg, T.~Sainath, M.~Picheny, and K.~Cho, ``Towards disentangled speech representations,'' in \emph{Proc. Interspeech}, 2022, pp. 3603--3607.

\bibitem{gao2018representation}
J.~Gao, D.~He, X.~Tan, T.~Qin, L.~Wang, and T.~Liu, ``Representation degeneration problem in training natural language generation models,'' in \emph{Proc. ICLR}, 2019.

\bibitem{xu2023decorrelating}
L.~Xu, K.~D. Mueller, J.~Liss, and V.~Berisha, ``Decorrelating language model embeddings for speech-based prediction of cognitive impairment,'' in \emph{Proc. ICASSP}.\hskip 1em plus 0.5em minus 0.4em\relax IEEE, 2023, pp. 1--5.

\bibitem{perero2019modeling}
J.~M. Perero-Codosero, F.~Espinoza-Cuadros, J.~Ant{\'o}n-Mart{\'\i}n, M.~A. Barbero-Alvarez, and L.~A. Hern{\'a}ndez-G{\'o}mez, ``Modeling obstructive sleep apnea voices using deep neural network embeddings and domain-adversarial training,'' \emph{IEEE Journal of Selected Topics in Signal Processing}, vol.~14, no.~2, pp. 240--250, 2019.

\bibitem{park2023adversarial}
D.~Park, Y.~Yu, D.~Katabi, and H.~K. Kim, ``Adversarial continual learning to transfer self-supervised speech representations for voice pathology detection,'' \emph{IEEE Signal Processing Letters}, 2023.

\bibitem{hsu2018robustness}
Y.-T. Hsu, Z.~Zhu, C.-T. Wang, S.-H. Fang, F.~Rudzicz, and Y.~Tsao, ``Robustness against the channel effect in pathological voice detection,'' \emph{arXiv preprint arXiv:1811.10376}, 2018.

\bibitem{amiri2024test}
M.~Amiri and I.~Kodrasi, ``Test-time adaptation for automatic pathological speech detection in noisy environments,'' in \emph{Proc. European Signal Processing Conference, Lyon, France}, 2024.

\bibitem{chen2019develop}
P.-H.~C. Chen, Y.~Liu, and L.~Peng, ``How to develop machine learning models for healthcare,'' \emph{Nature materials}, vol.~18, no.~5, pp. 410--414, 2019.

\bibitem{schuller2014interspeech}
B.~Schuller, S.~Steidl, A.~Batliner, J.~Epps, F.~Eyben, F.~Ringeval, E.~Marchi, and Y.~Zhang, ``The interspeech 2014 computational paralinguistics challenge: Cognitive \& physical load, multitasking,'' in \emph{Proc. Interspeech}, 2014.

\bibitem{berisha2021digital}
V.~Berisha, C.~Krantsevich, P.~R. Hahn, S.~Hahn, G.~Dasarathy, P.~Turaga, and J.~Liss, ``Digital medicine and the curse of dimensionality,'' \emph{NPJ digital medicine}, vol.~4, no.~1, p. 153, 2021.

\bibitem{robin2020evaluation}
J.~Robin, J.~E. Harrison, L.~D. Kaufman, F.~Rudzicz, W.~Simpson, and M.~Yancheva, ``Evaluation of speech-based digital biomarkers: review and recommendations,'' \emph{Digital Biomarkers}, vol.~4, no.~3, pp. 99--108, 2020.

\bibitem{quinn2022three}
T.~P. Quinn, S.~Jacobs, M.~Senadeera, V.~Le, and S.~Coghlan, ``The three ghosts of medical ai: Can the black-box present deliver?'' \emph{Artificial intelligence in medicine}, vol. 124, p. 102158, 2022.

\bibitem{montavon2019layer}
G.~Montavon, A.~Binder, S.~Lapuschkin, W.~Samek, and K.-R. M{\"u}ller, ``Layer-wise relevance propagation: an overview,'' \emph{Explainable AI: interpreting, explaining and visualizing deep learning}, pp. 193--209, 2019.

\bibitem{zhou2016learning}
B.~Zhou, A.~Khosla, A.~Lapedriza, A.~Oliva, and A.~Torralba, ``Learning deep features for discriminative localization,'' in \emph{Proceedings of the IEEE conference on computer vision and pattern recognition}, 2016, pp. 2921--2929.

\bibitem{ribeiro2016should}
M.~T. Ribeiro, S.~Singh, and C.~Guestrin, ``" why should i trust you?" explaining the predictions of any classifier,'' in \emph{Proceedings of the 22nd ACM SIGKDD international conference on knowledge discovery and data mining}, 2016, pp. 1135--1144.

\bibitem{rudin2019stop}
C.~Rudin, ``Stop explaining black box machine learning models for high stakes decisions and use interpretable models instead,'' \emph{Nature machine intelligence}, vol.~1, no.~5, pp. 206--215, 2019.

\bibitem{alvarez2018towards}
D.~Alvarez~Melis and T.~Jaakkola, ``Towards robust interpretability with self-explaining neural networks,'' \emph{Advances in neural information processing systems}, vol.~31, 2018.

\bibitem{stegmann2021estimation}
G.~M. Stegmann, S.~Hahn, C.~J. Duncan, S.~B. Rutkove, J.~Liss, J.~M. Shefner, and V.~Berisha, ``Estimation of forced vital capacity using speech acoustics in patients with als,'' \emph{Amyotrophic Lateral Sclerosis and Frontotemporal Degeneration}, vol.~22, no. sup1, pp. 14--21, 2021.

\bibitem{stegmann2022automated}
G.~Stegmann, S.~Hahn, S.~Bhandari, K.~Kawabata, J.~Shefner, C.~J. Duncan, J.~Liss, V.~Berisha, and K.~Mueller, ``Automated semantic relevance as an indicator of cognitive decline: Out-of-sample validation on a large-scale longitudinal dataset,'' \emph{Alzheimer's \& Dementia: Diagnosis, Assessment \& Disease Monitoring}, vol.~14, no.~1, p. e12294, 2022.

\bibitem{holmlund2020applying}
T.~B. Holmlund, C.~Chandler, P.~W. Foltz, A.~S. Cohen, J.~Cheng, J.~C. Bernstein, E.~P. Rosenfeld, and B.~Elvev{\aa}g, ``Applying speech technologies to assess verbal memory in patients with serious mental illness,'' \emph{NPJ digital medicine}, vol.~3, no.~1, p.~33, 2020.

\bibitem{semenova2022existence}
L.~Semenova, C.~Rudin, and R.~Parr, ``On the existence of simpler machine learning models,'' in \emph{Proceedings of the 2022 ACM Conference on Fairness, Accountability, and Transparency}, 2022, pp. 1827--1858.

\bibitem{achiam2023gpt}
J.~Achiam, S.~Adler, S.~Agarwal, L.~Ahmad, I.~Akkaya, F.~L. Aleman, D.~Almeida, J.~Altenschmidt, S.~Altman, S.~Anadkat \emph{et~al.}, ``Gpt-4 technical report,'' \emph{arXiv preprint arXiv:2303.08774}, 2023.

\bibitem{touvron2023llama}
H.~Touvron, T.~Lavril, G.~Izacard, X.~Martinet, M.-A. Lachaux, T.~Lacroix, B.~Rozi{\`e}re, N.~Goyal, E.~Hambro, F.~Azhar \emph{et~al.}, ``Llama: Open and efficient foundation language models,'' \emph{arXiv preprint arXiv:2302.13971}, 2023.

\bibitem{nori2023capabilities}
H.~Nori, N.~King, S.~M. McKinney, D.~Carignan, and E.~Horvitz, ``Capabilities of gpt-4 on medical challenge problems,'' \emph{arXiv preprint arXiv:2303.13375}, 2023.

\bibitem{kasai2023evaluating}
J.~Kasai, Y.~Kasai, K.~Sakaguchi, Y.~Yamada, and D.~Radev, ``Evaluating gpt-4 and chatgpt on japanese medical licensing examinations,'' \emph{arXiv preprint arXiv:2303.18027}, 2023.

\bibitem{rosol2023evaluation}
M.~Roso{\l}, J.~S. Gasior, J.~{\L}aba, K.~Korzeniewski, and M.~M{\l}y{\'n}czak, ``Evaluation of the performance of gpt-3.5 and gpt-4 on the polish medical final examination,'' \emph{Scientific Reports}, vol.~13, no.~1, p. 20512, 2023.

\bibitem{jin2024hidden}
Q.~Jin, F.~Chen, Y.~Zhou, Z.~Xu, J.~M. Cheung, R.~Chen, R.~M. Summers, J.~F. Rousseau, P.~Ni, M.~J. Landsman \emph{et~al.}, ``Hidden flaws behind expert-level accuracy of multimodal gpt-4 vision in medicine,'' \emph{ArXiv}, pp. arXiv--2401, 2024.

\bibitem{bektacs2024chatgpt}
M.~Bekta{\c{s}}, J.~K. Pereira, F.~Daams, and D.~L. van~der Peet, ``Chatgpt in surgery: a revolutionary innovation?'' \emph{Surgery today}, vol.~54, no.~8, pp. 964--971, 2024.

\bibitem{fink2023potential}
M.~A. Fink, A.~Bischoff, C.~A. Fink, M.~Moll, J.~Kroschke, L.~Dulz, C.~P. Heu{\ss}el, H.-U. Kauczor, and T.~F. Weber, ``Potential of chatgpt and gpt-4 for data mining of free-text ct reports on lung cancer,'' \emph{Radiology}, vol. 308, no.~3, p. e231362, 2023.

\bibitem{wang2023text}
C.~Wang, S.~Liu, A.~Li, and J.~Liu, ``Text dialogue analysis for primary screening of mild cognitive impairment: Development and validation study,'' \emph{Journal of Medical Internet Research}, vol.~25, p. e51501, 2023.

\bibitem{fei2024evaluating}
X.~Fei, Y.~Tang, J.~Zhang, Z.~Zhou, I.~Yamamoto, and Y.~Zhang, ``Evaluating cognitive performance: Traditional methods vs. chatgpt,'' \emph{Digital Health}, vol.~10, p. 20552076241264639, 2024.

\bibitem{botelho24_interspeech}
C.~Botelho, J.~Mendonça, A.~Pompili, T.~Schultz, A.~Abad, and I.~Trancoso, ``Macro-descriptors for alzheimer's disease detection using large language models,'' in \emph{Proc. Interspeech}, 2024, pp. 1975--1979.

\bibitem{hristidis2023chatgpt}
V.~Hristidis, N.~Ruggiano, E.~L. Brown, S.~R.~R. Ganta, and S.~Stewart, ``Chatgpt vs google for queries related to dementia and other cognitive decline: comparison of results,'' \emph{Journal of Medical Internet Research}, vol.~25, p. e48966, 2023.

\bibitem{wang2023can}
Z.~Wang, R.~Li, B.~Dong, J.~Wang, X.~Li, N.~Liu, C.~Mao, W.~Zhang, L.~Dong, J.~Gao \emph{et~al.}, ``Can llms like gpt-4 outperform traditional ai tools in dementia diagnosis? maybe, but not today,'' \emph{arXiv preprint arXiv:2306.01499}, 2023.

\bibitem{perlis2023application}
R.~H. Perlis, ``Application of gpt-4 to select next-step antidepressant treatment in major depression,'' \emph{MedRxiv}, 2023.

\bibitem{mohamad2023chatgpt}
E.~Mohamad, C.~Boutoleau-Bretonni{\`e}re, and G.~Chapelet, ``Chatgpt's dance with neuropsychological data: a case study in alzheimer’s disease,'' \emph{Ageing Research Reviews}, p. 102117, 2023.

\bibitem{us2022artificial}
U.~Food, D.~Administration \emph{et~al.}, ``Artificial intelligence and machine learning (ai/ml)-enabled medical devices,'' \emph{AI/ML-Enabled Medical Devices}, 2022.

\bibitem{vela2022temporal}
D.~Vela, A.~Sharp, R.~Zhang, T.~Nguyen, A.~Hoang, and O.~S. Pianykh, ``Temporal quality degradation in ai models,'' \emph{Scientific Reports}, vol.~12, no.~1, p. 11654, 2022.

\bibitem{wong2021quantification}
A.~Wong, J.~Cao, P.~G. Lyons, S.~Dutta, V.~J. Major, E.~{\"O}tle{\c{s}}, and K.~Singh, ``Quantification of sepsis model alerts in 24 us hospitals before and during the covid-19 pandemic,'' \emph{JAMA Network Open}, vol.~4, no.~11, pp. e2\,135\,286--e2\,135\,286, 2021.

\bibitem{wang2024performance}
B.~Wang, M.~Dohopolski, T.~Bai, J.~Wu, R.~Hannan, N.~Desai, A.~Garant, D.~Yang, D.~Nguyen, M.-H. Lin \emph{et~al.}, ``Performance deterioration of deep learning models after clinical deployment: a case study with auto-segmentation for definitive prostate cancer radiotherapy,'' \emph{Machine Learning: Science and Technology}, vol.~5, no.~2, p. 025077, 2024.

\bibitem{cao2023comparative}
J.~Cao, A.~Ganesh, J.~Cai, R.~Southwell, E.~M. Perkoff, M.~Regan, K.~Kann, J.~H. Martin, M.~Palmer, and S.~D'Mello, ``A comparative analysis of automatic speech recognition errors in small group classroom discourse,'' in \emph{Proceedings of the 31st ACM Conference on User Modeling, Adaptation and Personalization}, 2023, pp. 250--262.

\bibitem{vaidya2024demographic}
A.~Vaidya, R.~J. Chen, D.~F. Williamson, A.~H. Song, G.~Jaume, Y.~Yang, T.~Hartvigsen, E.~C. Dyer, M.~Y. Lu, J.~Lipkova \emph{et~al.}, ``Demographic bias in misdiagnosis by computational pathology models,'' \emph{Nature Medicine}, vol.~30, no.~4, pp. 1174--1190, 2024.

\bibitem{straw2020artificial}
I.~Straw and C.~Callison-Burch, ``Artificial intelligence in mental health and the biases of language based models,'' \emph{PloS one}, vol.~15, no.~12, p. e0240376, 2020.

\bibitem{lu2021deploying}
C.~Lu, K.~Chang, P.~Singh, S.~Pomerantz, S.~Doyle, S.~Kakarmath, C.~Bridge, and J.~Kalpathy-Cramer, ``Deploying clinical machine learning? consider the following...'' \emph{arXiv preprint arXiv:2109.06919}, 2021.

\bibitem{yu2018request}
S.~Yu, X.~Wang, and J.~C. Pr{\'\i}ncipe, ``Request-and-reverify: Hierarchical hypothesis testing for concept drift detection with expensive labels,'' pp. 3033--3039, 2018.

\bibitem{ginart2022mldemon}
T.~Ginart, M.~J. Zhang, and J.~Zou, ``Mldemon: Deployment monitoring for machine learning systems,'' in \emph{International conference on artificial intelligence and statistics}.\hskip 1em plus 0.5em minus 0.4em\relax PMLR, 2022, pp. 3962--3997.

\bibitem{ovadia2019can}
Y.~Ovadia, E.~Fertig, J.~Ren, Z.~Nado, D.~Sculley, S.~Nowozin, J.~Dillon, B.~Lakshminarayanan, and J.~Snoek, ``Can you trust your model's uncertainty? evaluating predictive uncertainty under dataset shift,'' \emph{Advances in neural information processing systems}, vol.~32, 2019.

\bibitem{koch2024distribution}
L.~M. Koch, C.~F. Baumgartner, and P.~Berens, ``Distribution shift detection for the postmarket surveillance of medical ai algorithms: a retrospective simulation study,'' \emph{NPJ Digital Medicine}, vol.~7, no.~1, p. 120, 2024.

\bibitem{ma2022test}
W.~Ma, C.~Chen, S.~Zheng, J.~Qin, H.~Zhang, and Q.~Dou, ``Test-time adaptation with calibration of medical image classification nets for label distribution shift,'' in \emph{International Conference on Medical Image Computing and Computer-Assisted Intervention}.\hskip 1em plus 0.5em minus 0.4em\relax Springer, 2022, pp. 313--323.

\bibitem{davis2020detection}
S.~E. Davis, R.~A. Greevy~Jr, T.~A. Lasko, C.~G. Walsh, and M.~E. Matheny, ``Detection of calibration drift in clinical prediction models to inform model updating,'' \emph{Journal of biomedical informatics}, vol. 112, p. 103611, 2020.

\bibitem{leschanowsky24_spsc}
A.~Leschanowsky and S.~Das, ``Examining the interplay between privacy and fairness for speech processing: A review and perspective,'' in \emph{4th Symposium on Security and Privacy in Speech Communication}, 2024, pp. 1--11.

\bibitem{gelfer2005relative}
M.~P. Gelfer and V.~A. Mikos, ``The relative contributions of speaking fundamental frequency and formant frequencies to gender identification based on isolated vowels,'' \emph{Journal of voice}, vol.~19, no.~4, pp. 544--554, 2005.

\bibitem{reubold2010vocal}
U.~Reubold, J.~Harrington, and F.~Kleber, ``Vocal aging effects on f0 and the first formant: A longitudinal analysis in adult speakers,'' \emph{Speech communication}, vol.~52, no. 7-8, pp. 638--651, 2010.

\bibitem{berg2017speaking}
M.~Berg, M.~Fuchs, K.~Wirkner, M.~Loeffler, C.~Engel, and T.~Berger, ``The speaking voice in the general population: normative data and associations to sociodemographic and lifestyle factors,'' \emph{Journal of Voice}, vol.~31, no.~2, pp. 257--e13, 2017.

\bibitem{bertelsen2018sociodemographic}
C.~Bertelsen, S.~Zhou, E.~R. Hapner, and M.~M. Johns, ``Sociodemographic characteristics and treatment response among aging adults with voice disorders in the united states,'' \emph{JAMA Otolaryngology--Head \& Neck Surgery}, vol. 144, no.~8, pp. 719--726, 2018.

\bibitem{clopper2011effects}
C.~G. Clopper and R.~Smiljanic, ``Effects of gender and regional dialect on prosodic patterns in american english,'' \emph{Journal of phonetics}, vol.~39, no.~2, pp. 237--245, 2011.

\bibitem{feng2024towards}
S.~Feng, B.~M. Halpern, O.~Kudina, and O.~Scharenborg, ``Towards inclusive automatic speech recognition,'' \emph{Computer Speech \& Language}, vol.~84, p. 101567, 2024.

\bibitem{hutiri2022bias}
W.~T. Hutiri and A.~Y. Ding, ``Bias in automated speaker recognition,'' in \emph{Proceedings of the 2022 ACM Conference on Fairness, Accountability, and Transparency}, 2022, pp. 230--247.

\bibitem{kaushal2020geographic}
A.~Kaushal, R.~Altman, and C.~Langlotz, ``Geographic distribution of us cohorts used to train deep learning algorithms,'' \emph{Jama}, vol. 324, no.~12, pp. 1212--1213, 2020.

\bibitem{wolf2019huggingface}
T.~Wolf, L.~Debut, V.~Sanh, J.~Chaumond, C.~Delangue, A.~Moi, P.~Cistac, T.~Rault, R.~Louf, M.~Funtowicz \emph{et~al.}, ``Huggingface's transformers: State-of-the-art natural language processing,'' \emph{arXiv preprint arXiv:1910.03771}, 2019.

\bibitem{li2023defend}
K.~Li, C.~Baird, and D.~Lin, ``Defend data poisoning attacks on voice authentication,'' \emph{IEEE Transactions on Dependable and Secure Computing}, 2023.

\bibitem{oliynyk2023know}
D.~Oliynyk, R.~Mayer, and A.~Rauber, ``I know what you trained last summer: A survey on stealing machine learning models and defences,'' \emph{ACM Computing Surveys}, vol.~55, no. 14s, pp. 1--41, 2023.

\bibitem{VERDE20212624}
\BIBentryALTinterwordspacing
L.~Verde, F.~Marulli, and S.~Marrone, ``Exploring the impact of data poisoning attacks on machine learning model reliability,'' \emph{Procedia Computer Science}, vol. 192, pp. 2624--2632, 2021, knowledge-Based and Intelligent Information \& Engineering Systems: Proceedings of the 25th International Conference KES2021. [Online]. Available: \url{https://www.sciencedirect.com/science/article/pii/S1877050921017695}
\BIBentrySTDinterwordspacing

\bibitem{kwon2019selective}
H.~Kwon, Y.~Kim, H.~Yoon, and D.~Choi, ``Selective audio adversarial example in evasion attack on speech recognition system,'' \emph{IEEE Transactions on Information Forensics and Security}, vol.~15, pp. 526--538, 2019.

\bibitem{zelasko2021adversarial}
P.~{\.Z}elasko, S.~Joshi, Y.~Shao, J.~Villalba, J.~Trmal, N.~Dehak, and S.~Khudanpur, ``Adversarial attacks and defenses for speech recognition systems,'' \emph{arXiv preprint arXiv:2103.17122}, 2021.

\bibitem{pizzi2023introducing}
K.~Pizzi, F.~Boenisch, U.~Sahin, and K.~B{\"o}ttinger, ``Introducing model inversion attacks on automatic speaker recognition,'' \emph{arXiv preprint arXiv:2301.03206}, 2023.

\bibitem{shah21_interspeech}
M.~A. Shah, J.~Szurley, M.~Mueller, A.~Mouchtaris, and J.~Droppo, ``{Evaluating the Vulnerability of End-to-End Automatic Speech Recognition Models to Membership Inference Attacks},'' in \emph{Proc. Interspeech}, 2021, pp. 891--895.

\bibitem{TOMASHENKO2022101362}
\BIBentryALTinterwordspacing
N.~Tomashenko, X.~Wang, E.~Vincent, J.~Patino, B.~M.~L. Srivastava, P.-G. Noé, A.~Nautsch, N.~Evans, J.~Yamagishi, B.~O’Brien, A.~Chanclu, J.-F. Bonastre, M.~Todisco, and M.~Maouche, ``The voiceprivacy 2020 challenge: Results and findings,'' \emph{Computer Speech \& Language}, vol.~74, p. 101362, 2022. [Online]. Available: \url{https://www.sciencedirect.com/science/article/pii/S0885230822000080}
\BIBentrySTDinterwordspacing

\bibitem{rahman24_spsc}
M.~U. Rahman, M.~Larson, L.~ten Bosch, and C.~Tejedor-García, ``Scenario of use scheme: Threat modelling for speaker privacy protection in the medical domain,'' in \emph{4th Symposium on Security and Privacy in Speech Communication}, 2024, pp. 21--25.

\bibitem{ghosh24_interspeech}
S.~Ghosh, M.~Jouaiti, A.~Das, Y.~Sinha, T.~Polzehl, I.~Siegert, and S.~Stober, ``Anonymising elderly and pathological speech: Voice conversion using ddsp and query-by-example,'' in \emph{Proc. Interspeech}, 2024, pp. 4438--4442.

\bibitem{10.1145/3411495.3421355}
\BIBentryALTinterwordspacing
R.~Aloufi, H.~Haddadi, and D.~Boyle, ``Privacy-preserving voice analysis via disentangled representations,'' in \emph{Proceedings of the 2020 ACM SIGSAC Conference on Cloud Computing Security Workshop}, ser. CCSW'20.\hskip 1em plus 0.5em minus 0.4em\relax New York, NY, USA: Association for Computing Machinery, 2020, p. 1–14. [Online]. Available: \url{https://doi.org/10.1145/3411495.3421355}
\BIBentrySTDinterwordspacing

\bibitem{orabi2018deep}
A.~H. Orabi, P.~Buddhitha, M.~H. Orabi, and D.~Inkpen, ``Deep learning for depression detection of twitter users,'' in \emph{Proceedings of the fifth workshop on computational linguistics and clinical psychology: from keyboard to clinic}, 2018, pp. 88--97.

\bibitem{petti2023much}
U.~Petti, S.~Baker, A.~Korhonen, and J.~Robin, ``How much speech data is needed for tracking language change in alzheimer’s disease? a comparison of random length, 5-min, and 1-min spontaneous speech samples,'' \emph{Digital Biomarkers}, vol.~7, no.~1, pp. 157--166, 2023.

\bibitem{halpern2023automatic}
B.~M. Halpern, S.~Feng, R.~van Son, M.~van~den Brekel, and O.~Scharenborg, ``Automatic evaluation of spontaneous oral cancer speech using ratings from naive listeners,'' \emph{Speech Communication}, vol. 149, pp. 84--97, 2023.

\end{thebibliography}

\end{document}